\documentclass[preprint,3p]{elsarticle}

\makeatletter
\def\ps@pprintTitle{%
 \let\@oddhead\@empty
 \let\@evenhead\@empty
 \def\@oddfoot{\centerline{\thepage}}%
 \let\@evenfoot\@oddfoot}
\makeatother

\usepackage{graphics}
\usepackage{graphicx}
\usepackage{subfig}
\usepackage{epsfig}
\usepackage{amsmath,amssymb,esint}
\usepackage{amsthm}
\usepackage{float}   
\usepackage{caption}
\usepackage{multirow}
\usepackage[colorlinks]{hyperref}

\usepackage{tikz}
\usetikzlibrary{calc}
\usetikzlibrary{arrows.meta}
\tikzset{>={Latex[width=1mm,length=1mm]}}
\usetikzlibrary{shapes.geometric, arrows}

\biboptions{sort&compress}
\usepackage[ddmmyyyy,nodayofweek]{datetime}

\biboptions{authoryear}

\begin{document}

\begin{frontmatter}

\title{Numerical modelling of shock-bubble interactions using a
pressure-based algorithm without Riemann solvers}

\author{Fabian Denner\corref{cor1}}
\ead{fabian.denner@ovgu.de}
\author{Berend G.M.~van Wachem\corref{cor2}}

\address{Chair of Mechanical Process Engineering,
Otto-von-Guericke-Universit\"{a}t Magdeburg,\\ Universit\"atsplatz 2, 39106
Magdeburg, Germany \\ \vspace{5mm}  \vspace{-10mm}}

\cortext[cor1]{Corresponding author: }

\begin{abstract}
The interaction of a shock wave with a bubble features in many engineering and emerging technological applications, and has been used widely to test new numerical methods for compressible interfacial flows. Recently, density-based algorithms with pressure-correction methods as well as fully-coupled pressure-based algorithms have been established as promising alternatives to classical density-based algorithms based on Riemann solvers. The current paper investigates the predictive accuracy of fully-coupled pressure-based algorithms without Riemann solvers in modelling the interaction of shock waves with one-dimensional and two-dimensional bubbles in gas-gas and liquid-gas flows. For a gas bubble suspended in another gas, the mesh resolution and the applied advection schemes are found to only have a minor influence on the bubble shape and position, as well as the behaviour of the dominant shock waves and rarefaction fans. For a gas bubble suspended in a liquid, however, the mesh resolution has a critical influence on the shape, the position and the post-shock evolution of the bubble, as well as the pressure and temperature distribution.
\end{abstract}
\end{frontmatter}

\section{Introduction}
The interaction of a shock wave with a bubble is a process of broad academic and engineering interest, with applications in combustion and detonation \citep{Michael2019}, medical applications, {\em e.g.}~shock-wave lithotripsy \citep{Johnsen2007, Pan2018}, in geophysics \citep{Delale2013} and in microfluidics \citep{Ando2012, Ohl2016}, featuring a rich variety of fluid dynamic and thermodynamic phenomena, such as compression and expansion waves, strong local heating, cavitation, evaporation and condensation, as well as the production of vorticity and turbulence \citep{Ranjan2011, Delale2013}. Shock-bubble interaction in a gas-gas flow is also used to study the interaction of shocks with gas inhomogeneities, whereas the shock-induced collapse of gas bubbles in liquids is of direct relevance to cavitating flows \citep{Delale2013, Ohl2016, Fuster2019}. As a result, the fluid dynamics of shock-bubble interactions have been studied extensively, both experimentally \citep{Haas1987, Layes2003, Layes2005, Ranjan2007, Zhai2011} and computationally \citep{Quirk1996, Bagabir2001, Johnsen2007, Niederhaus2008, Niederhaus2008a, Johnsen2009, Zhai2011, Hejazialhosseini2013, Xiang2017, Pan2018, Yoo2018, Michael2019}. In addition, shock-bubble interactions have also been widely used as a canonical reference system to test and scrutinise new numerical schemes, see {\em e.g.}~\citep{Saurel1999a, Allaire2002, Hu2004, Nourgaliev2006, Chang2007, Terashima2009, Kokh2010, Shukla2010, Shukla2014, Wong2017, Denner2018b}.

Computational fluid dynamics has assumed an increasingly prominent role for the study and analysis of compressible interfacial flows, and especially for the study of shock-bubble interactions, over the past decades, as a result of rapidly advancing developments of the relevant numerical algorithms as well as the substantial computational resources routinely available nowadays. The numerical modelling of compressible interfacial flows thereby requires a consistent numerical treatment of the fluid interface that retains the main features of the solution, in particular the propagation of pressure waves \citep{Abgrall2003, Coralic2014, Denner2018b}. However, the typically sharp change in Mach number at the fluid interface and the associated change in dominant physical mechanisms lead to distinct, and often contrasting, numerical requirements, which complicate the accurate and robust modelling of compressible fluid phenomena, such as the interaction of shock waves with gas bubbles. For instance, while the numerical algorithm has to ensure that the density is independent of the pressure and recovers a divergence-free velocity field in the incompressible limit \citep{Chorin1993, Hauke1998}, the accurate prediction of shock waves requires a conservative discretisation of the governing conservation laws \citep{Hou1994}.

Contemporary numerical methods for compressible interfacial flows are typically predicated on {\em density-based} algorithms, where the governing conservation equations are solved for the density, momentum and total energy of the flow \citep{Baer1986, Allaire2002, Murrone2005, Coralic2014}. In these models, an exact or approximate Riemann solver is usually applied to evaluate the fluxes, with HLLC-type solvers \citep{Toro1994} having gained particular popularity for interfacial flows \citep{Shyue2006, Tokareva2010, Tian2011, Coralic2014}. The Ghost-Fluid method (GFM) \citep{Fedkiw1999} has established itself as a promising alternative to solving a Riemann problem \citep{Fedkiw1999a, Terashima2009, Bo2014}, with recent extensions to improve the stability of simulations with strong shock-interface interactions and compressible gas-liquid flows \citep{Liu2003, Wang2006a, Liu2017}. While density-based algorithms are naturally suited for compressible flows, they are poorly suited for low-Mach number flows \citep{Chorin1967, Karimian1994, Wesseling2001, Cordier2012}, where the coupling of pressure and density vanishes. Traditionally, tailored pre-conditioning techniques have been applied to extend density-based methods to low-Mach number flows \citep{Turkel1993, Turkel2006}, which are however computationally very expensive for transient problems. This has been motivating recent work on combining density-based methods with segregated pressure-correction algorithms \citep{Xiao2004, Moguen2012, Fuster2018, Moguen2019} and {\em hybrid density/pressure-based} algorithms \citep{vanderHeul2003, Park2005}, in which the continuity equation is solved for density but the energy equation is reformulated as an equation for pressure. An additional difficulty for interfacial flows associated with density-based methods is that the pressure field has to be reconstructed based on the applied thermodynamic closure model, which has proved to be a considerable difficulty in interfacial cells where two bulk phases coexist \citep{Abgrall2001, Allaire2002, Murrone2005}.

{\em Pressure-based} algorithms for compressible flows, in which the continuity equation is solved for pressure, are less prominent than their density-based counterparts. Deriving stable and efficient numerical schemes for the transonic regime, and formulating consistent shock-capturing schemes, is known to be difficult for pressure-based algorithms \citep{Wesseling2001}. However, because pressure plays an important role in all Mach number regimes, {\em i.e.}~the pressure-velocity coupling dominates at low Mach numbers and the pressure-density coupling dominantes at high Mach numbers \citep{VanDoormaal1987, Moukalled2016}, pressure-based algorithms potentially offer a distinct advantage for applications in all Mach number regimes or in which the Mach number varies strongly, such as interfacial flows. Although, starting with the seminal work of \citet{Harlow1971a}, a variety of pressure-based algorithms for compressible single-phase flows has been proposed, notably \citep{VanDoormaal1987, Demirdzic1993, Karimian1994, Xiao2017}, it was only recently that \citet{Denner2018b} proposed a conservative pressure-based algorithm for compressible interfacial flows at all speeds. This algorithm was proposed in conjunction with a new interface discretisation, the acoustically-conservative interface discretisation (ACID), that retains the acoustic features of the flow, which facilitates a rational definition of fluid properties in interfacial cells and which does not require Riemann solvers to compute the fluxes through the fluid interface. \citet{Denner2018b} showed that such an algorithm yields unique definitions of the speed of sound and the Rankine-Hugoniot conditions in the interface region, and demonstrated a reliable prediction of acoustic and shock waves even in interfacial flows with acoustic impedance matching and shock impedance matching.

To model shock-bubble interactions, density-based algorithms in conjunction with Riemann-type solvers have been applied in most published studies to date \citep{Quirk1996, Bagabir2001, Nourgaliev2006, Johnsen2007, Niederhaus2008, Hejazialhosseini2013, Xiang2017}. While exact Riemann solvers are prohibitively time consuming, approximate Riemann solvers require an {\em a priori} approximation of the characteristic wave speeds, which ensues a substantial complexity of the numerical algorithms \citep{Saurel2018}. Thereby a strong dependence of the solution on the spatial resolution and the applied numerical schemes has been generally observed. For instance, a distinct feature of the shock-bubble interaction in gas-gas systems observed  in numerical simulations of shock-bubble interactions, are instabilities ({\em cf.}~Richtmyer-Meshkov instability \citep{Brouillette2002}) forming on the interface as the shock wave passes.  Noticeably, the shape and evolution of these interface instabilities depend strongly on the applied numerical methods \citep{Johnsen2006, Denner2018b, Saurel2018}, {\em e.g.}~the interface treatment or the advection schemes. In addition, these interface instabilities feature ever smaller lengthscales with an increasing spatial resolution of the simulation \citep{Wong2017, Denner2018b}, although this is to be expected if surface tension, viscous stresses and heat conduction are neglected, a common assumption, supported by experimental observations \citep{Layes2003, Zhai2011}, with reference to the small timescales considered in typical numerical simulations and the associated marginal influence of these effects, since there is no physical means to regulate or dissipate the small-scale flow features. However, the influence of the spatial resolution of the computational mesh and of the choice of discretisation schemes on the predictive quality of the modelling of shock-bubble interactions have not yet been studied comprehensively, especially in the context of pressure-based algorithms without Riemann solvers. It is, moreover, as yet unclear what influence such differences in interface instabilities have on the development and evolution of shock waves and rarefaction fans during and after the shock-bubble interaction, which is important for many of the associated engineering applications, especially in medical and bioengineering applications.

This article investigates the modelling of shock-bubble interactions using the pressure-based algorithm proposed by \citet{Denner2018b}, where the fluxes are evaluated with the ACID method and no Riemann solvers are applied. The aim is to identify the minimum spatial resolution requirements for a converged solution with respect to the primary flow quantities, as well as the influence of the discretisation scheme and of interface instabilities on the predictive accuracy of the main flow features, for shock-bubble interactions in both gas-gas and liquid-gas flows. As test-cases the shock interaction with a one-dimensional helium-bubble in air, a one-dimensional air-bubble in water, a two-dimensional R22-bubble in air and a two-dimensional air-bubble in water are considered. While the shock-bubble interaction in gas-gas flows is not very sensitive to the employed discretisation schemes or the resolution of the computational mesh, the presented results demonstrate a very strong dependency of the primary flow quantities, especially temperature, on the spatial resolution of the computational mesh during the interaction of a shock wave with an air bubble suspended in water.

The governing equations are introduced in Section \ref{sec:governingEq} and the numerical framework is presented in Section \ref{sec:numericalFramework}. The results of this study are presented and discussed in Section \ref{sec:results}, and conclusions are drawn in Section \ref{sec:conclusions}.

\section{Governing equations}
\label{sec:governingEq}

The conservation laws governing fluid flow at all speeds, assuming viscous stresses and heat conduction are neglected, are the Euler equations, consisting of the conservation of mass
\begin{equation}
\frac{\partial \rho}{\partial t} + \frac{\partial \rho u_i}{\partial x_i} = 0 
, \label{eq:continuity}
\end{equation}
the conversation of momentum 
\begin{equation}
\frac{\partial \rho u_j}{\partial t} + \frac{\partial \rho u_i u_j}{\partial
x_i} = - \frac{\partial p}{\partial x_j} , \label{eq:momentum}  
\end{equation}
and the conservation of energy 
\begin{equation}
\frac{\partial \rho h}{\partial t} + \frac{\partial \rho u_i h}{\partial x_i} =
\frac{\partial p}{\partial t}   , \label{eq:energy}
\end{equation}
where $t$ is time,
$\boldsymbol{u}$ is the velocity vector, $p$ is pressure, $\rho$ is the density
and  $h = c_p \, T + \boldsymbol{u}^2/2$ is the specific total enthalpy, with $c_p$ the
specific isobaric heat capacity and $T$ the temperature. Gravity and surface tension are neglected in this study.

The stiffened-gas model \citep{Harlow1971, Saurel2007} is applied to define the thermodynamic properties of the fluid and close the governing conservation laws. The density-pressure relationship is defined by the stiffened-gas equation of state (EOS)
\begin{equation}
\rho =  \frac{p + \gamma_0 \Pi_0}{R_0 \, T} , \label{eq:rhoComp}
\end{equation}
where $\Pi_0$ is a fluid-dependent pressure constant, $R_0= c_{p,0}-c_{v,0}$ is the specific heat capacity and $\gamma_0 = c_{p,0}/c_{v,0}$ is the heat capacity ratio, with the reference specific isobaric heat capacity $c_{p,0}$ and the reference specific isochoric heat capacity $c_{v,0} $. The speed of sound is given as
\begin{equation}
a = \sqrt{\gamma_0 \, \frac{p + \Pi_0}{\rho}} = \sqrt{(\gamma_0 -1) \, c_p \, T}
\label{eq:soundSpeed}
\end{equation}
and the specific isobaric heat capacity is \citep{Denner2018b}
\begin{equation}
c_p = c_{p,0} \frac{p + \Pi_0}{p + \gamma_0 \Pi_0}   . \label{eq:cpComp}
\end{equation}
For $\Pi_0=0$,  the stiffened-gas EOS reverts to the ideal-gas EOS, and the fluid is calorically perfect with $c_p = c_{p,0}$.

The Volume-of-Fluid (VOF) method \citep{Hirt1981} is adopted to capture the fluid interface between two immiscible bulk phases. To this end, the VOF method applies a colour function field $\psi$, defined as 
\begin{equation}
\psi (\boldsymbol{x}) =
\begin{cases}
0 & \text{if} \ \boldsymbol{x} \in \Omega_\textup{a} \\
1 & \text{if} \ \boldsymbol{x} \in \Omega_\textup{b},
\end{cases} \label{eq:colourDef}
\end{equation}
where $\Omega_\textup{a}$ and $\Omega_\textup{b}$ are the subdomains occupied by fluid a and b, respectively, and $\Omega = \Omega_\textup{a} \cup \Omega_\textup{b}$ is the computational domain. The interface is located in every cell where $0 < \psi < 1$. Because the interface is a material front propagating with the flow \citep{Denner2018b}, the colour function $\psi$ is advected with the underlying fluid velocity by the advection equation
\begin{equation}
\frac{\partial \psi}{\partial t} + u_i \frac{\partial \psi}{\partial x_i}
 = 0 .
\label{eq:vofAdvection}
\end{equation}

\section{Numerical framework}
\label{sec:numericalFramework}
The numerical framework is based on the fully-coupled pressure-based algorithm of \citet{Denner2018b}, which is predicated on a finite-volume discretisation with collocated variable arrangement.

\subsection{Temporal and spatial discretisation}
\label{sec:discretisation}
The First-Order Backward Euler scheme (BDF1) and the Second-Order Backward Euler scheme (BDF2) are used to discretise the transient terms of the governing flow equations. The BDF1 scheme applied to the integrated transient term of a general flow variable $\phi$ is given for cell $P$ as
\begin{equation}
\iiint_{V_P} \left. \frac{\partial \phi}{\partial t} \right|_P \, dV \approx
\frac{\phi_P - \phi_P^{(t-\Delta t)}}{\Delta t} \, V_P , 
\end{equation}
and the BDF2 scheme is defined as
\begin{equation}
\iiint_{V_P} \left. \dfrac{\partial \phi}{\partial t} \right|_P \, dV \approx
\frac{3 \phi_P^{(t)} -
4 \phi_P^{(t-\Delta t)} + \phi_P^{(t-2\Delta t)}}{2 \Delta t} V_P, 
\label{eq:sobeRaw2}
\end{equation}
where $\Delta t$ is the time-step, superscript $(t-\Delta t$) denotes values of the previous time-level, superscript $(t-2\Delta t)$ denotes values of the previous-previous time-level and $V_P$ is the volume of mesh cell $P$. As previously suggested by \citet{Denner2018b}, for consistency all transient terms of the governing equations (\ref{eq:continuity})-(\ref{eq:energy}) are discretised with the same scheme.

Applying the divergence theorem, assuming the surface of the control volume has
a finite number of flat faces $f$ and applying the midpoint rule, the
discretised advection terms of Eqs.~(\ref{eq:continuity})-(\ref{eq:energy}) are
given as 
\begin{align}
\iiint_{V_P} \left. \frac{\partial \rho u_i \phi}{\partial x_i} \right|_P \,
\textup{d}V \approx \sum_f \tilde{\rho}_f \vartheta_f \tilde{\phi}_f A_f ,
\end{align}
where $\vartheta_f = \boldsymbol{u}_f \cdot \boldsymbol{n}_f$ is the advecting velocity of face $f$ (see Section \ref{sec:advectingVel}), $\boldsymbol{n}_f$ is the outward-pointing unit normal vector of face $f$ and $A_f$ is the area of face $f$. The advected variable $\phi$ at face $f$ is interpolated from the adjacent cell-centred values using a total variation diminishing (TVD) interpolation scheme \citep{Denner2015a}, with which the face value is given as
\begin{equation}
\tilde{\phi}_f = \phi_U + \frac{\xi_f}{2} (\phi_D-\phi_U) \ ,
\end{equation} 
where subscripts $U$ and $D$ denote the upwind and downwind cells, respectively, and $\xi_f$ is the flux limiter. In this study, the first-order upwind scheme ($\xi_f = 0$), the Minmod scheme and the Superbee scheme \citep{Roe1986} are considered.

\subsection{Advecting velocity}
\label{sec:advectingVel}
The momentum-weighted interpolation (MWI) is applied to define an advecting velocity $\vartheta_f = \boldsymbol{u}_f \cdot \boldsymbol{n}_f$ at cell faces, which is used in the discretised advection terms of the governing equations. Following the unified formulation of the MWI proposed by \citet{Bartholomew2018}, the advecting velocity $\vartheta_f$ at face $f$ is given as
\begin{equation}
\vartheta_f  = \overline{u}_{i,f}  {n}_{i,f} - \hat{{d}}_f
\left[\left. \frac{\partial p}{\partial x_i} \right|_f - \frac{\rho_f^\ast}{2} 
\left(\left.
\frac{1}{\rho_P} \frac{\partial p}{\partial x_i} \right|_P +
\left. \frac{1}{\rho_Q} \frac{\partial p}{\partial x_i}
\right|_Q \right) \right] {n}_{i,f} +  \hat{{d}}_f  \frac{\rho^{\ast (t-\Delta
t_1)}_f}{\Delta t} \left(\vartheta^{(t-\Delta t)}_f -
\overline{u}_{i,f}^{(t-\Delta t)} {n}_{i,f} \right) ,
\label{eq:advVel}
\end{equation} 
where subscript $Q$ denotes the neighbour cell of $P$ adjacent to face $f$, the interpolated face velocities $\overline{\boldsymbol{u}}_f$ and $\overline{\boldsymbol{u}}_f^{(t-\Delta t)}$ are obtained by linear interpolation, and the pressure gradient normal to face $f$ is discretised as
\begin{equation}
\left. \frac{\partial p}{\partial x_i} \right|_f n_{i,f} \approx
\frac{p_Q-p_P}{\Delta x}.
\end{equation}
The face density $\rho_f^\ast$ is interpolated by a harmonic average \citep{Bartholomew2018} and the coefficient $\hat{{d}}_f$ follows from the coefficients associated with the advection term (and shear stress term if viscosity is considered) of the momentum equations, as detailed in \citep{Bartholomew2018}.

MWI provides a robust pressure-velocity coupling for incompressible and low Mach number flows by applying a low-pass filter acting on the third derivative of pressure \citep{Bartholomew2018}, thus avoiding pressure-velocity decoupling due to the collocated variable arrangement. As a result of the additional terms required to ensure a robust pressure-velocity coupling, the MWI introduces an unphysical dissipation of kinetic energy, which however diminishes with $\Delta x^3$ and is independent of $\Delta t$ \citep{Bartholomew2018}. The transient term of Eq.~(\ref{eq:advVel}) ensures a time-step independent contribution of the MWI in conjunction with the coefficient $\hat{{d}}_f$ of the pressure term \citep{Bartholomew2018} and including the transient term is important for a correct temporal evolution of pressure waves \citep{Moguen2015a, Bartholomew2018}.

\subsection{Discretised governing conservation laws}
\label{sec:discGoverningEq}

The discretised continuity equation (\ref{eq:continuity}) for cell $P$, applying the BDF1 scheme for clarity of presentation, is given as
\begin{equation}
\frac{\rho_P^{(n+1)} - \rho_P^{(t-\Delta t)}}{\Delta t} V_P + \sum_f
\tilde{\rho}^{(n)}_f \, \vartheta_f^{(n+1)} +
\tilde{\rho}^{(n+1)}_f \, \vartheta_f^{(n)} - \tilde{\rho}^{(n)}_f \,
\vartheta_f^{(n)} \, A_f = 0 ,
\label{eq:continuityDisc}
\end{equation}
where the superscript $(n)$ denotes known values of the most recent available solution and superscript $(n+1)$ denotes quantities which are solved implicitly. The advection term is linearised by a Newton linearisation to facilitate a smooth transition from low to high Mach number regions  \citep{Karimian1994, Kunz1999, Xiao2017}. Following previous studies \citep{Denner2018b, Denner2018c}, the semi-implicit formulation of the advecting velocity is given as
\begin{equation}
\begin{split}
\vartheta_f^{(n+1)} & = \overline{u}_{i,f}^{(n+1)}  {n}_{i,f} - \hat{{d}}_f
\left[ \frac{p_Q^{(n+1)} - p_P^{(n+1)}}{\Delta x} -  \frac{\rho_f^\ast}{2}
\left( \left. \frac{1}{\rho_P}\frac{\partial p}{\partial x_i}
\right|_P^{(n)} + \left. \frac{1}{\rho_Q} \frac{\partial p}{\partial
x_i} \right|_Q^{(n)} \right) {n}_{i,f}  \right] \\ 
 & +  \hat{{d}}_f \, \frac{\rho^{\ast (t-\Delta t)}_f}{\Delta t}
\left(\vartheta^{(t-\Delta t)}_f - \overline{u}_{i,f}^{(t-\Delta t)}
{n}_{i,f} \right)  
\label{eq:advVelImp}
\end{split}
\end{equation} 
and the pressure-implicit formulation of the density is given as
\begin{equation}
\rho^{(n+1)} = \frac{p^{(n+1)} + \gamma_0 \Pi_0}{R_0 \, T}.
\label{eq:rhoFullImp}
\end{equation}

The discretised momentum equations (\ref{eq:momentum}) of cell $P$ are given, with both the transient term and the advection term linearised by a Newton linearisation \citep{Denner2018b, Denner2018c}, as
\begin{equation}
\begin{split}
& \frac{\rho_P^{(n)} \, u_{j,P}^{(n+1)} + \rho_P^{(n+1)} \, u_{j,P}^{(n)}
- \rho_P^{(n)} \, u_{j,P}^{(n)} - \rho^{(t-\Delta t)}_P \, u^{(t-\Delta
t)}_{j,P}}{\Delta t} V_P  \\ & + \sum_f \left(\tilde{\rho}_f^{(n)} 
\vartheta_f^{(n)} \tilde{u}_{j,f}^{(n+1) }+ \tilde{\rho}_f^{(n)} 
\vartheta_f^{(n+1)} \tilde{u}_{j,f}^{(n)} +  \tilde{\rho}_f^{(n+1)} 
\vartheta_f^{(n)} \tilde{u}_{j,f}^{(n)} - 2 \tilde{\rho}_f^{(n)} 
\vartheta_f^{(n)} \tilde{u}_{j,f}^{(n)} \right) \, A_f 
 = - \sum_f \overline{p}_f^{(n+1)} \, {n}_{j,f} \, A_f 
\end{split} \label{eq:momentumDisc}
\end{equation}
with $\rho^{(n+1)}_P$ given by Eq.~(\ref{eq:rhoFullImp}) and $\vartheta^{(n+1)}_f$ given by Eq.~(\ref{eq:advVelImp}). Similarly, the discretised energy equation (\ref{eq:energy}) of cell $P$ is given as
\begin{equation}
\begin{split}
& \frac{\rho_P^{(n)} \, h_P^{(n+1)} + \rho_P^{(n+1)} \, h_P^{(n)}
- \rho_P^{(n)} \, h_P^{(n)} - \rho^{(t-\Delta t)}_P \, h^{(t-\Delta
t)}_{P}}{\Delta t} V_P  \\ & + \sum_f \left(\tilde{\rho}_f^{(n)} 
\vartheta_f^{(n)} \tilde{h}_{f}^{(n+1)}+ \tilde{\rho}_f^{(n)} 
\vartheta_f^{(n+1)} \tilde{h}_{f}^{(n)} + \tilde{\rho}_f^{(n+1)} 
\vartheta_f^{(n)} \tilde{h}_{f}^{(n)} - 2 \tilde{\rho}_f^{(n)} 
\vartheta_f^{(n)} \tilde{h}_{f}^{(n)} \right) \, A_f  = \frac{p_{P}^{(n+1)} -
p^{(t-\Delta t)}_P }{\Delta t} V_P .
\end{split} \label{eq:energyDisc}
\end{equation}

\subsection{Interface advection}
\label{sec:compressiveVOF}
The VOF advection equation (\ref{eq:vofAdvection}) is discretised using a compressive VOF method \citep{Denner2014e, Denner2018b}. Following \citet{Denner2018b}, Eq.~(\ref{eq:vofAdvection}) is reformulated as
\begin{equation}
\frac{\partial \psi}{\partial t} + \frac{\partial u_i \psi}{\partial x_i} - \psi \frac{\partial u_i}{\partial x_i} = 0 \ . \label{eq:vofAdvComp}
\end{equation} 
Using the Crank-Nicolson scheme for the discretisation of the transient term, the semi-discretised form of Eq.~(\ref{eq:vofAdvComp}) becomes
\begin{equation}
\frac{\psi_P - \psi_P^{(t-\Delta t_\psi)}}{\Delta t_\psi} \, V_P + \sum_f
\frac{\psi_f + \psi_f^{(t-\Delta t_\psi)}}{2} \, \vartheta_f A_f -
\frac{\psi_P + \psi_P^{(t-\Delta t_\psi)}}{2}  \sum_f
\vartheta_f \, A_f = 0 ,
\label{eq:vofAdvCompDisc}
\end{equation}
where $\Delta t_\psi$ is the time-step applied to advect the colour function $\psi$. The advection of the colour function is discretised using the same advecting velocity $\vartheta_f$ as for all advection terms of the governing equations. The face value $\psi_f$ is interpolated using the CICSAM scheme \citep{Ubbink1999}, taking into account the orientation of the interface and the available flux volume. Excellent volume conservation has previously been demonstrated for this compressive VOF method for incompressible \citep{Denner2014e} and compressible flows \citep{Denner2018b}. 

\subsection{Coupling of the bulk phases}
\label{sec:acid}
The discretised governing equations presented in Section \ref{sec:discGoverningEq} are extended to interfacial flows using the acoustically-conservative interface discretisation (ACID) \citep{Denner2018b}. The ACID method assumes that, for the purpose of discretising the governing conservation laws for a given cell, all cells in its finite-volume stencil are assigned the same colour function value, {\em i.e.}~the colour function is assumed to be constant in the entire finite-volume stencil. The relevant thermodynamic properties that are discontinuous at the interface, {\em i.e.}~density and enthalpy, are evaluated based on the constant colour function field in the applied finite-volume stencil. This recovers the contact discontinuity associated with the interface \citep{Anderson2003, Denner2018b} and enables the application of the fully-conservative discretisation scheme presented in Section \ref{sec:discGoverningEq}, identical to the one applied for single-phase flows.

\subsubsection{Fluid properties}
\label{sec:fluidProp}
A hydrodynamic and thermodynamic consistent definition of the fluid properties for interfacial flows requires special consideration. The density of the fluid is defined based on the colour function $\psi$ and the densities of the bulk phases as
\begin{equation}
\rho = (1 - \psi) \, \rho_\textup{a} +  \psi \,
\rho_\textup{b} , \label{eq:densityFluid}
\end{equation}
where the partial densities $\rho_\textup{a}$ and $\rho_\textup{b}$ of the bulk phases are given by Eq.~(\ref{eq:rhoComp}). This linear interpolation of the density is required for the discrete conservation of mass, momentum and energy and is equivalent to an isobaric closure assumption for compressible interfacial flows \citep{Allaire2002, Shyue2006}. The heat capacity ratio also follows from the isobaric closure assumption as
\begin{equation}
\frac{1}{\gamma - 1} =\frac{1-\psi}{\gamma_{0,\textup{a}}
- 1} +  \frac{\psi}{\gamma_{0,\textup{b}} - 1} .
\label{eq:gammaMinusOne}
\end{equation}
The specific isobaric heat capacity is defined by a mass-weighted interpolation \citep{Denner2018b}, which is essential for the conservation of the total energy, given as
\begin{equation}
c_p = \frac{(1 - \psi) \, \rho_\textup{a} \, c_{p,\textup{a}} 
+ \psi \, \rho_\textup{b} \, c_{p,\textup{b}}}{\rho} \ ,
\label{eq:cpFluid}
\end{equation}
where the partial densities $\rho_\textup{a}$ and $\rho_\textup{b}$ are given by Eq.~(\ref{eq:rhoComp}), density $\rho$ is given by Eq.~(\ref{eq:densityFluid}), and the partial specific isobaric heat capacities $c_{p,\textup{a}}$ and $c_{p,\textup{b}}$ are given by Eq.~(\ref{eq:cpComp}). As shown by \citet{Denner2018b}, the speed of sound is defined throughout the domain based on Eq.~(\ref{eq:soundSpeed}) as $a = \sqrt{(\gamma - 1) \, c_p \, T}$, and the material-dependent pressure constant  of the stiffened-gas model is given as $\Pi = [(\gamma-1) \, \rho \, c_{p} \, T /\gamma] - p$, with the density $\rho$ given by Eq.~(\ref{eq:densityFluid}), the specific isobaric heat capacity $c_p$ given by Eq.~(\ref{eq:cpFluid}), and $(\gamma -1)$ as well as $\gamma$ given by Eq.~(\ref{eq:gammaMinusOne}).

\subsubsection{Density treatment}
\label{sec:densityTreatment}
Under the assumption that the colour function $\psi$ is constant throughout the finite-volume stencil of cell $P$, the density interpolated to face $f$ from the adjacent cell centre is given as
\begin{equation}
\tilde{\rho}_f =  \rho_U^\star  + \frac{\xi_{f}}{2} \left( \rho_D^\star
- \rho_U^\star \right) \ . \label{eq:densityFaceDiscNew}
\end{equation} 
The density $\rho_U$ at the upwind cell $U$ and $\rho_D$ at the downwind cell $D$ are given based on the colour function value of cell $P$ by Eq.~(\ref{eq:densityFluid}), so that
\begin{equation}
\rho_U^\star = \rho_{\textup{a},U} +  \psi_P \, \left(\rho_{\textup{b},U} -
\rho_{\textup{a},U} \right)  \label{eq:rhoUstar}
\end{equation} 
and 
\begin{equation}
\rho_D^\star = \rho_{\textup{a},D} + 
\psi_P \, \left(\rho_{\textup{b},D} -
\rho_{\textup{a},D} \right) .  \label{eq:rhoDstar}
\end{equation}
The density at previous time-levels is evaluated in a similar fashion based on the colour function value of cell $P$, with \citep{Denner2018b}
\begin{equation}
\rho_P^{(t-\Delta t)} = \rho_{\textup{a},P}^{(t-\Delta t)} + 
\psi_P \, \left(\rho_{\textup{b},P}^{(t-\Delta t)} -
\rho_{\textup{a},P}^{(t-\Delta t)}\right) \label{eq:acidDensityT1}
\end{equation}
and likewise for  $\rho_P^{(t-2 \Delta t)}$, if required.

\subsubsection{Enthalpy treatment}
\label{sec:enthalpyTreatment}
The specific total enthalpy at face $f$ is given, again assuming the colour function $\psi$ is constant throughout the finite-volume stencil of cell $P$, as \citep{Denner2018b}
\begin{equation}
\tilde{h}_f = \frac{1}{\tilde{\rho}_f} \left[\rho_U^\star h_U^\star +
 \frac{\xi_{f}}{2} \left(\rho_D^\star h_D^\star - \rho_U^\star h_U^\star
\right) \right] , \label{eq:enthalpyACIDCorr}
\end{equation}
with $\tilde{\rho}_f$ given by Eq.~(\ref{eq:densityFaceDiscNew}), where the specific total enthalpy of the upwind and downwind cells are given as
\begin{align}
h_U^\star & = c_{p,U}^\star \, T_U + \frac{1}{2} \boldsymbol{u}^{2}_U, \\
h_D^\star & = c_{p,D}^\star \, T_D + \frac{1}{2} \boldsymbol{u}^{2}_D  ,
\end{align}
respectively, $\rho_U^\star$ is given by Eq.~(\ref{eq:rhoUstar}) and $\rho_D^\star$ is given by Eq.~(\ref{eq:rhoDstar}). The specific isobaric heat capacities $c_{p,U}^\star$ and $c_{p,D}^\star$ are defined by Eq.~(\ref{eq:cpFluid}) with $\psi_P$ as
\begin{equation}
c_{p,U}^\star = \frac{\rho_{\textup{a},U} \, c_{p,\textup{a},U} + \psi_P \,
(\rho_{\textup{b},U} \, c_{p,\textup{b},U} - \rho_{\textup{a},U} \,
c_{p,\textup{a},U})}{\rho_U^\star}
\end{equation}
and
\begin{equation}
c_{p,D}^\star = \frac{\rho_{\textup{a},D} \, c_{p,\textup{a},D} + \psi_P \,
(\rho_{\textup{b},D} \, c_{p,\textup{b},D}-\rho_{\textup{a},D} \,
c_{p,\textup{a},D})}{\rho_D^\star} .
\end{equation}
Since the specific total enthalpy is a primary solution variable, a deferred correction approach as proposed by \citet{Denner2018b} is applied to enforce Eq.~(\ref{eq:enthalpyACIDCorr}).

The specific total enthalpy at the previous time-levels follow analogously as
\citep{Denner2018b}
\begin{equation}
h_P^{(t-\Delta t)} = c_{p,P}^{\star,(t-\Delta t)} \, T_P^{(t-\Delta t)} +
\dfrac{1}{2} \boldsymbol{u}^{(t-\Delta t),2}_P
\end{equation}
with
\begin{equation}
c_{p,P}^{\star,(t-\Delta t)} = \frac{\rho_{\textup{a},P}^{(t-\Delta t)} \,
c_{p,\textup{a},P}^{(t-\Delta t)}  + \psi_P 
\left(\rho_{\textup{b},P}^{(t-\Delta t)} \, c_{p,\textup{b},P}^{(t-\Delta
t)}-\rho_{\textup{a},P}^{(t-\Delta t)} \, c_{p,\textup{a},P}^{(t-\Delta
t)}\right)}{\rho_{P}^{(t-\Delta t)}} ,
\end{equation}
and likewise for $h_P^{(t-2 \Delta t)}$ and $c_{p,P}^{\star,(t-2 \Delta t)}$,
if required.

\subsection{Solution procedure}
The discretised governing equations presented in Section \ref{sec:discGoverningEq} are solved simultaneously in a single linear system of equations \citep{Denner2018b, Denner2018c}, $\boldsymbol{A} \boldsymbol{\chi} =  \boldsymbol{b}$, with $\boldsymbol{A}$ being the coefficient matrix of size $5N \times 5N$, $\boldsymbol{\chi} \equiv (\boldsymbol{u}, p, h)^T$ is the solution vector of length $5N$ of the primary solution variables and $\boldsymbol{b}$ is the right-hand side vector containing all known contributions, where $N$ is the number of mesh cells of the three-dimensional computational mesh. The solution procedure performs nonlinear iterations in which the linear system of governing equations is solved using the {\em Block-Jacobi} preconditioner and the {\em BiCGSTAB} solver of the software library PETSc \citep{petsc-user-ref}, as described in detail by \citet{Denner2018c}.

\section{Results}
\label{sec:results}

The presented results focus on the spatial resolution requirements and discretisation necessary for the accurate prediction of shock-bubble interactions using a pressure-based algorithm. As already comprehensively demonstrated by \citet{Denner2018b}, the applied numerical algorithm captures shock waves and rarefaction fans accurately in single-phase flows and interfacial flows, with a robust convergence under mesh refinement and a precise prediction of the Rankine-Hugoniot relations also in interfacial cells.

\subsection{One-dimensional helium-bubble in air}
\label{sec:gasGas1D}

\begin{figure}[t]
\begin{center}
\subfloat[Velocity $u$]
{\includegraphics[width=0.31\textwidth]{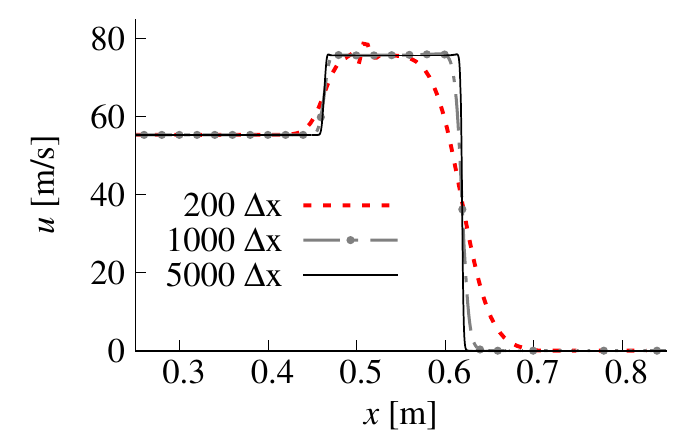}}
\quad
\subfloat[Pressure $\Delta p = p - p_\textup{II}$]
{\includegraphics[width=0.31\textwidth]{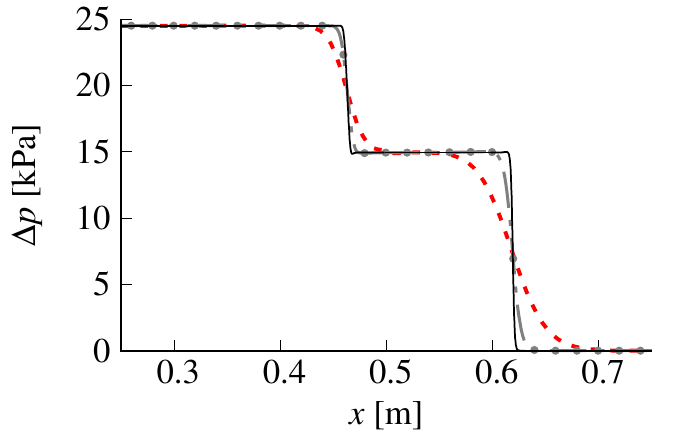}}
\quad
\subfloat[Temperature $T$]
{\includegraphics[width=0.31\textwidth]{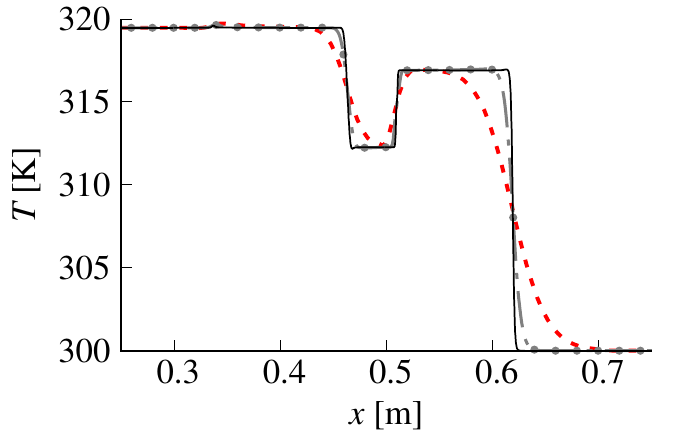}}
\caption{Profiles of the velocity $u$, pressure $\Delta p = p - p_\textup{II}$ and temperature $T$ of the interaction of a shock wave with $M_\textup{s} = 1.1$ with a one-dimensional helium-bubble in air on meshes with different mesh spacings $\Delta x$ at time $t=6.5 \times 10^{-4} \, \textup{s}$.}
\label{fig:HeliumAirt6p5em4M1p1}
\end{center}
\end{figure}

The interaction of a shock wave travelling in air with Mach number $M_\textup{s} =1.1$ and interacting with a helium bubble in a one-dimensional domain with a length of $1 \, \textup{m}$ is considered. The initial post-shock conditions ($\textup{I}$) are
\begin{equation}
\begin{array}{ccccc}
u_\textup{I} = 55.33 \, \textup{m} \, \textup{s}^{-1}, & p_\textup{I} = 
1.245 \times 10^5 \, \textup{Pa}, & T_\textup{I} = 319.48 \, \textup{K} ,
\nonumber
\end{array}
\end{equation}
and the pre-shock conditions ($\textup{II}$) are
\begin{equation}
\begin{array}{ccccc}
u_\textup{II} = 0 \, \textup{m} \, \textup{s}^{-1}, & p_\textup{II} =
 10^5 \, \textup{Pa}, & T_\textup{II} = 300 \, \textup{K}.
\nonumber
\end{array}
\end{equation}
Air is taken to have a heat capacity ratio of $\gamma_\textup{0,Air} = 1.4$ and
a specific gas constant of $R_\textup{0,Air} = 288.0 \, \textup{J} \, \textup{kg}^{-1} \, \textup{K}^{-1}$, and helium is assumed to have a heat capacity ratio of $\gamma_\textup{0,He} = 1.648$ and a specific gas constant of $R_\textup{0,He} = 1581.2 \, \textup{J} \, \textup{kg}^{-1} \, \textup{K}^{-1}$. The shock is initially located at $x_0 = 0.3 \, \textup{m}$, the helium bubble occupies the interval $0.5 \leq x \leq 0.7$ and the applied time-step corresponds to a Courant number of $\textup{Co} = a_\textup{II,He} \Delta t/\Delta x = 0.44$.

The results, shown in Fig.~\ref{fig:HeliumAirt6p5em4M1p1}, are obtained on equidistant meshes with three mesh resolutions resolving the one-dimensional domain with $200$, $1000$ and $5000$ cells, which corresponds to a spatial resolution of $40$, $200$ and $1000$ cells for the initial length of the helium bubble, respectively. The distribution of pressure, velocity and temperature within the one-dimensional domain depends considerably on the applied mesh resolution. While this may be a problem for the accurate local prediction of pressure, velocity and temperature, as well as the related thermodynamic or chemical processes, the position of the primary shock wave as it travels through the bubble is, apart from the sharpness of the discontinuity, unaffected by the mesh resolution. In addition, the colour function $\psi$ and the density $\rho$ (both not shown) are in very good agreement on the different meshes. All quantities obtained with the finest mesh resolution, $1000$ cells for the initial length of the helium bubble, are in excellent agreement with the corresponding exact Riemann solution.

\subsection{One-dimensional air-bubble in water}
\label{sec:liquidGas1D}

\begin{figure}[t]
\begin{center}
\subfloat[Velocity $u$]
{\includegraphics[width=0.31\textwidth]{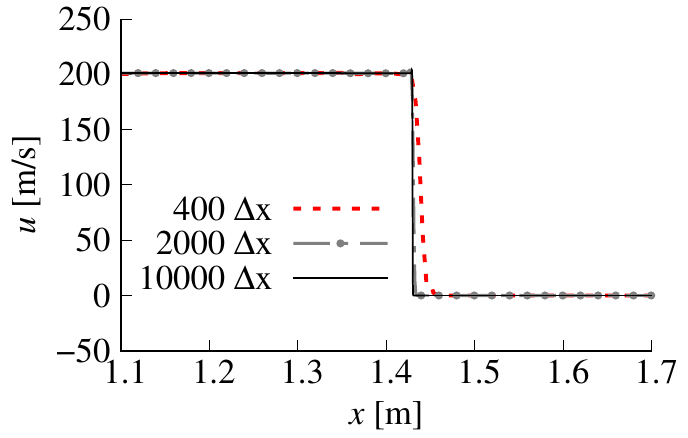}}
\quad
\subfloat[Density $\rho$]
{\includegraphics[width=0.31\textwidth]{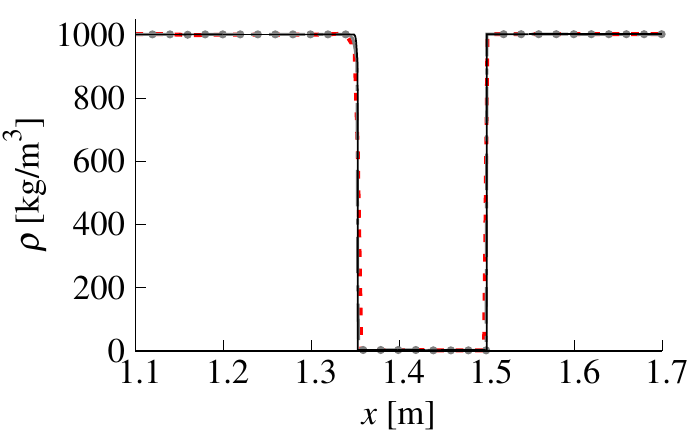}}
\quad
\subfloat[Temperature $T$]
{\includegraphics[width=0.31\textwidth]{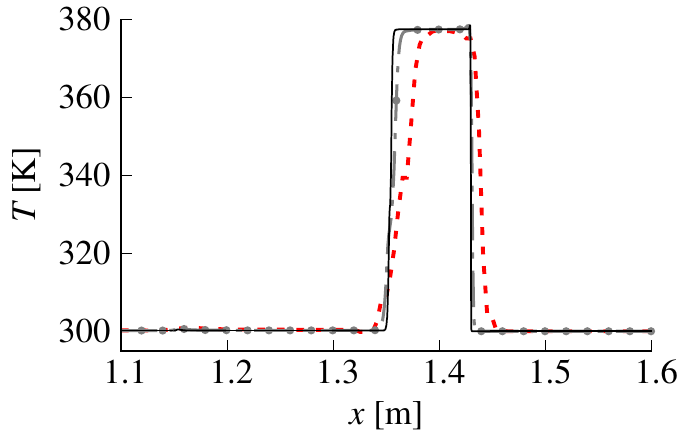}}
\caption{Profiles of the velocity $u$,  density $\rho$ and temperature $T$ of the interaction of a shock wave with $M_\textup{s} = 1.1$ with a one-dimensional air-bubble in water on meshes with different mesh spacings $\Delta x$ at time $t=4.0 \times 10^{-4} \, \textup{s}$.}
\label{fig:AirWatert4p0em4M1p1}
\end{center}
\end{figure}

\begin{figure}[t]
\begin{center}
\subfloat[$t=4.0 \times 10^{-4} \, \textup{s}$]
{\includegraphics[width=0.31\textwidth]{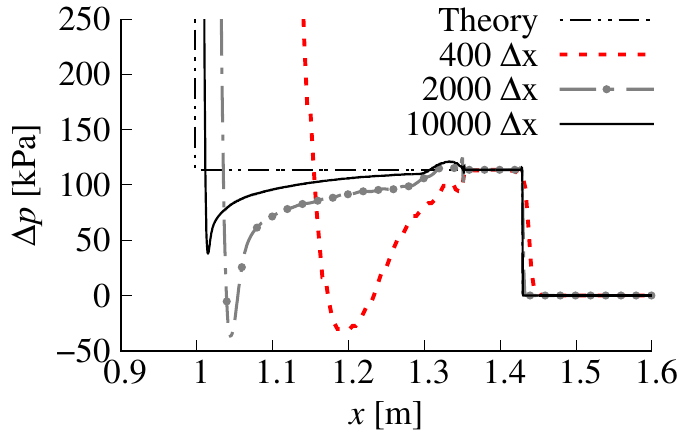}}
\quad
\subfloat[$t=6.5 \times 10^{-4} \, \textup{s}$]
{\includegraphics[width=0.31\textwidth]{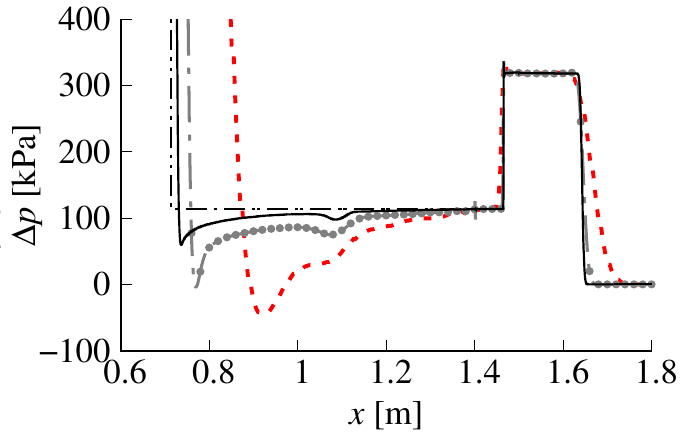}}
\caption{Profiles of the pressure $\Delta p = p - p_\textup{II}$ of the interaction of a shock wave with $M_\textup{s} = 1.1$ with a one-dimensional air-bubble in water on meshes with different mesh spacings $\Delta x$ at (a) time $t=4.0 \times 10^{-4} \, \textup{s}$ and (b) time $t=6.5 \times 10^{-4} \, \textup{s}$. The theoretical Riemann solution is shown as a reference.}
\label{fig:AirWaterPresM1p1}
\end{center}
\end{figure}

The interaction of a shock wave travelling in water with Mach number $M_\textup{s} =1.1$ and interacting with an air bubble in a one-dimensional domain with a length of $2 \, \textup{m}$ is considered. The initial post-shock conditions ($\textup{I}$) are
\begin{equation}
\begin{array}{ccccc}
u_\textup{I} = 100.45 \, \textup{m} \, \textup{s}^{-1}, & p_\textup{I} = 
1.487 \times 10^8 \, \textup{Pa}, & T_\textup{I} = 302.61 \, \textup{K} ,
\nonumber
\end{array}
\end{equation}
and the pre-shock conditions ($\textup{II}$)  are
\begin{equation}
\begin{array}{ccccc}
u_\textup{II} = 0 \, \textup{m} \, \textup{s}^{-1}, & p_\textup{II} =
 10^5 \, \textup{Pa}, & T_\textup{II} = 300 \, \textup{K}.
\nonumber
\end{array}
\end{equation}
Water is taken to have a heat capacity ratio of  $\gamma_\textup{0,Water} = 4.1$, a pressure constant of $\Pi_\textup{0,Water} = 4.4 \times 10^{8} \, \textup{Pa}$ and a specific gas constant of $R_\textup{0,Water} = 6000 \, \textup{J} \, \textup{kg}^{-1} \, \textup{K}^{-1}$, and air is taken to have a heat capacity ratio of $\gamma_\textup{0,Air} = 1.4$, a pressure constant of $\Pi_\textup{0,Air} =0 \, \textup{Pa}$ and a specific gas constant of $R_\textup{0,Air} = 288.0 \, \textup{J} \, \textup{kg}^{-1} \, \textup{K}^{-1}$. The shock is initially located at $x_0 = 1.1 \, \textup{m}$, the air bubble occupies the interval $1.3 \leq x \leq 1.5$ and the applied time-step corresponds to $\textup{Co} = a_\textup{II,Water} \Delta t/\Delta x = 0.45$.

The results are obtained on equidistant meshes with three mesh resolutions, resolving the one-dimensional domain with $400$, $2000$ and $10000$ cells, which corresponds to a spatial resolution of $40$, $200$ and $1000$ cells for the initial length of the air bubble, respectively, as considered in the previous section for the helium bubble in air. As observed in Fig.~\ref{fig:AirWatert4p0em4M1p1}, the density is principally in good agreement on all three meshes, whereas the temperature distribution inside the bubble appears to be especially sensitive to the mesh resolution, with visible differences between the results obtained on the coarsest mesh compared to the results obtained on the two meshes with higher resolution. Furthermore, the pressure upstream of the water-air interface after the shock wave has passed exhibits a considerable dependency on the mesh resolution, as seen in Fig.~\ref{fig:AirWaterPresM1p1}. The pressure is significantly underpredicted compared to the exact Riemann solution at both time-instances shown in Fig.~\ref{fig:AirWaterPresM1p1}, with an underprediction of approximately $1.5 \times  10^5 \, \textup{Pa}$ on the coarsest mesh and approximately $0.6 \times  10^5 \, \textup{Pa}$ on the finest mesh. Despite these differences in pressure and temperature, the position of the primary shock wave as it travels through the bubble is, apart from the sharpness of the discontinuity, still in very good agreement on the meshes corresponding to $200$ and $1000$ cells for the initial length of the air bubble. On the coarsest considered mesh, corresponding to $40$ cells for the initial length of the air bubble, however, the position of the shock wave has an offset in the downstream direction, which may be attributed to an overprediction of the speed of sound as the shock wave passes the interface.

\subsection{Two-dimensional R22-bubble in air}
\label{sec:R22Air2D}

\begin{figure}[t]
\begin{center}
\includegraphics{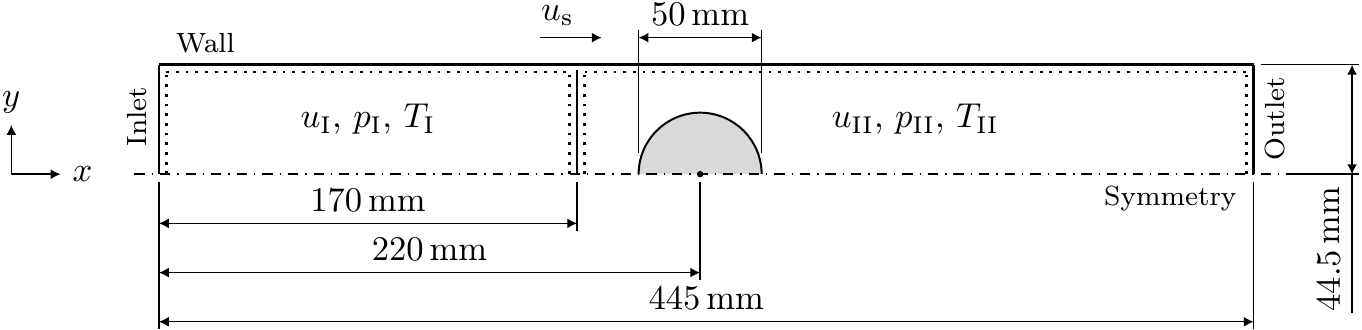}
\caption{Schematic illustration of the computational setup of the
two-dimensional R22 bubble in air interacting with a shock wave with Mach number $M_\textup{s} = 1.22$. The shock wave is initially located at $x = 0.17 \, \textup{m}$ and travels from left to right. The shaded area represents the bubble with a diameter of $d_0 = 0.05 \, \textup{m}$, with the bubble centre initially located at $x = 0.22 \, \textup{m}$.}
\label{fig:shockBubbleSchematic}
\end{center}
\end{figure}

The interaction of a shock wave with $M_\textup{s} = 1.22$ in air with a circular R22 bubble is simulated, a shock-bubble interaction which has previously been studied experimentally \citep{Haas1987} and numerically \citep{Quirk1996, Niederhaus2008}. The computational setup is schematically illustrated in Fig.~\ref{fig:shockBubbleSchematic}. The shock wave is initially situated at $x=0.17 \, \textup{m}$ and travels from left to right at speed $u_\textup{s}$. The shock wave separates the post-shock region ($\textup{I}$) and the pre-shock region ($\textup{II}$), which are initialised with
\begin{equation}
\begin{array}{ccc}
u_\textup{I} = 125.65 \, \textup{m} \, \textup{s}^{-1}, & p_\textup{I} =
1.59060 \times 10^5 \, \textup{Pa}, & T_\textup{I} = 402.67 \, \textup{K}, \\
u_\textup{II} = 0  \, \textup{m} \, \textup{s}^{-1}, & p_\textup{II} = 1.01325
\times 10^5 \, \textup{Pa}, & T_\textup{II} = 351.82 \, \textup{K}. \nonumber
\end{array}
\end{equation}
Air is taken to have a heat capacity ratio of $\gamma_\textup{0,Air} = 1.4$ and a specific gas constant of $R_\textup{0,Air} = 288.0 \, \textup{J} \, \textup{kg}^{-1} \, \textup{K}^{-1}$, and R22 is assumed to have a heat capacity ratio of $\gamma_\textup{0,R22} = 1.249$ and a specific gas constant of \citep{Denner2018b} $R_\textup{0,R22} = 90.885 \, \textup{J} \, \textup{kg}^{-1} \, \textup{K}^{-1}$. The applied computational mesh is equidistant and Cartesian, and the applied time-step corresponds to a Courant number of $\textup{Co} = a_\textup{Air,II} \Delta t / \Delta x = 0.38$.

\begin{figure}[t!]
\begin{center}
\includegraphics[width=0.16\textwidth]{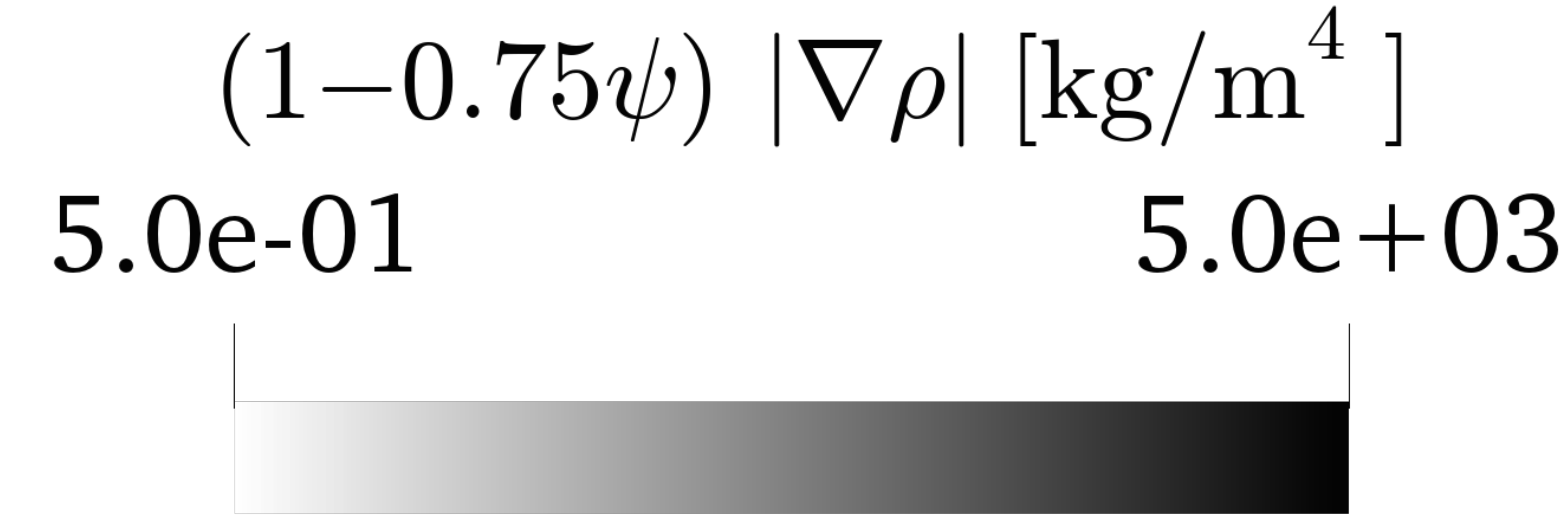} \qquad
\includegraphics[width=0.16\textwidth]{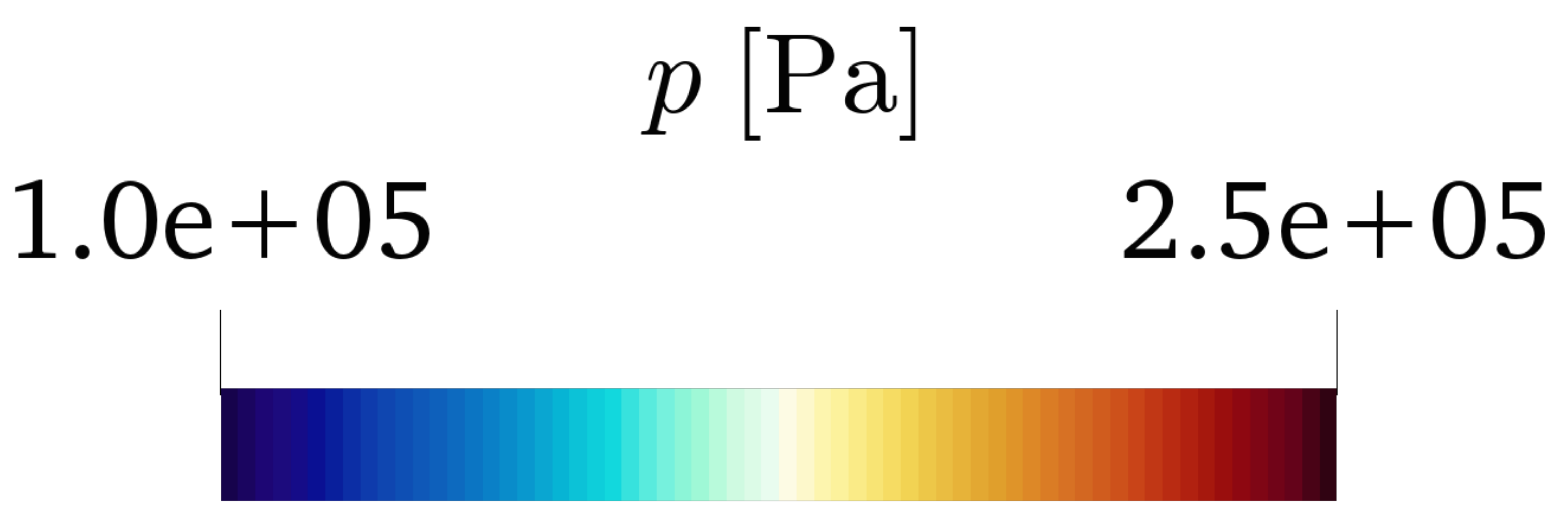}\\ 
\subfloat[$\Delta x = d_0/200$]
{\includegraphics[width=0.325\textwidth]{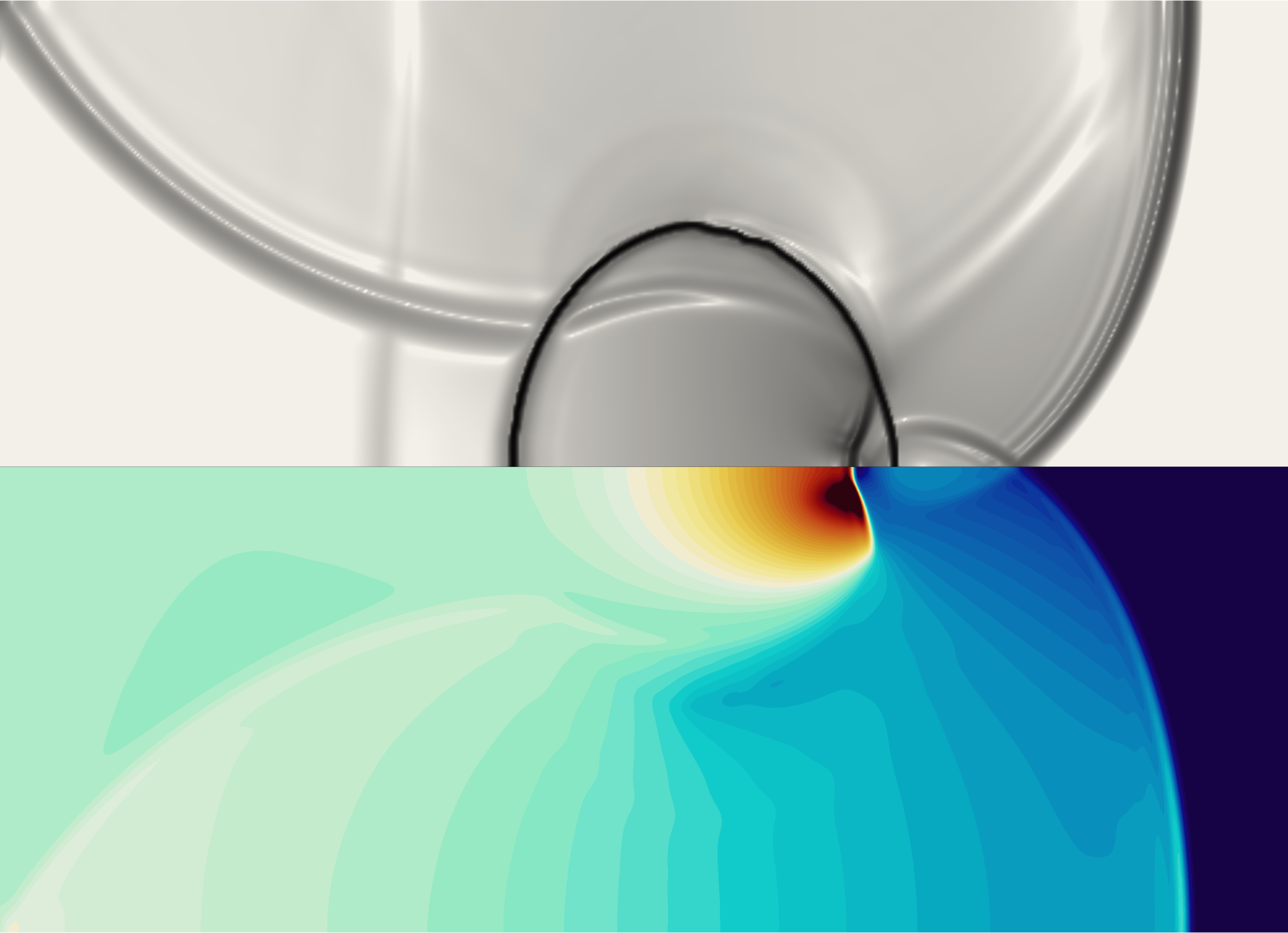}}
\
\subfloat[$\Delta x = d_0/300$]
{\includegraphics[width=0.325\textwidth]{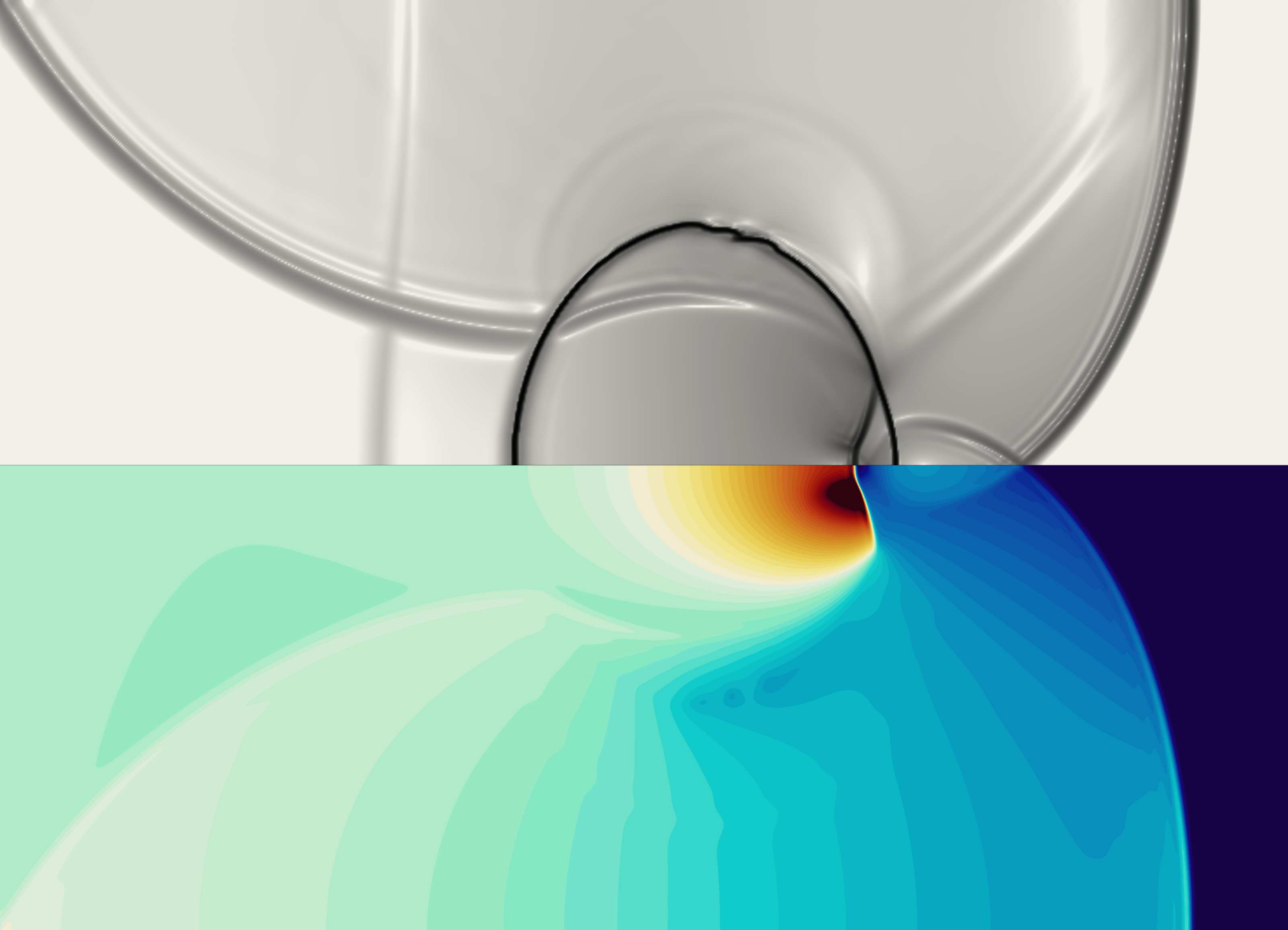}}
\
\subfloat[$\Delta x = d_0/500$]
{\includegraphics[width=0.325\textwidth]{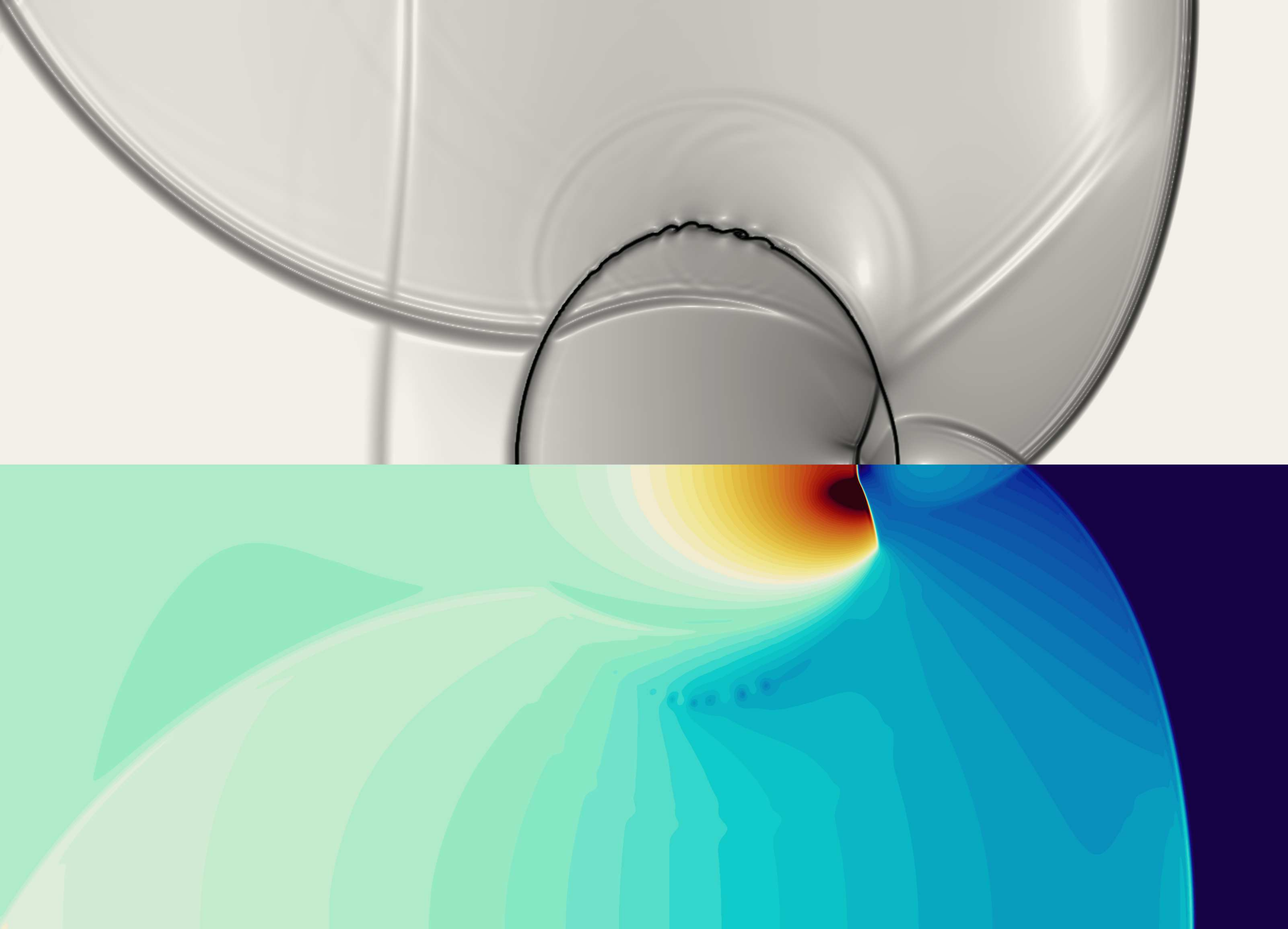}}
\caption{Contours of the density gradient $(1-0.75\psi) |\boldsymbol{\nabla} \rho|$ (upper half) and the pressure $p$ (lower half) of the two-dimensional shock-bubble interaction of the R22 bubble in air on a Cartesian mesh with different mesh resolutions $\Delta x$ at $\tau = t \, a_\textup{R22,II} / d_0 = 0.68$, using the Minmod scheme.}
\label{fig:R22DxTau0p68}
\end{center}
\end{figure}

\begin{figure}[t]
\begin{center}
\includegraphics[width=0.16\textwidth]{figures/R22_rhoScale.pdf} \qquad
\includegraphics[width=0.16\textwidth]{figures/R22_pScale.pdf}\\ 
\subfloat[$\Delta x = d_0/200$]
{\includegraphics[width=0.325\textwidth]{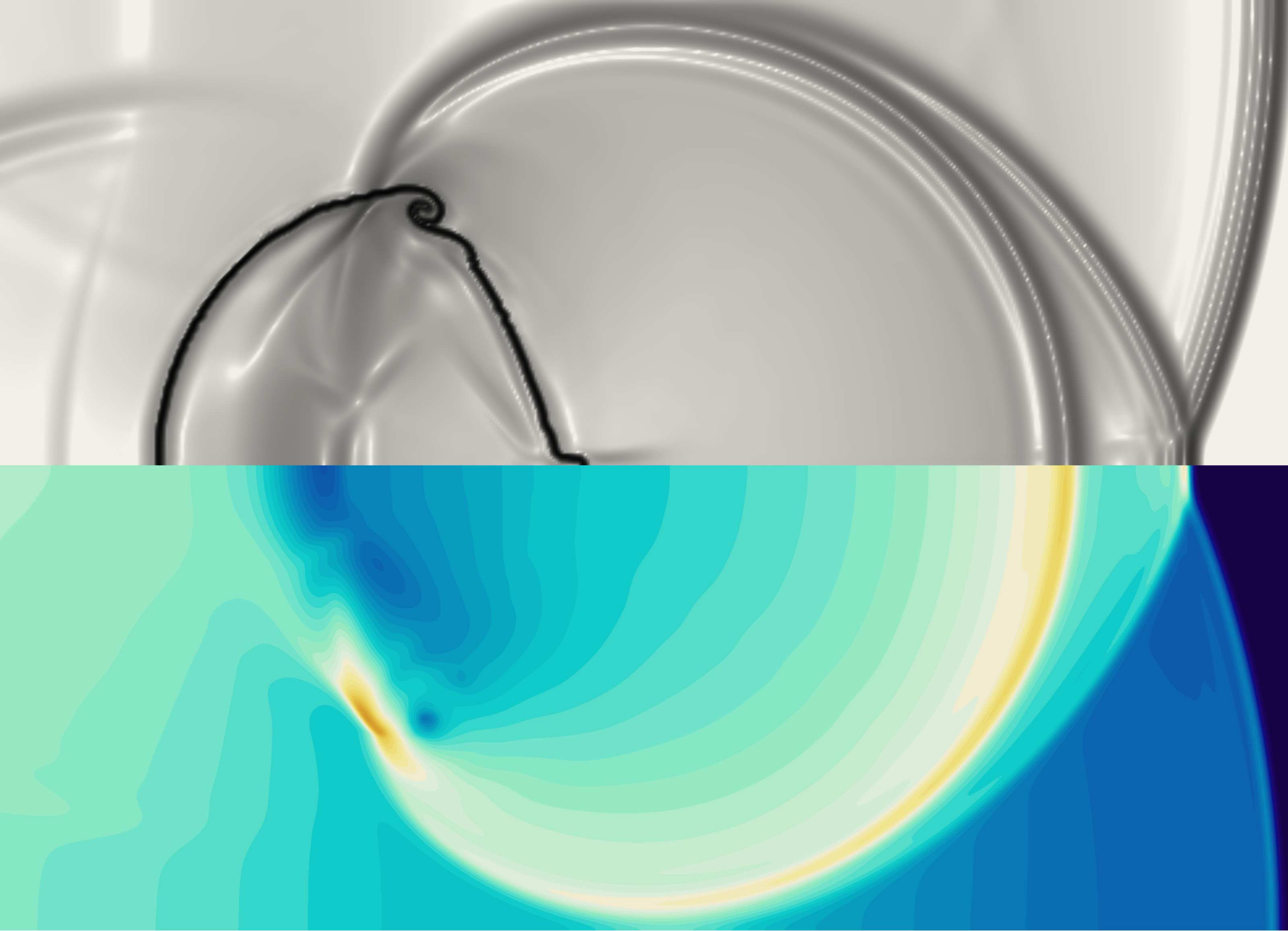}}
\
\subfloat[$\Delta x = d_0/300$]
{\includegraphics[width=0.325\textwidth]{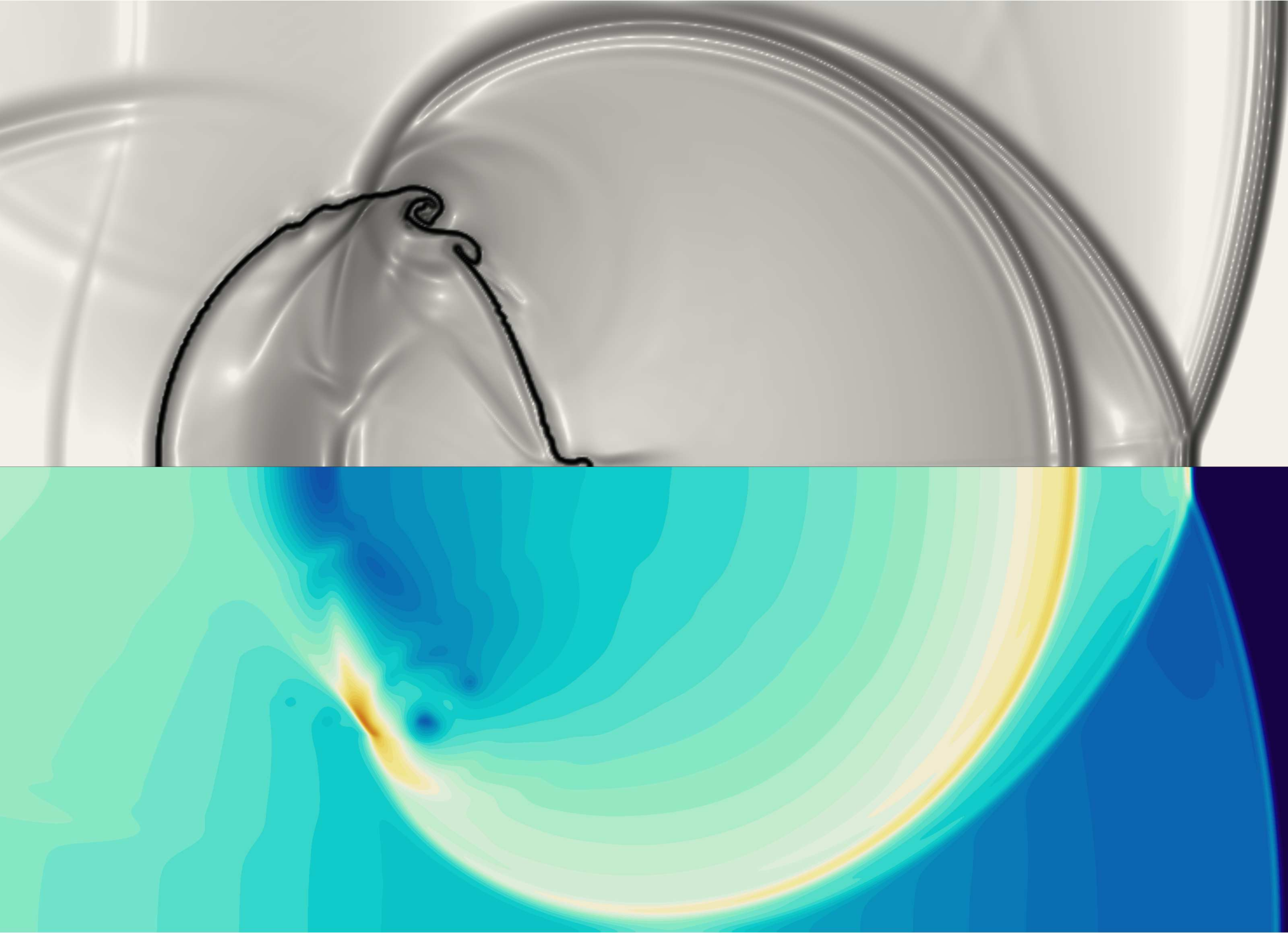}}
\
\subfloat[$\Delta x = d_0/500$]
{\includegraphics[width=0.325\textwidth]{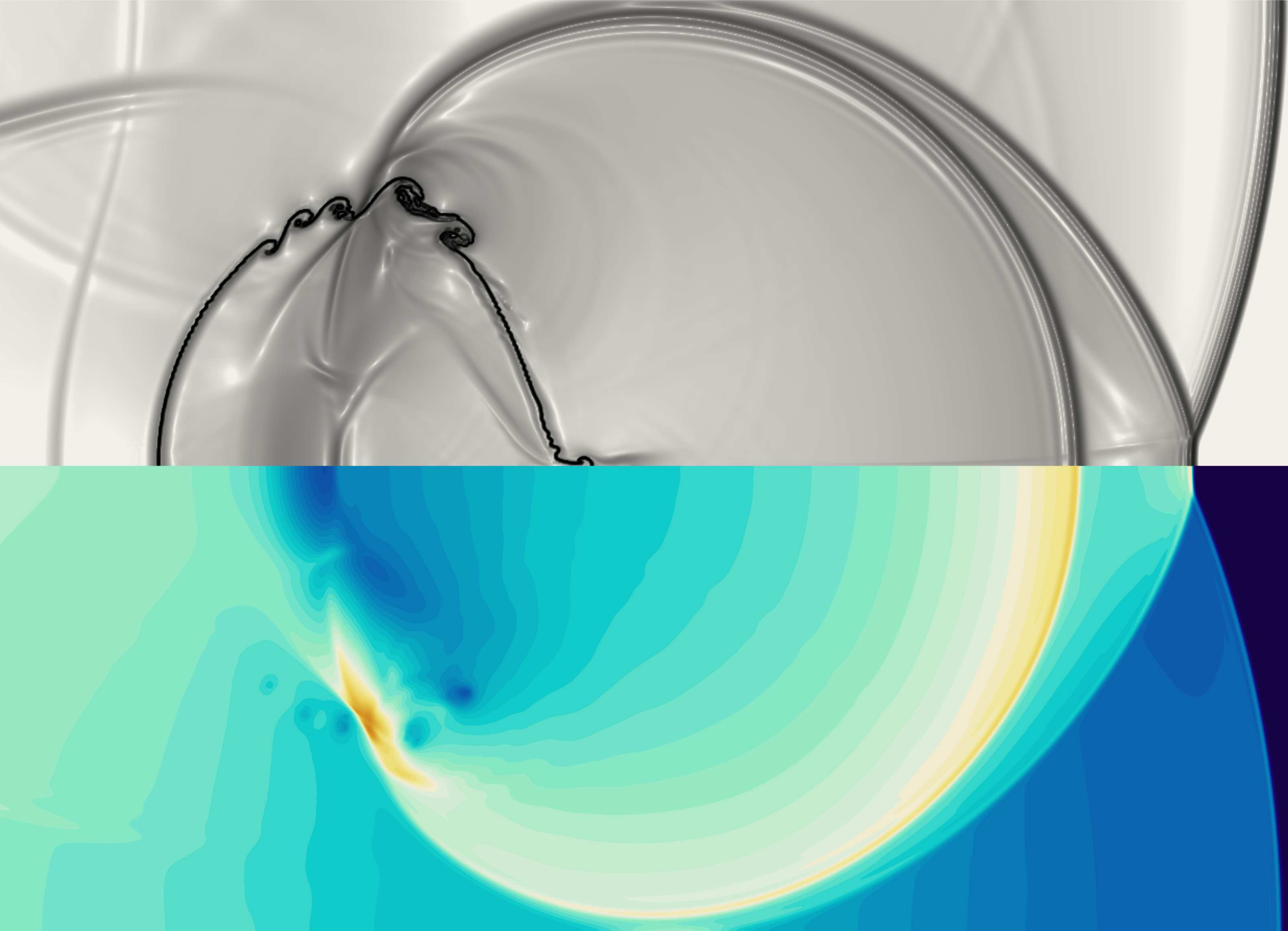}}
\caption{Contours of the density gradient $(1-0.75\psi) |\boldsymbol{\nabla} \rho|$ (upper half) and the pressure $p$ (lower half) of the two-dimensional shock-bubble interaction of the R22 bubble in air on a Cartesian mesh with different mesh resolutions $\Delta x$ at $\tau = t \, a_\textup{R22,II} / d_0 = 1.15$, using the Minmod scheme.}
\label{fig:R22DxTau1p15}
\end{center}
\end{figure}

\begin{figure}[t!]
\begin{center}
\subfloat[$\tau = 0.68$]
{\includegraphics[width=0.31\textwidth]{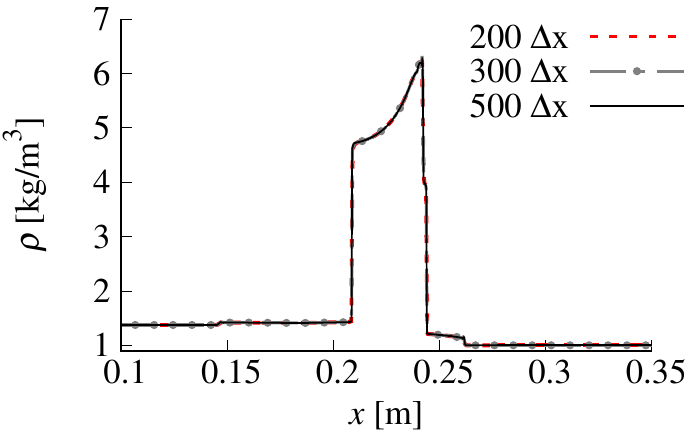}}
\quad
\subfloat[$\tau = 0.89$] 
{\includegraphics[width=0.31\textwidth]{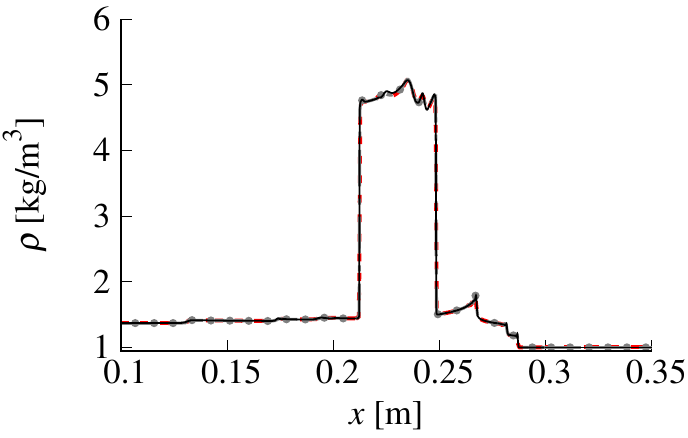}}
\quad
\subfloat[$\tau = 1.15$] 
{\includegraphics[width=0.31\textwidth]{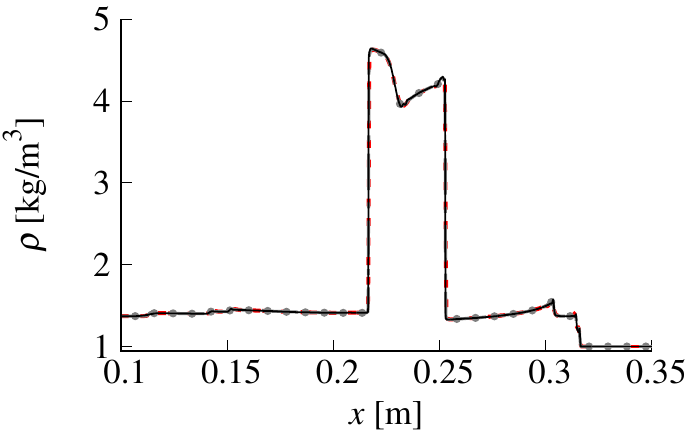}}
\caption{Profiles of the density $\rho$ along the $x$-axis at $y= 0.005 \, \textup{m}$ of the two-dimensional shock-bubble interaction of the R22 bubble in air on Cartesian meshes with different mesh spacings $\Delta x$ at different dimensionless times $\tau = t \, a_\textup{R22,II} / d_0$, using the Minmod scheme.}
\label{fig:R22profileDx005rho}
\end{center}
\end{figure}

Figures \ref{fig:R22DxTau0p68} and \ref{fig:R22DxTau1p15} show the contours of the density gradient and the pressure distribution at different dimensionless times $\tau = t \, a_\textup{R22,II} / d_0$ for equidistant Cartesian meshes with a mesh resolution of $200$, $300$ and $500$ cells per initial bubble diameter $d_0$. While the overall shape as well as the position of the bubble predicted on the different meshes are largely the same, interface instabilities with smaller structures develop as the mesh resolution increases. As mentioned in the introduction, this is to be expected, yet a coherent and sufficiently accurate description of the magnitude and frequency with which these instabilities occur in reality is presently not available. These interface instabilities generate acoustic waves, as seen in Figs.~\ref{fig:R22DxTau0p68} and \ref{fig:R22DxTau1p15}, which however do not affect the position, shape and strength of the dominant flow structures, {\em i.e.}~shock waves and rarefaction fans. In general it is noticeable in Figs.~\ref{fig:R22DxTau0p68} and \ref{fig:R22DxTau1p15}, that the overall impact of the mesh resolution on the observed flow features is minor, apart from the interface instabilities developing as a result of the passing shock wave and the resolution of the shock waves and rarefaction fans. This observation is supported by the density profiles along the $x$-axis (direction of travel of the primary shock wave) shown in Fig.~\ref{fig:R22profileDx005rho}, which exhibit very little differences for the three considered mesh resolutions.

\begin{figure}[t]
\begin{center}
\includegraphics[width=0.16\textwidth]{figures/R22_rhoScale.pdf} \qquad
\includegraphics[width=0.16\textwidth]{figures/R22_pScale.pdf}\\ 
\subfloat[Upwind]
{\includegraphics[width=0.325\textwidth]{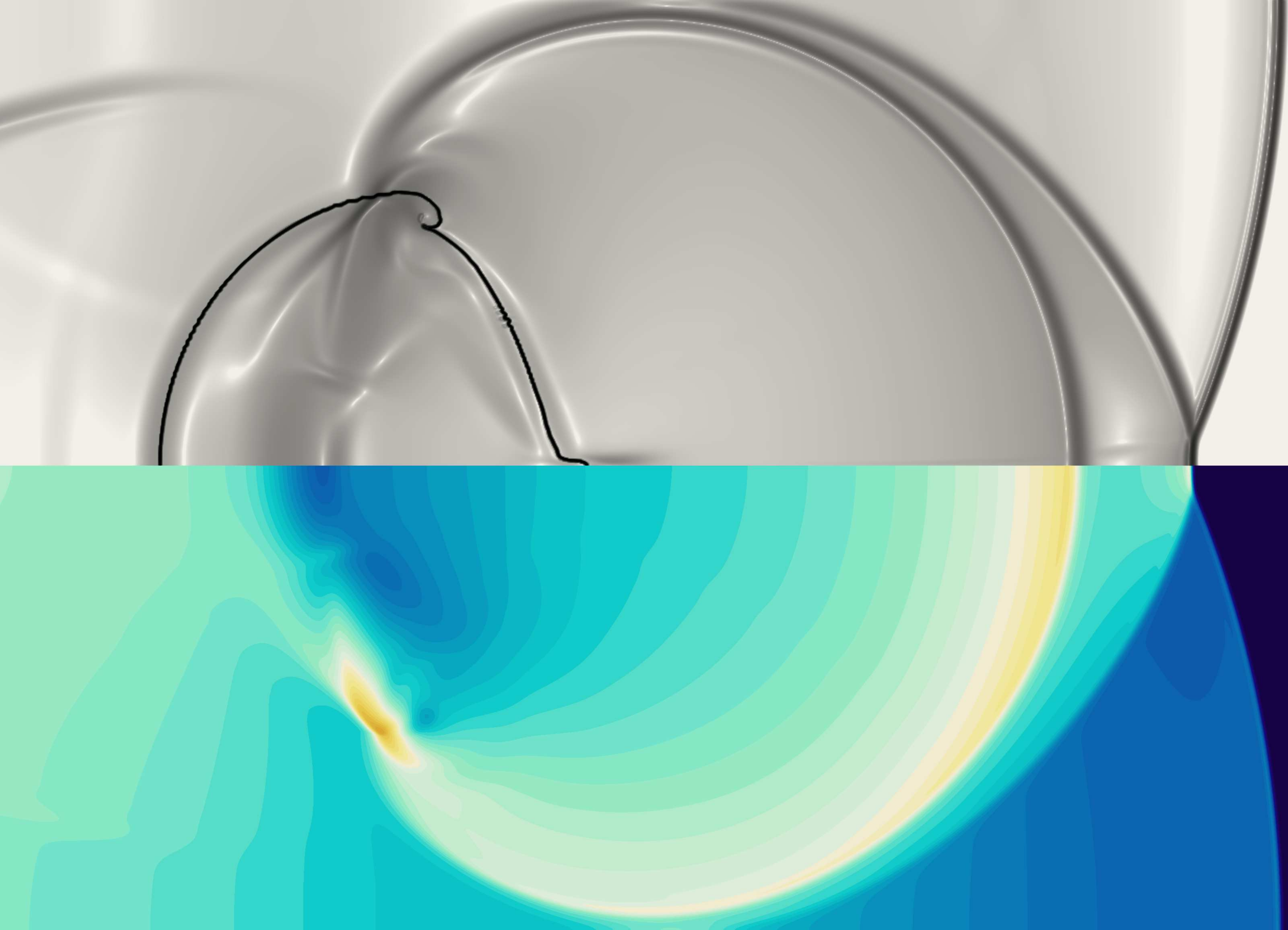}}
\
\subfloat[Minmod]
{\includegraphics[width=0.325\textwidth]{figures/R22_500dx_Minmod_tau1p15_2.pdf}}
\
\subfloat[Superbee]
{\includegraphics[width=0.325\textwidth]{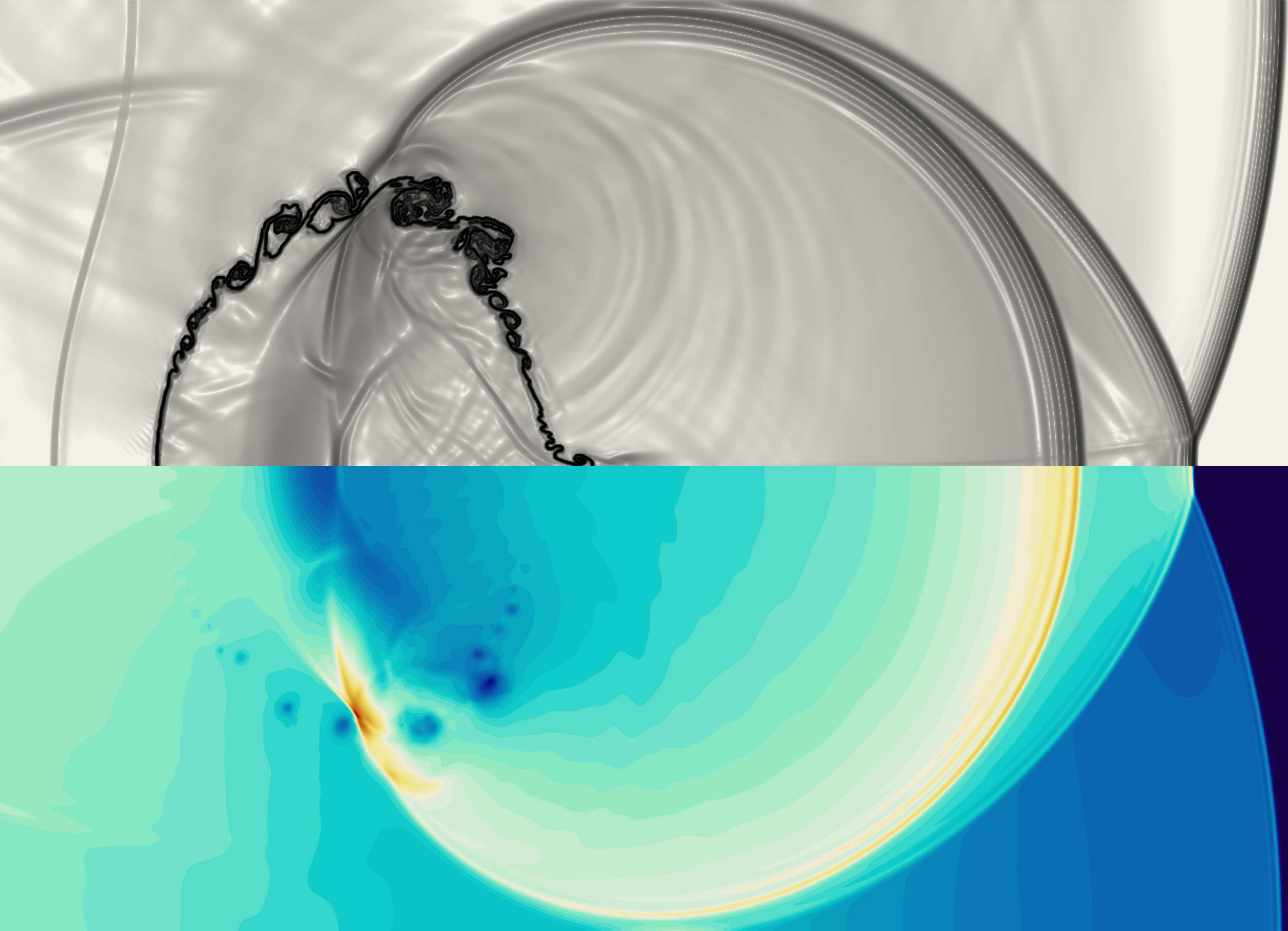}}
\caption{Contours of the density gradient $(1-0.75\psi) |\boldsymbol{\nabla} \rho|$ (upper half) and the pressure $p$ (lower half) of the two-dimensional shock-bubble interaction of the R22 bubble in air on a Cartesian mesh with $\Delta x = d_0/500$ at $\tau = t \, a_\textup{R22,II} / d_0 = 1.15$, using the first-order upwind scheme, the Minmod scheme and Superbee scheme.}
\label{fig:R22SchemeTau1p15}
\end{center}
\end{figure}

\begin{figure}[ht!]
\begin{center}
\subfloat[$\tau = 0.68$]
{\includegraphics[width=0.31\textwidth]{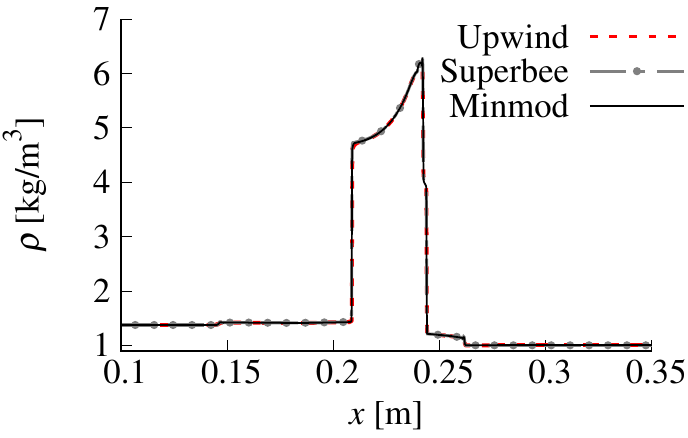}}
\quad
\subfloat[$\tau = 0.89$] 
{\includegraphics[width=0.31\textwidth]{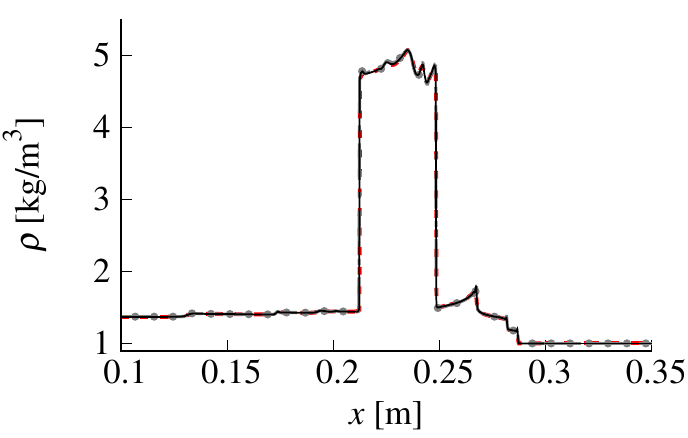}}
\quad
\subfloat[$\tau = 1.15$] 
{\includegraphics[width=0.31\textwidth]{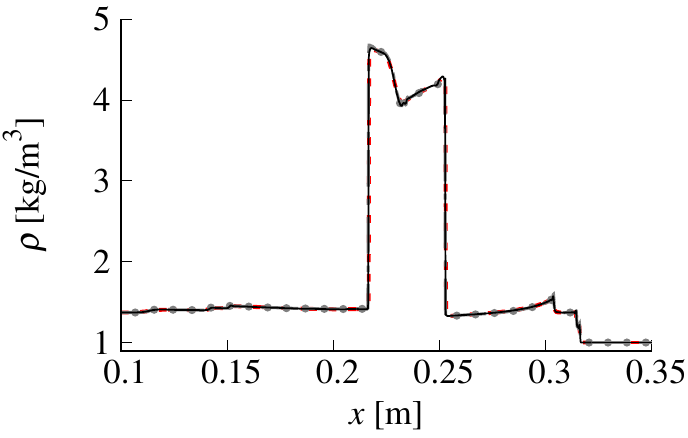}}
\caption{Profiles of the density $\rho$ along the $x$-axis at $y= 0.005 \, \textup{m}$ of the two-dimensional shock-bubble interaction of the R22 bubble in air on Cartesian meshes with mesh spacing $\Delta x = d_0/500$ at different dimensionless times $\tau = t \, a_\textup{R22,II} / d_0$, using the first-order upwind scheme, the Minmod scheme and the Superbee scheme.}
\label{fig:R22profileScheme005rho}
\end{center}
\end{figure}

Considering different TVD differencing schemes for the discretisation of advected variables (see Section \ref{sec:discretisation}) leads to very similar observations as mesh refinement; applying a more compressive differencing scheme facilitates and increases the generation of interface instabilities. Figure \ref{fig:R22SchemeTau1p15} shows the contours of the density gradient and the pressure distribution at dimensionless time $\tau = 1.15$ on a mesh with a mesh resolution of $500$ cells per initial bubble diameter $d_0$, using (in order of increasing compression) the first-order upwind scheme, the Minmod scheme and the Superbee scheme. The interface advection is unaffected by this choice and identical for all these cases, discretised as described in Section \ref{sec:compressiveVOF}. Applying different TVD advection schemes only influences the development of interface instabilities, while the position and overall shape of the R22 bubble is largely the same. The strong instabilities observed at the interface when the Superbee scheme is applied, and the ensuing acoustic waves, can be clearly observed in Fig.~\ref{fig:R22SchemeTau1p15}, yet it is also apparent that the different resolution of the discontinuities and the interface instabilities developing with the Minmod and Superbee schemes have very little influence on the position and strength of the dominant shock waves and rarefaction fans. The density profiles along the $x$-axis in
Fig.~\ref{fig:R22profileScheme005rho} support this observation.

\subsection{Two-dimensional air-bubble in water}
\label{sec:AirWater2D}

\begin{figure}[t!]
\begin{center}
\includegraphics{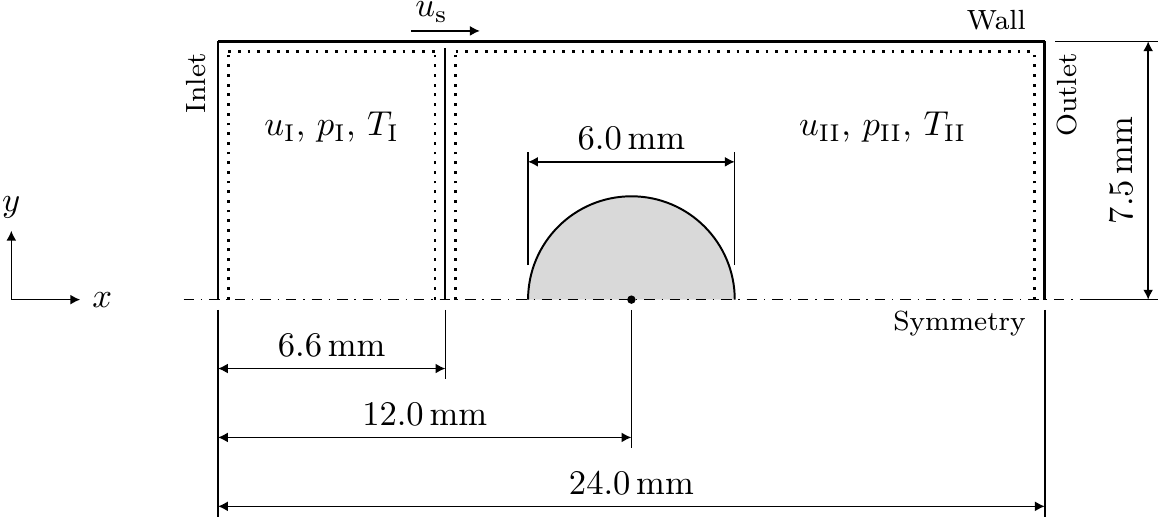}
\caption{Schematic illustration of the computational setup of the
air bubble in water interacting with a shock wave with Mach number $M_\textup{s} = 1.72$. The shock wave is initially located at $x_{\textup{s},0} = 6.6 \times 10^{-3} \, \textup{m}$ and travels from left to right. The shaded area represents the air bubble with a diameter of $d_0 = 6.0 \times 10^{-3} \,
\textup{m}$, with the bubble centre initially located at $x_{\textup{b},0} = 12.0 \times 10^{-3} \, \textup{m}$.}
\label{fig:bubbleCollapseSchematic}
\end{center}
\end{figure}

The interaction of a shock wave with $M_\textup{s} = 1.72$ in water with a circular air bubble is simulated, as considered previously in other studies \citep{Nourgaliev2006, Shukla2014, Haimovich2017, Goncalves2019}. The computational setup is schematically illustrated in Fig.~\ref{fig:bubbleCollapseSchematic}. The shock is initially situated at $x= 6.6 \times 10^{-3} \, \textup{m}$ and travels from left to right at speed $u_\textup{s}$. The shock wave separates the post-shock region ($\textup{I}$) and the pre-shock region ($\textup{II}$), which are initialised with
\begin{equation}
\begin{array}{ccc}
u_\textup{I} = 685.25 \, \textup{m} \, \textup{s}^{-1}, & p_\textup{I} =
1.91530 \times 10^5 \, \textup{Pa}, & T_\textup{I} = 381.80 \, \textup{K}, \\
u_\textup{II} = 0  \, \textup{m} \, \textup{s}^{-1}, & p_\textup{II} = 10^5 \,
\textup{Pa}, & T_\textup{II} = 293.15 \, \textup{K}. \nonumber
\end{array}
\end{equation}
Water is taken to have a heat capacity ratio of  $\gamma_\textup{0,Water} = 4.1$, a pressure constant of $\Pi_\textup{0,Water} = 4.4 \times 10^{8} \, \textup{Pa}$ and a specific gas constant of $R_\textup{0,Water} = 6000 \, \textup{J} \, \textup{kg}^{-1} \, \textup{K}^{-1}$, and air is taken to have a heat capacity ratio of $\gamma_\textup{0,Air} = 1.4$, a pressure constant of $\Pi_\textup{0,Air} =0 \, \textup{Pa}$ and a specific gas constant of $R_\textup{0,Air} = 288.0 \, \textup{J} \, \textup{kg}^{-1} \, \textup{K}^{-1}$. The applied computational mesh is equidistant and Cartesian, and the applied time-step corresponds to a Courant number of $\textup{Co} = a_\textup{Water,II} \Delta t / \Delta x = 0.11$.

\begin{figure}[t]
\begin{center}
\includegraphics[width=0.143\textwidth]{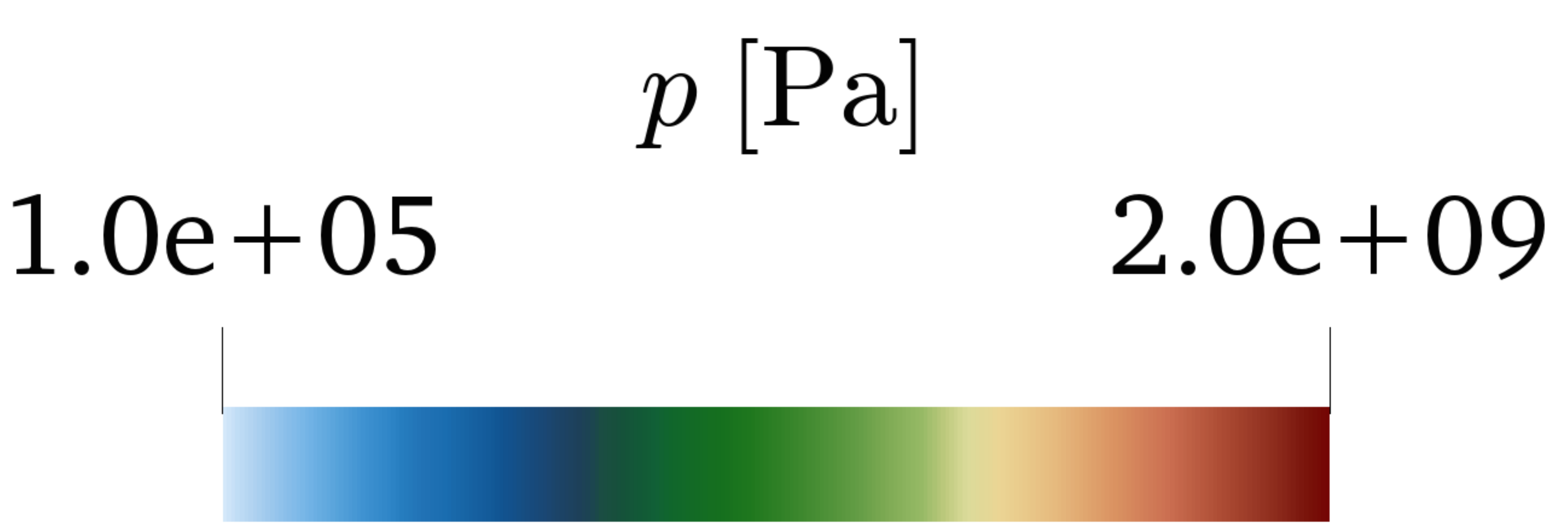}
\ \
\includegraphics[width=0.143\textwidth]{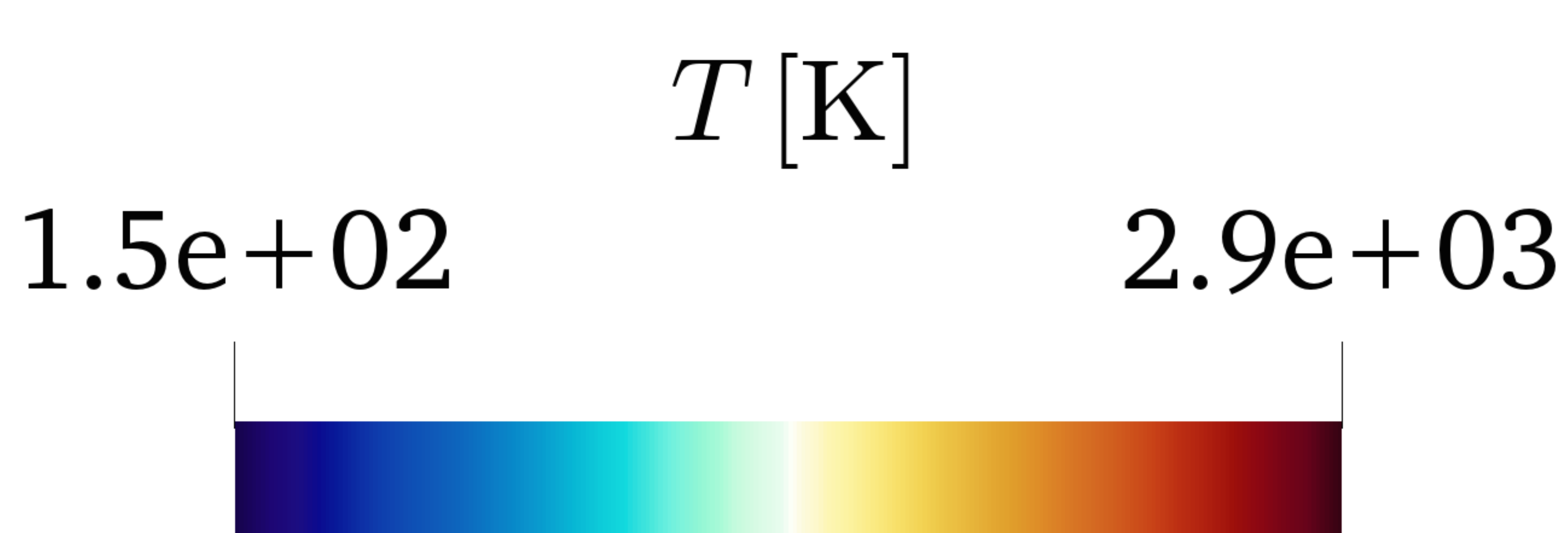}
\quad
\includegraphics[width=0.143\textwidth]{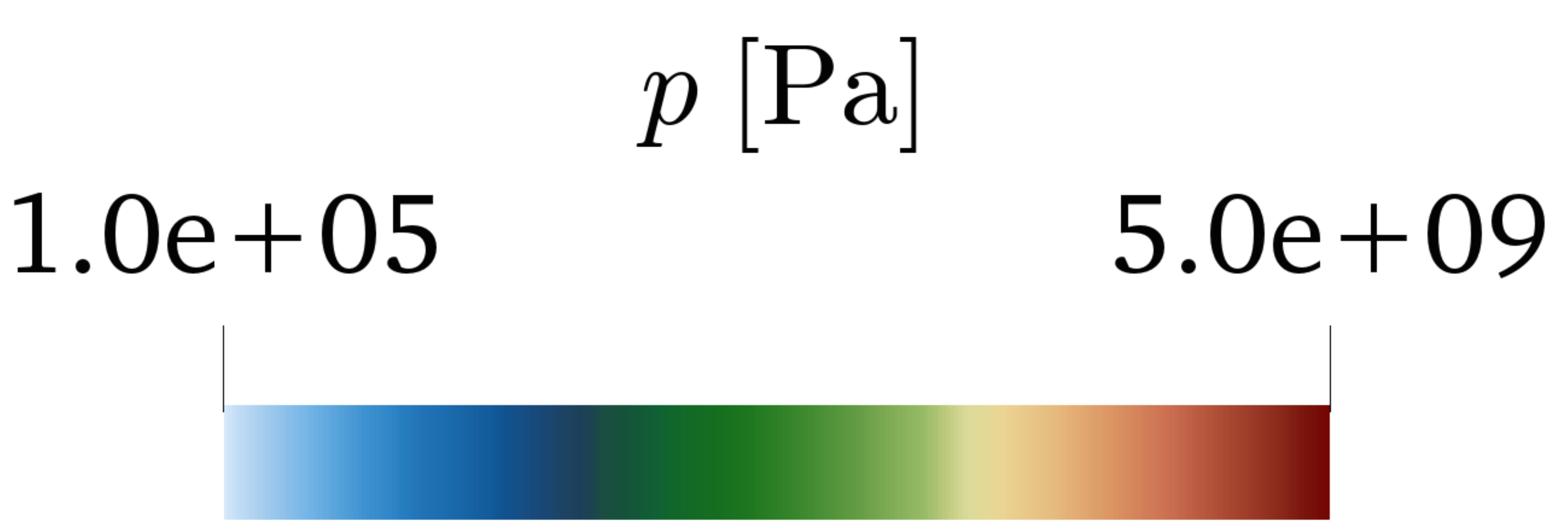}
\ \
\includegraphics[width=0.143\textwidth]{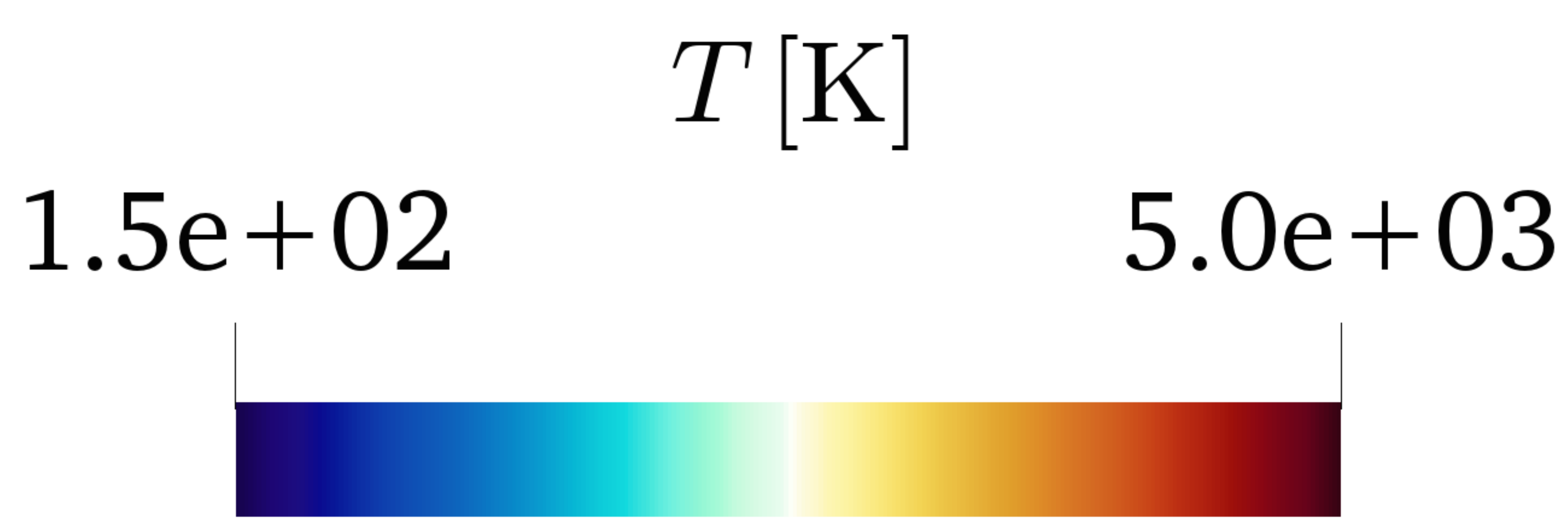}
\quad
\includegraphics[width=0.143\textwidth]{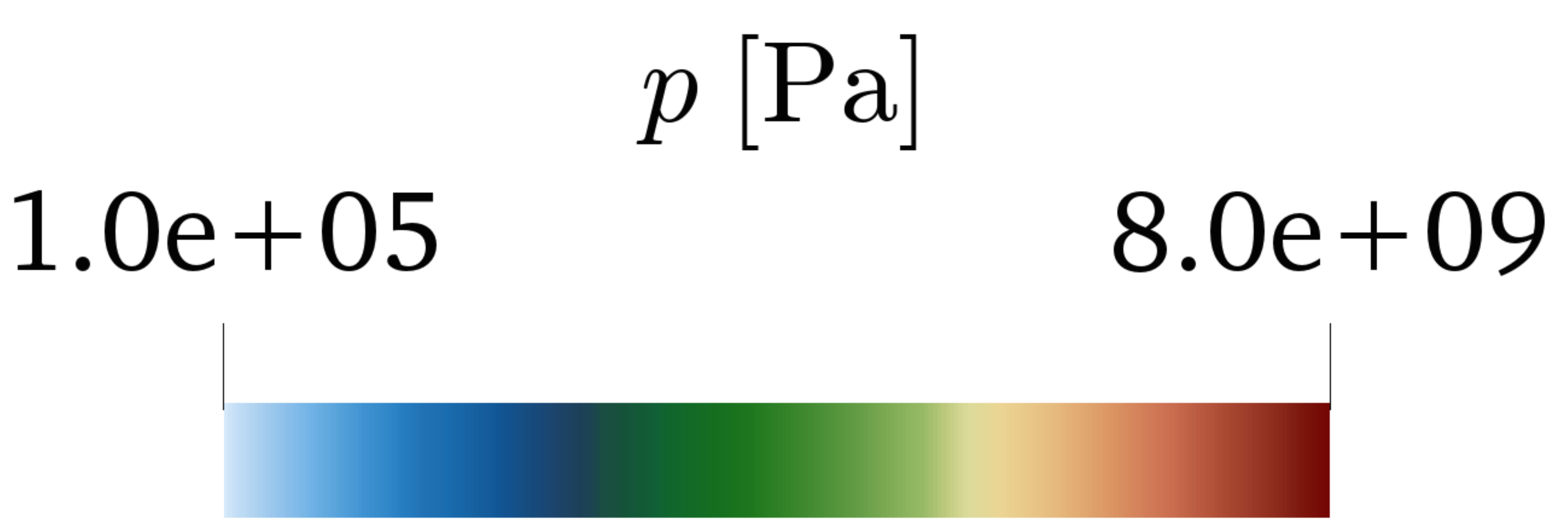}
\ \
\includegraphics[width=0.143\textwidth]{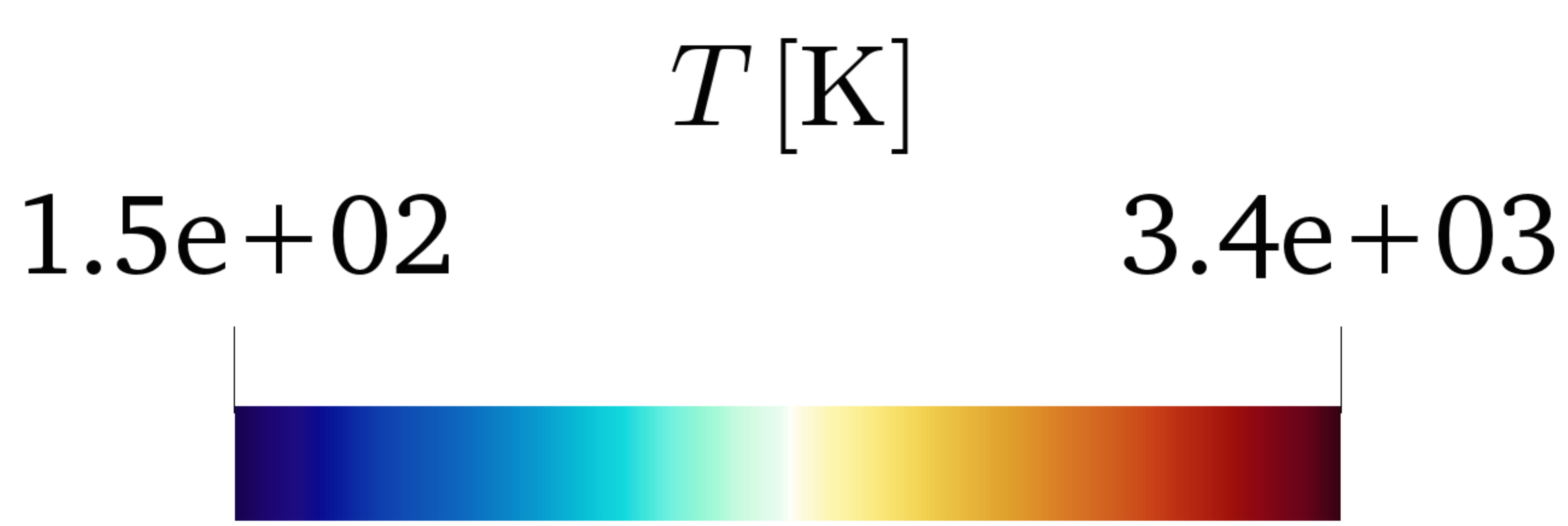}
\\
\subfloat[$t=3.0 \, \mu \textup{s}$]
{\includegraphics[trim= 6cm 14cm 15cm 4cm, clip=true,
width=0.325\textwidth]{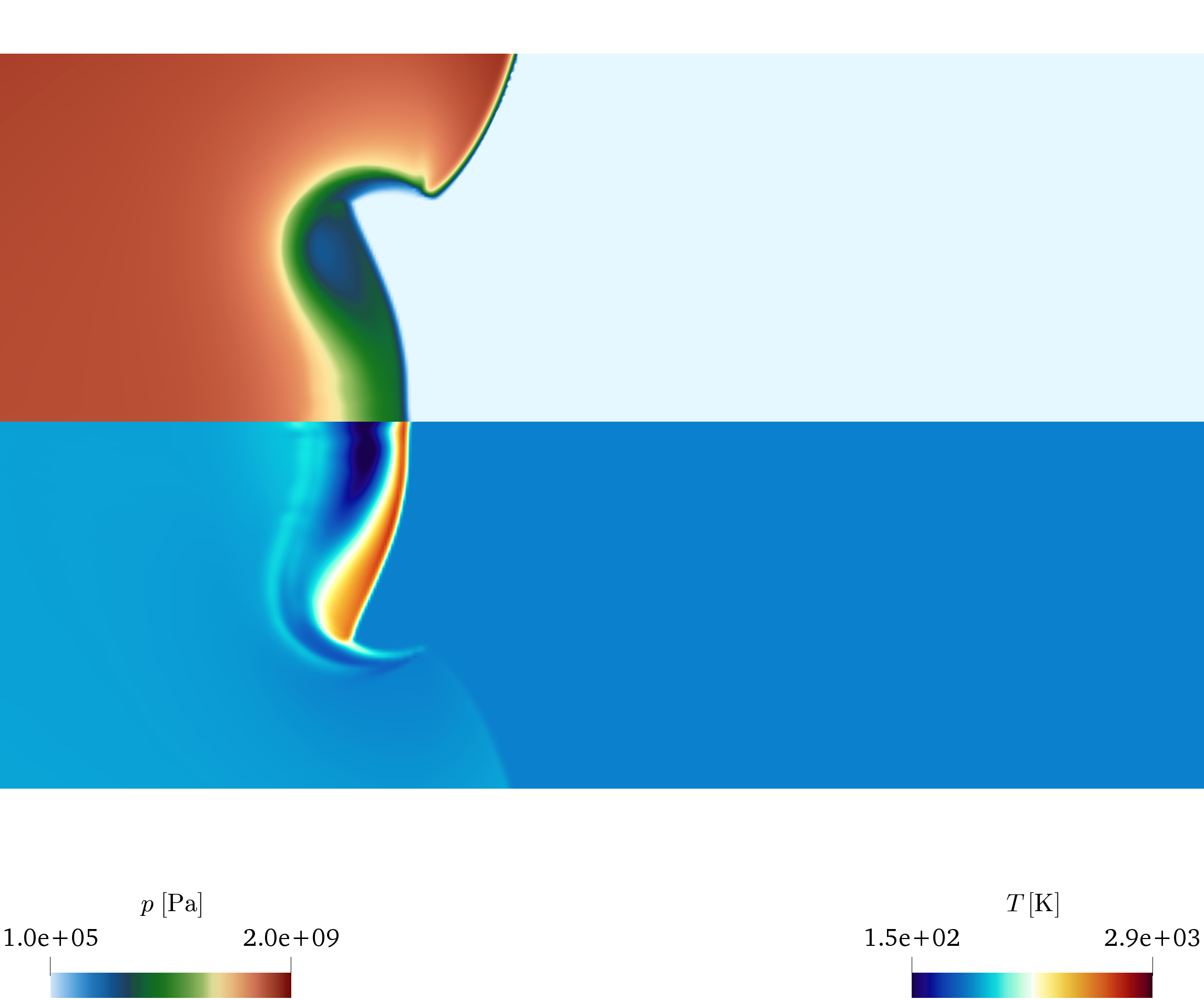}}
\
\subfloat[$t=3.8 \, \mu \textup{s}$]
{\includegraphics[trim= 6cm 14cm 15cm 4cm, clip=true,
width=0.325\textwidth]{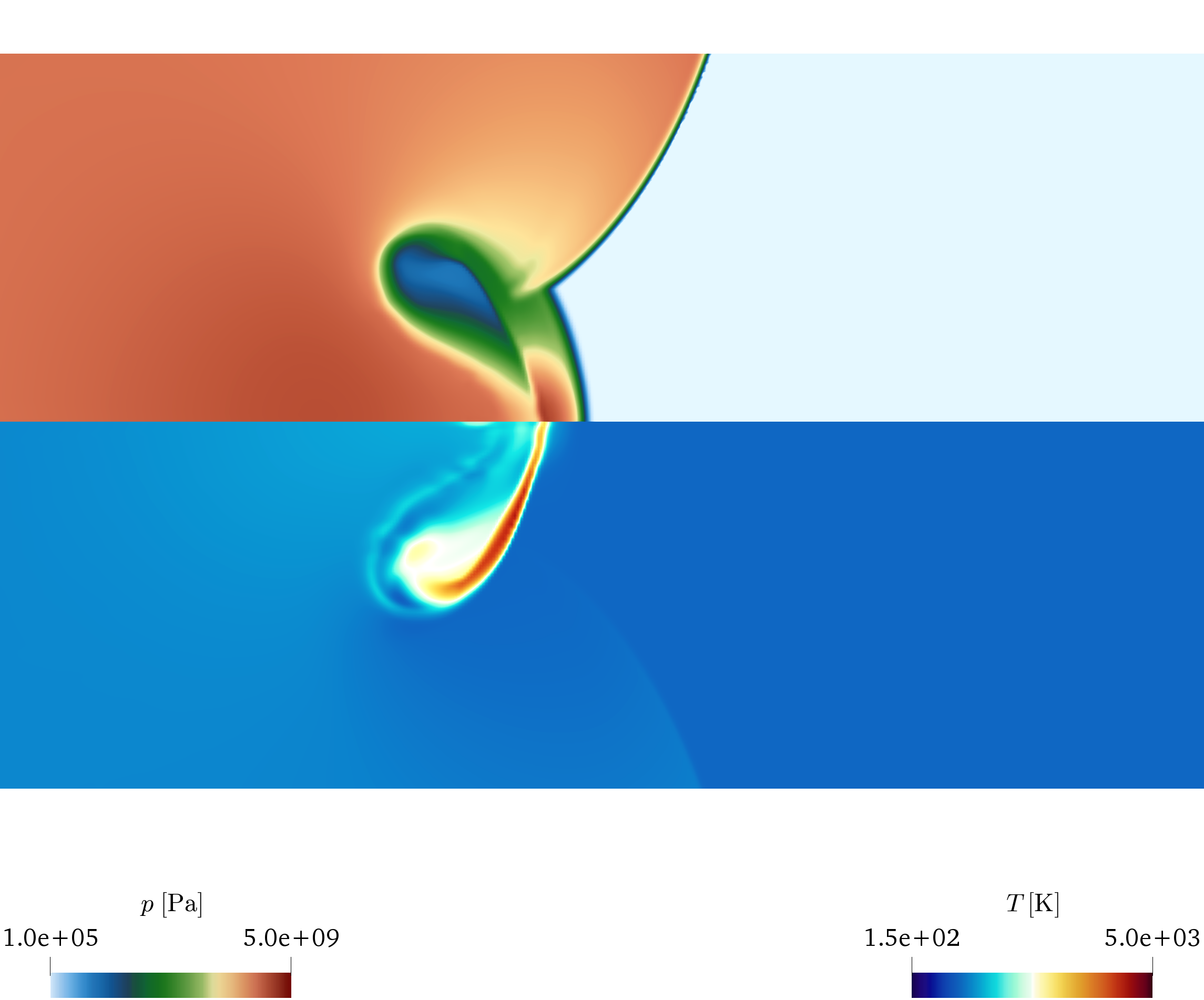}}
\
\subfloat[$t=4.5 \, \mu \textup{s}$]
{\includegraphics[trim= 6cm 14cm 15cm 4cm, clip=true,
width=0.325\textwidth]{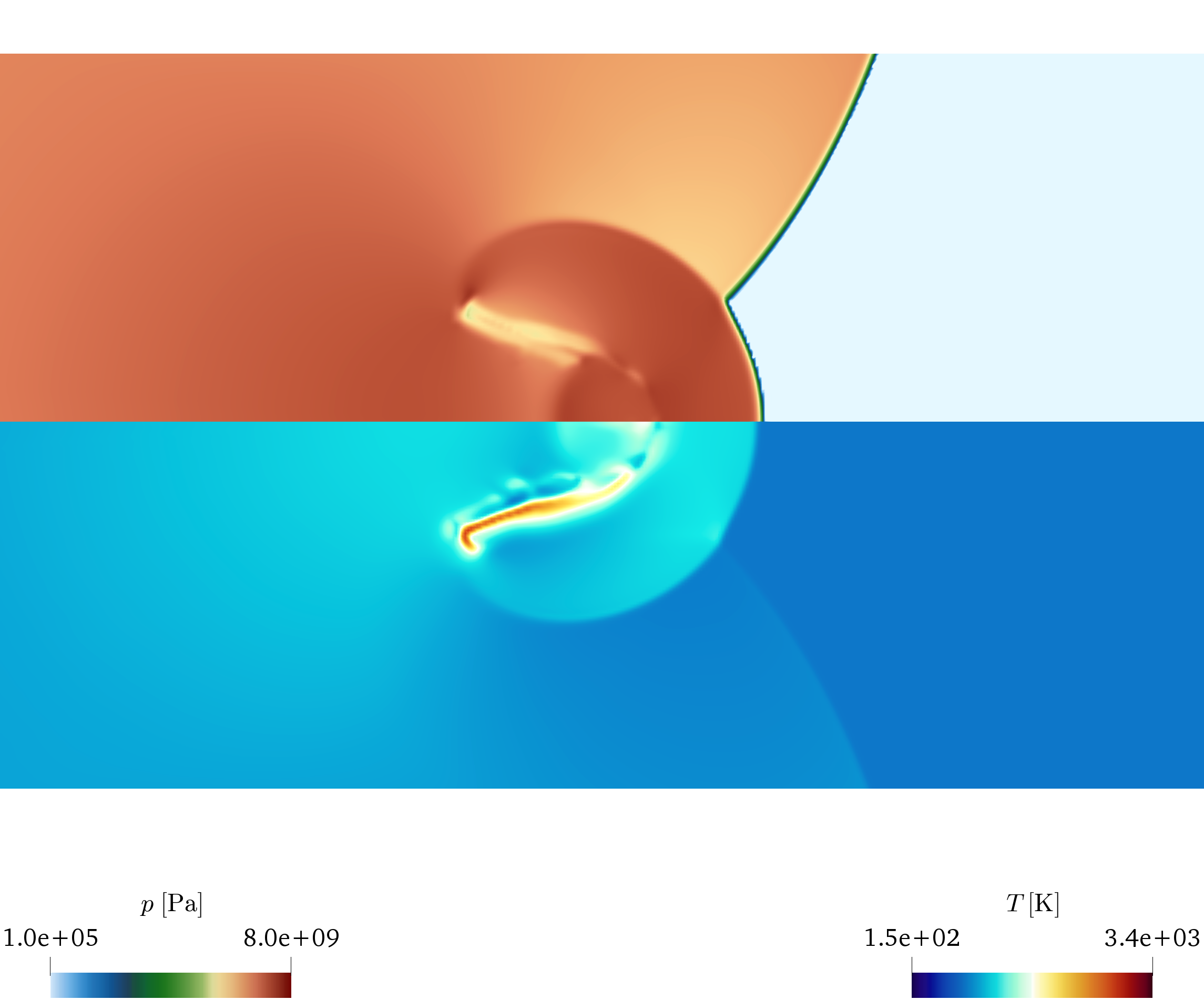}}
\caption{Contours of the pressure $p$ (upper half) and the temperature $T$ (lower half) of the two-dimensional shock-bubble interaction of the air bubble in water on a Cartesian mesh with $\Delta x = d_0/200$ at different times $t$. Both the pressure scale and the temperature scale are logarithmic.}
\label{fig:bubbleCollapseDx200}
\end{center}
\end{figure}

\begin{figure}[h!]
\begin{center}
\includegraphics[width=0.143\textwidth]{figures/BubbleCollapse_pScale_3p0mus.pdf}
\ \
\includegraphics[width=0.143\textwidth]{figures/BubbleCollapse_TScale_3p0mus.pdf}
\quad
\includegraphics[width=0.143\textwidth]{figures/BubbleCollapse_pScale_3p8mus.pdf}
\ \
\includegraphics[width=0.143\textwidth]{figures/BubbleCollapse_TScale_3p8mus.pdf}
\quad
\includegraphics[width=0.143\textwidth]{figures/BubbleCollapse_pScale_4p5mus.pdf}
\ \
\includegraphics[width=0.143\textwidth]{figures/BubbleCollapse_TScale_4p5mus.pdf}
\\
\subfloat[$t=3.0 \, \mu \textup{s}$]
{\includegraphics[trim= 6cm 14cm 15cm 4cm, clip=true,
width=0.325\textwidth]{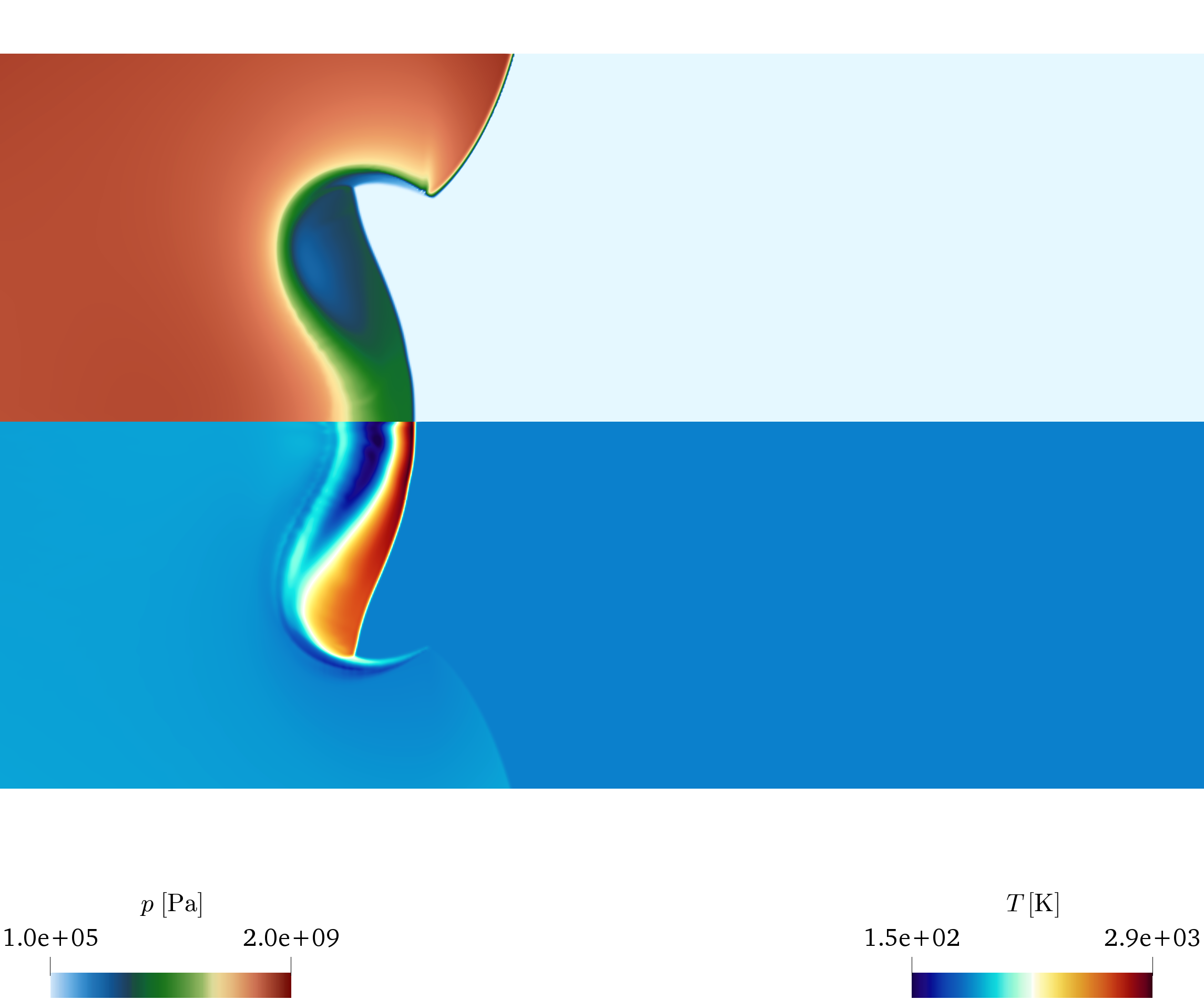}}
\
\subfloat[$t=3.8 \, \mu \textup{s}$]
{\includegraphics[trim= 6cm 14cm 15cm 4cm, clip=true,
width=0.325\textwidth]{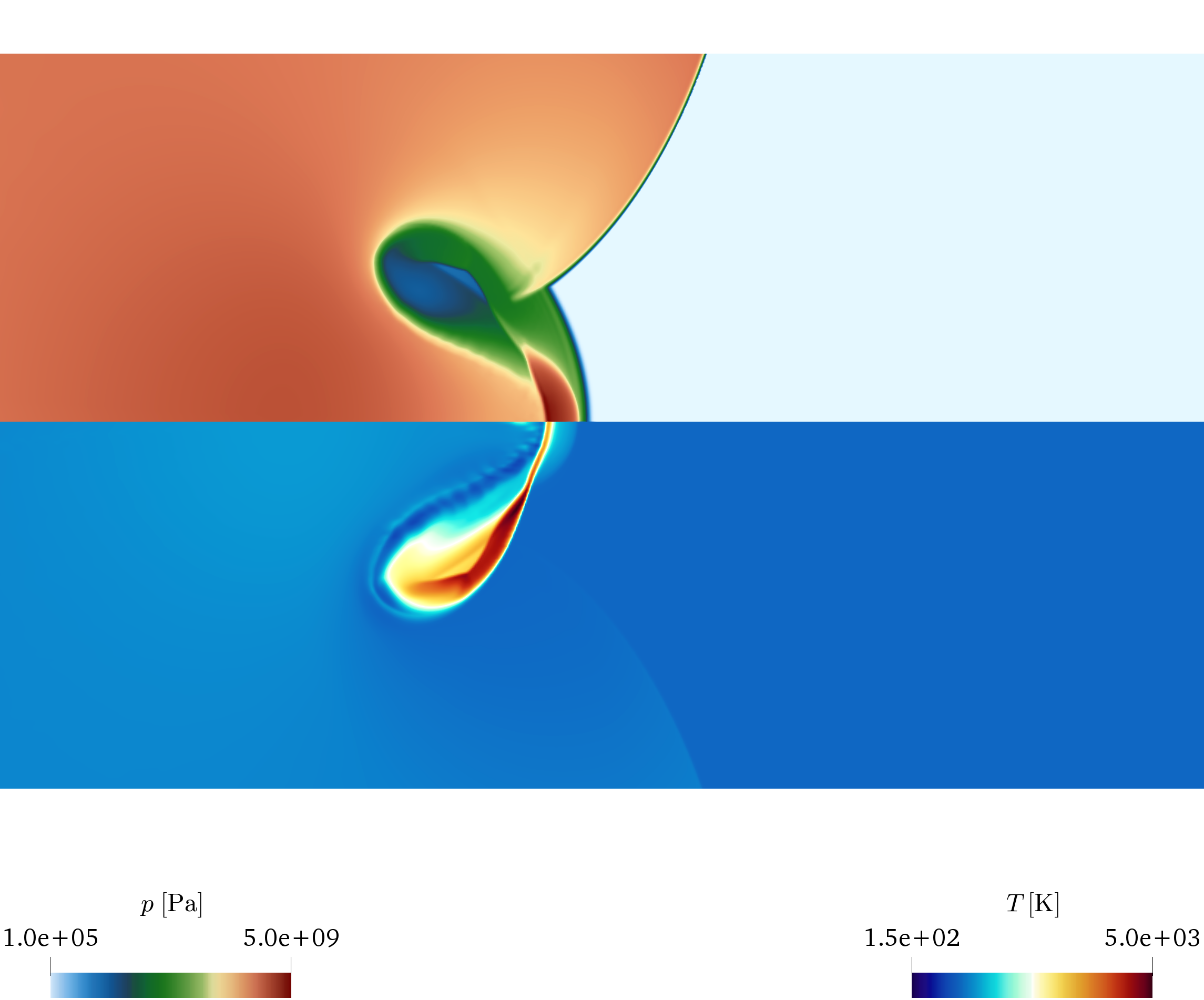}}
\
\subfloat[$t=4.5 \, \mu \textup{s}$]
{\includegraphics[trim= 6cm 14cm 15cm 4cm, clip=true,
width=0.325\textwidth]{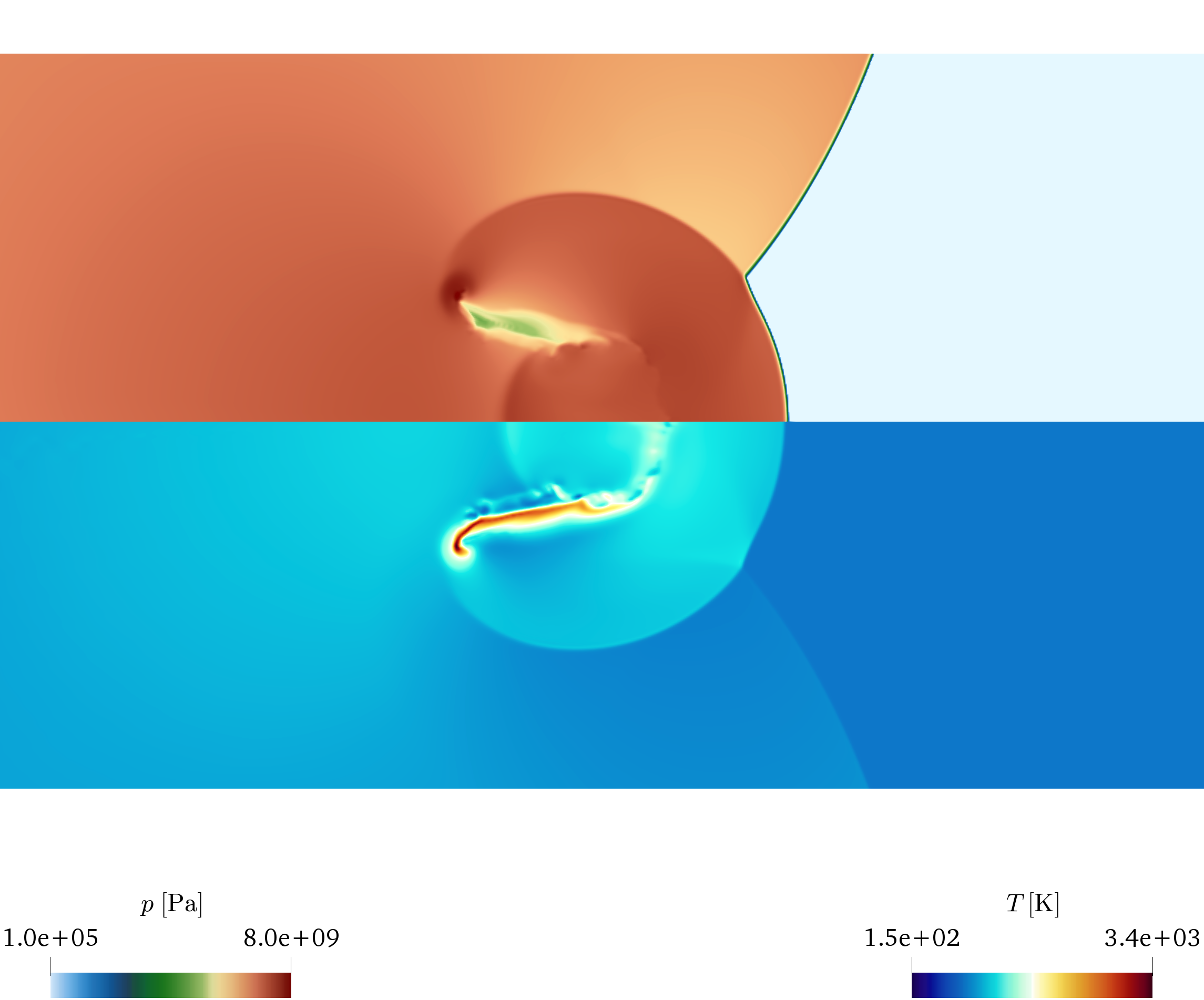}}
\caption{Contours of the pressure $p$ (upper half) and the temperature $T$ (lower half) of the two-dimensional shock-bubble interaction of the air bubble in water on a Cartesian mesh with $\Delta x = d_0/400$ at different times $t$. Both the pressure scale and the temperature scale are logarithmic.}
\label{fig:bubbleCollapseDx400}
\end{center}
\end{figure}

\begin{figure}[h!]
\begin{center}
\includegraphics[width=0.143\textwidth]{figures/BubbleCollapse_pScale_3p0mus.pdf}
\ \
\includegraphics[width=0.143\textwidth]{figures/BubbleCollapse_TScale_3p0mus.pdf}
\quad
\includegraphics[width=0.143\textwidth]{figures/BubbleCollapse_pScale_3p8mus.pdf}
\ \
\includegraphics[width=0.143\textwidth]{figures/BubbleCollapse_TScale_3p8mus.pdf}
\quad
\includegraphics[width=0.143\textwidth]{figures/BubbleCollapse_pScale_4p5mus.pdf}
\ \
\includegraphics[width=0.143\textwidth]{figures/BubbleCollapse_TScale_4p5mus.pdf}
\\
\subfloat[$t=3.0 \, \mu \textup{s}$]
{\includegraphics[trim= 6cm 14cm 15cm 4cm, clip=true,
width=0.325\textwidth]{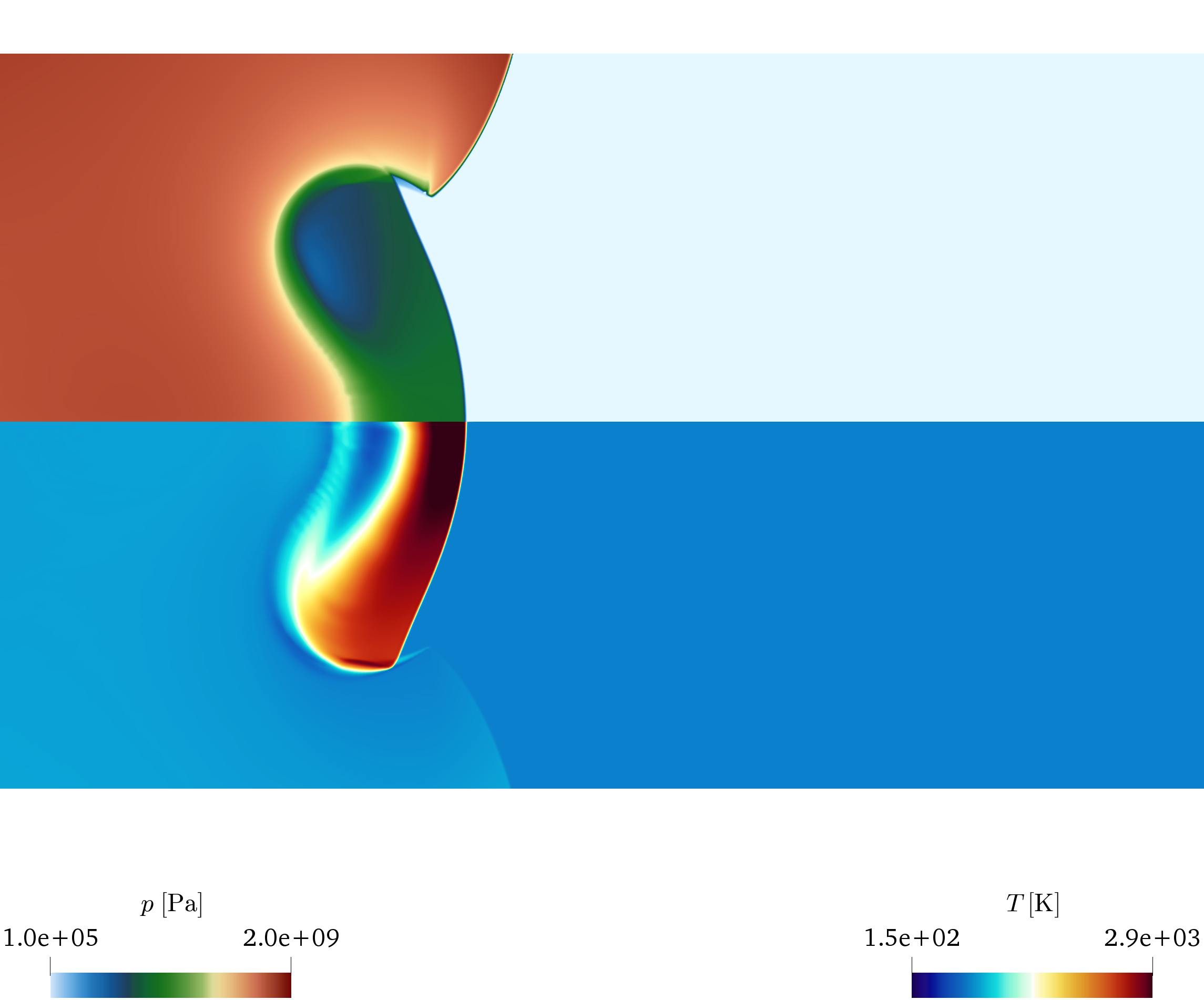}} \
\subfloat[$t=3.8 \, \mu \textup{s}$]
{\includegraphics[trim= 6cm 14cm 15cm 4cm, clip=true,
width=0.325\textwidth]{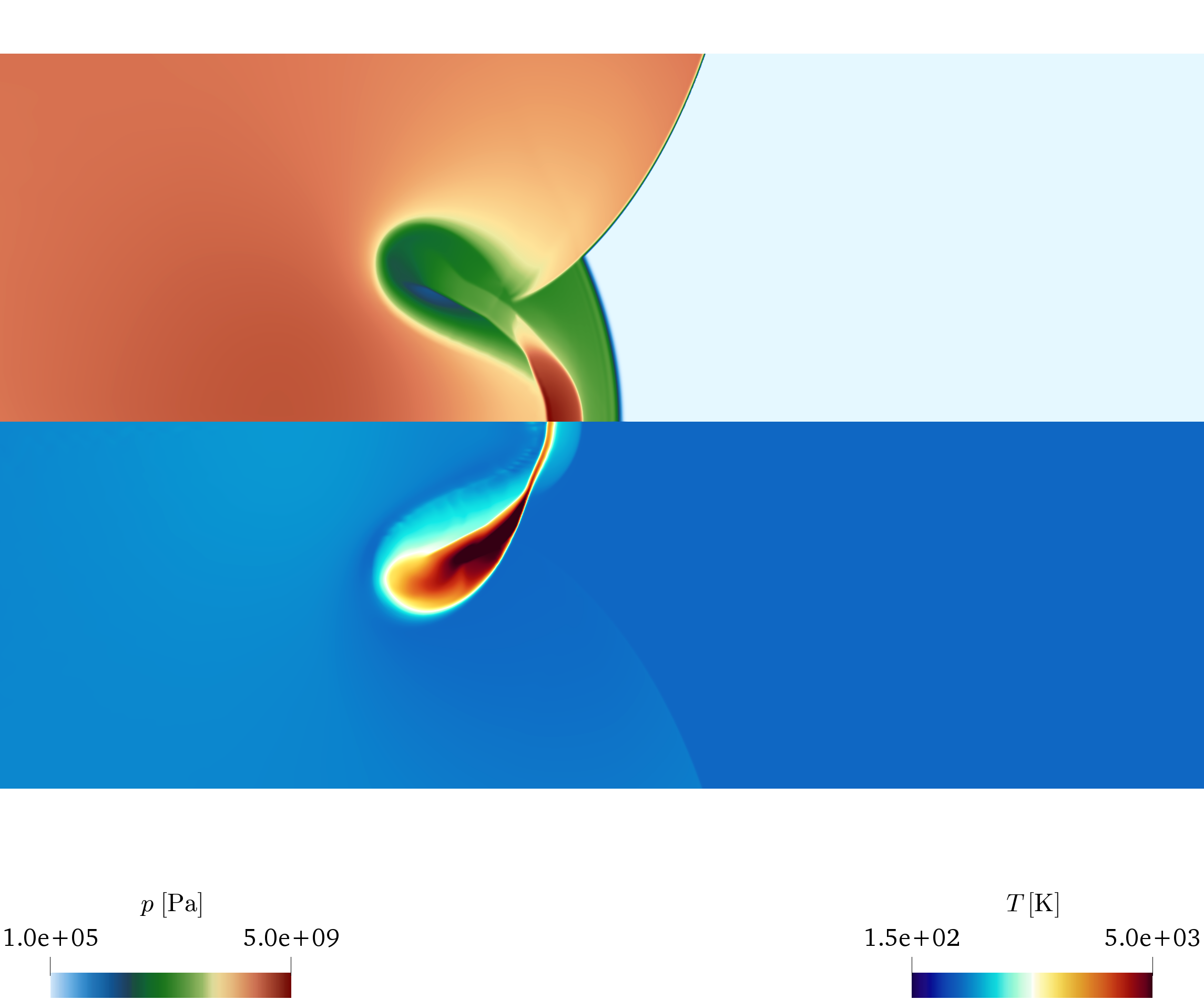}}
\
\subfloat[$t=4.5 \, \mu \textup{s}$]
{\includegraphics[trim= 6cm 14cm 15cm 4cm, clip=true,
width=0.325\textwidth]{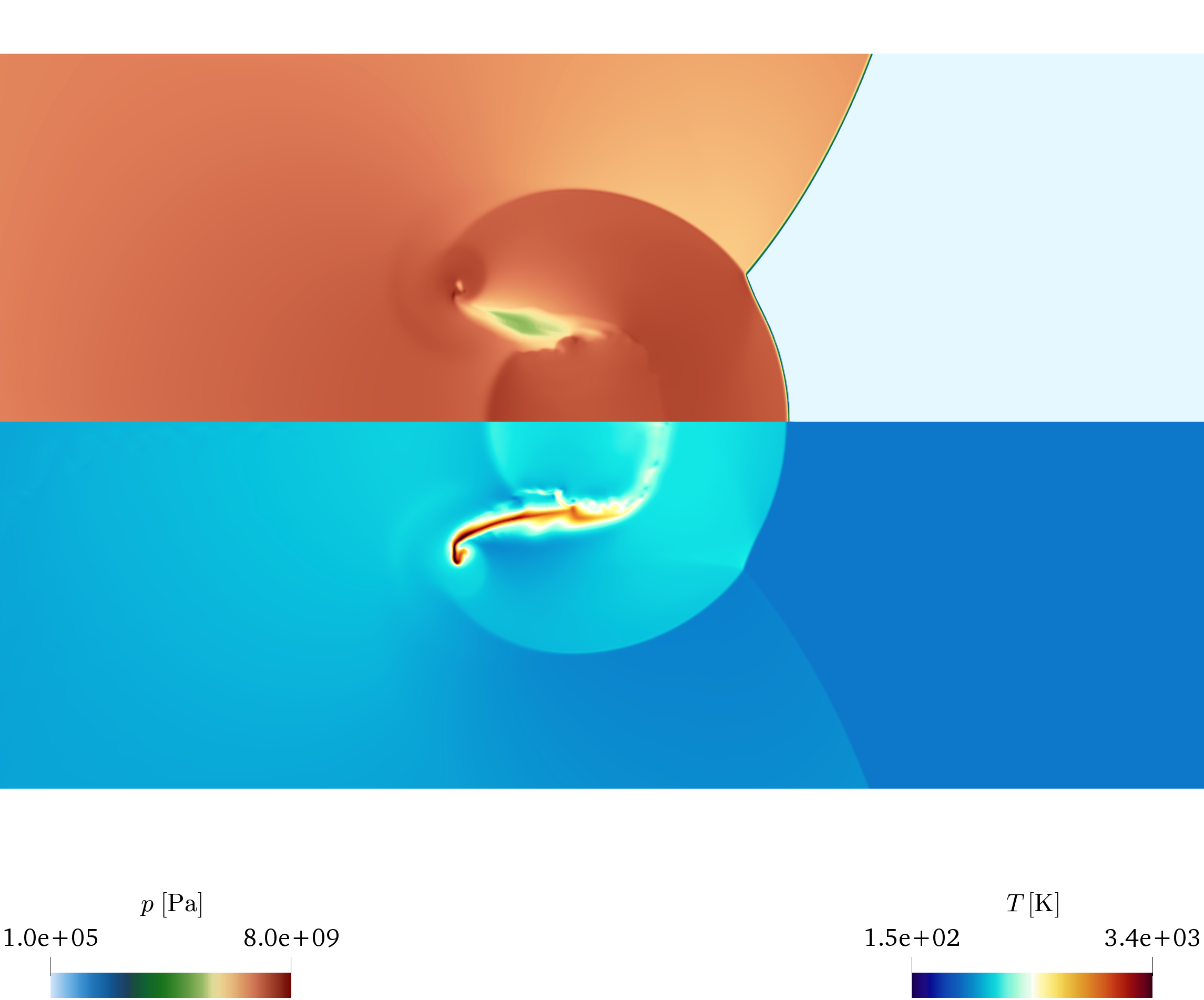}}
\caption{Contours of the pressure $p$ (upper half) and the temperature $T$ (lower half) of the two-dimensional shock-bubble interaction of the air bubble in water on a Cartesian mesh with $\Delta x = d_0/600$ at different times $t$. Both the pressure scale and the temperature scale are logarithmic.}
\label{fig:bubbleCollapseDx600}
\end{center}
\end{figure}

\begin{figure}[h!]
\begin{center}
\subfloat[Pressure $p$]
{\includegraphics[width=0.325\textwidth]{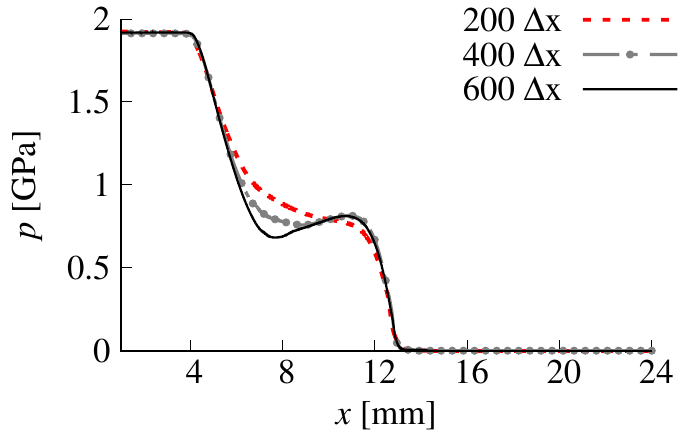}}
\
\subfloat[Density $\rho$]
{\includegraphics[width=0.325\textwidth]{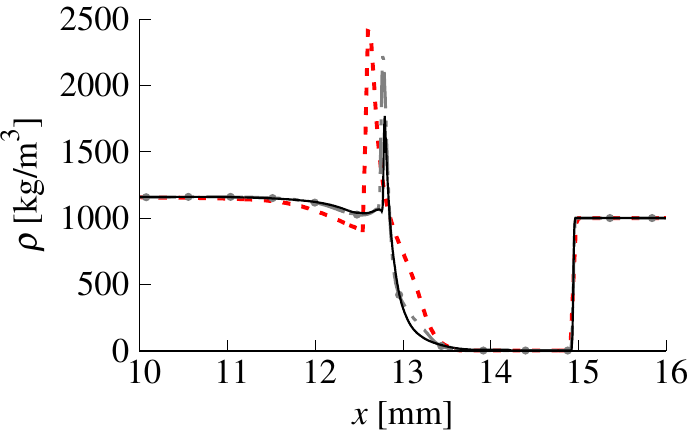}}
\
\subfloat[Temperature $T$]
{\includegraphics[width=0.325\textwidth]{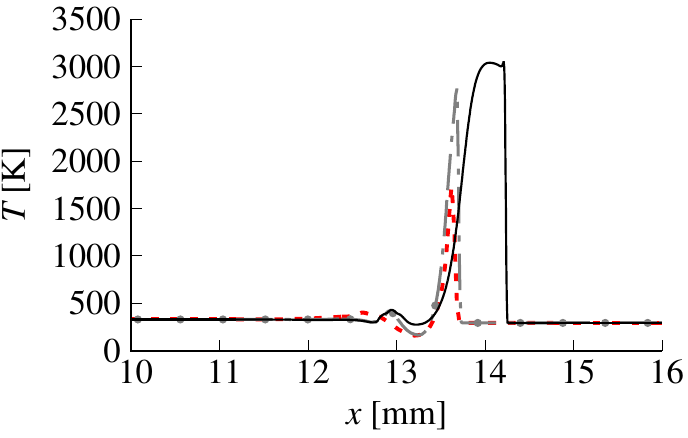}}
\caption{Profiles of pressure $p$, density $\rho$ and temperature $T$ and along the $x$-axis at $y= 6 \times 10^{-4} \, \textup{m}$ of the two-dimensional shock-bubble interaction of the air bubble in water on Cartesian meshes with different mesh spacings $\Delta x$ at time $t=3.0 \times 10^{-6} \, \textup{s}$.}
\label{fig:bubbleCollapseProfileDxt3p0mus}
\end{center}
\end{figure}

\begin{figure}[h!]
\begin{center}
\subfloat[Pressure $p$]
{\includegraphics[width=0.325\textwidth]{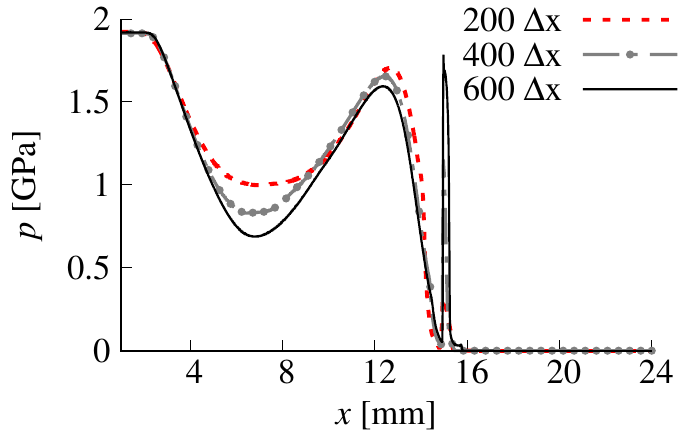}}
\
\subfloat[Density $\rho$]
{\includegraphics[width=0.325\textwidth]{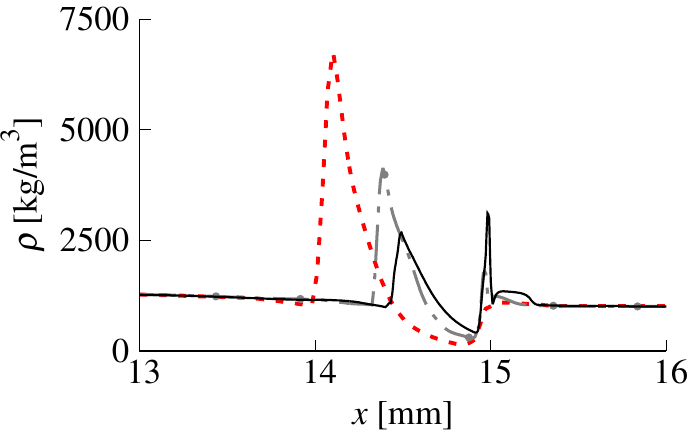}}
\
\subfloat[Temperature $T$]
{\includegraphics[width=0.325\textwidth]{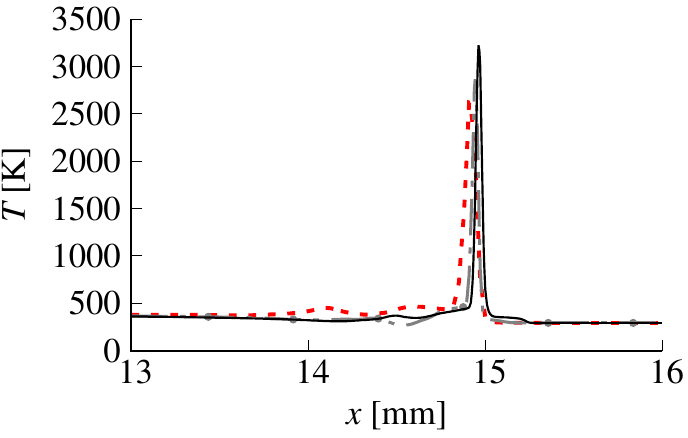}} 
\caption{Profiles of pressure $p$, density $\rho$ and temperature $T$ and along the $x$-axis at $y= 6 \times 10^{-4} \, \textup{m}$ of the two-dimensional shock-bubble interaction of the air bubble in water on Cartesian meshes with different mesh spacings $\Delta x$ at time $t=3.8 \times 10^{-6} \, \textup{s}$.}
\label{fig:bubbleCollapseProfileDxt3p8mus}
\end{center}
\end{figure}

\begin{figure}[h!]
\begin{center}
\subfloat[Pressure $p$]
{\includegraphics[width=0.325\textwidth]{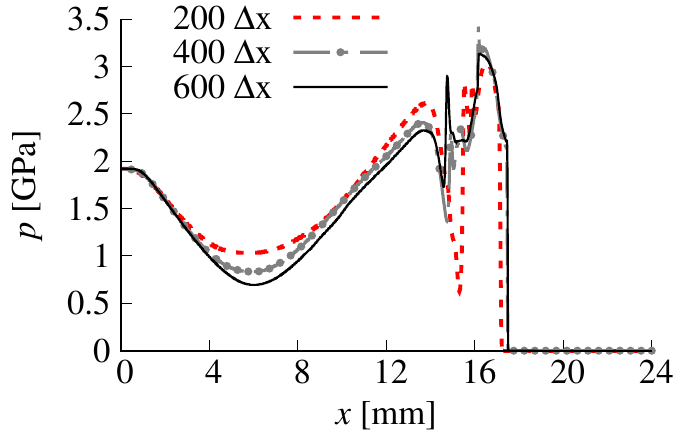}}
\
\subfloat[Density $\rho$]
{\includegraphics[width=0.325\textwidth]{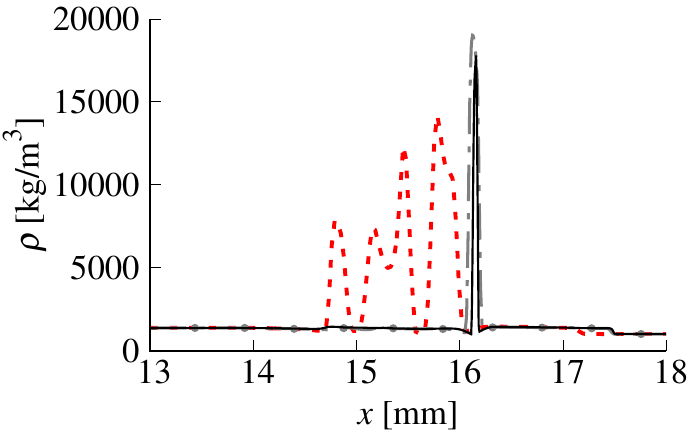}}
\
\subfloat[Temperature $T$]
{\includegraphics[width=0.325\textwidth]{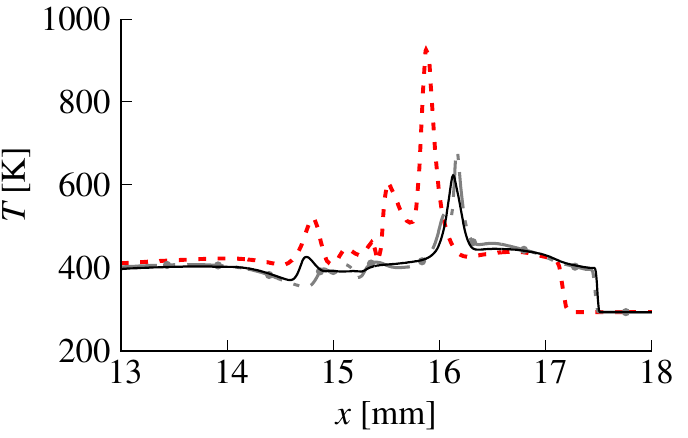}}
\caption{Profiles of pressure $p$, density $\rho$ and temperature $T$ and along the $x$-axis at $y= 6 \times 10^{-4} \, \textup{m}$ of the two-dimensional shock-bubble interaction of the air bubble in water on Cartesian meshes with different mesh spacings $\Delta x$ at time $t=4.5 \times 10^{-6} \, \textup{s}$.}
\label{fig:bubbleCollapseProfileDxt4p5mus}
\end{center}
\end{figure}

Figures \ref{fig:bubbleCollapseDx200}-\ref{fig:bubbleCollapseDx600} show the contours of the pressure and temperature distribution of the water-air system at different times $t$ for equidistant Cartesian meshes with a mesh resolution of $200$, $400$ and $600$ cells per initial bubble diameter $d_0$. Contrary to the rather similar solutions obtained on different meshes for the gas-gas shock-bubble interaction in Section \ref{sec:R22Air2D}, the evolution of the shock-bubble interaction of the air bubble in water is strongly dependent on the mesh resolution. In particular the temperature distribution inside the air bubble exhibits distinct differences on the considered meshes, with generally higher temperatures predicted when the mesh resolution is increased. These differences are especially pronounced when the primary shock wave travels through the bubble, {\em e.g.}~at $t=3.0 \, \mu \textup{s}$, where the higher temperature appears to influence the position of the shock wave considerably, as seen in Fig.~\ref{fig:bubbleCollapseProfileDxt3p0mus}c. Despite these differences in temperature distribution and position of the shock wave at $t=3.0 \, \mu \textup{s}$, which are much less pronounced in the pressure and density fields shown in Fig.~\ref{fig:bubbleCollapseProfileDxt3p0mus}, the results obtained on the meshes with $400$ and $600$ cells per initial bubble diameter $d_0$ are in reasonably good agreement at the later stages of the shock-bubble interaction, as seen in Figs.~\ref{fig:bubbleCollapseProfileDxt3p8mus} and  \ref{fig:bubbleCollapseProfileDxt4p5mus}. In fact, similar observations with respect to the mesh dependency for the same shock-bubble interaction were recently reported by \citet{Shukla2014} and \citet{Goncalves2019} using density-based methods. The mesh with $200$ cells per initial bubble diameter$d_0$, on the other hand, yields significantly different results compared to  the meshes with higher resolution, which affects the position of the shock wave as well as the shape of the bubble, as evident by comparing Fig.~\ref{fig:bubbleCollapseDx200}c with Fig.~\ref{fig:bubbleCollapseDx400}c, and by the density profiles in Fig.~\ref{fig:bubbleCollapseProfileDxt3p0mus}b, \ref{fig:bubbleCollapseProfileDxt3p8mus}b and \ref{fig:bubbleCollapseProfileDxt4p5mus}b.

\section{Conclusions}
\label{sec:conclusions}

In the current paper, the numerical modelling of shock-bubble interactions using the pressure-based algorithm proposed by \citet{Denner2018b}, where the fluxes are evaluated with the ACID method and no Riemann solvers are applied, has been investigated. While shock-bubble interactions in gas-gas flows are largely of academic interest, the interaction of shock waves with a bubble suspended in a liquid is encountered in many different engineering and emerging technological applications, especially in microfluidics and medical applications. Of course, an accurate prediction of shock-bubble interactions is, therefore, a prerequisite for numerical methods to be utilised in research and process development pertaining to these applications.

The presented results have demonstrated a very strong dependency of the interaction of a shock wave with an air bubble suspended in water on the spatial resolution of the computational mesh. In particular, the temperature field has been found to exhibit large differences between different mesh resolutions, which also contributes to differences in the propagation of the primary shock wave. To this end, for the shock interaction with the air bubble in water, the presented results suggest a mesh resolution of at least $400-600$ cells per initial diameter to yield reasonably converged results. However, even though the position of the shock wave as well as the pressure, temperature and density fields have been found to yield agreement for such a mesh resolution at later stages of the shock-bubble interaction, {\em i.e.}~after the shock-wave has passed the bubble, these quantities still exhibit considerable differences while the shock wave passes through the bubble. Considering the rapid and significant increase in pressure, temperature and density as the bubble is compressed when the shock wave passes, the accuracy of the ideal-gas model also warrants further study, since the evolution of the bubble collapse has been shown by the presented results to be strongly dependent on the quality and accuracy of the prediction of pressure and temperature.

Shock-bubble interactions also provide a convenient canonical reference system to test and scrutinise new numerical schemes for the simulation of compressible interfacial flows; shock-bubble interaction can be found in most publications that propose a new numerical scheme for compressible interfacial flows. These tests mostly focus on gas-gas flows, {\em e.g.}~the R22 bubble in air also considered in this study, while the computationally more expensive and challenging shock-bubble interaction in liquid-gas flows is frequently neglected. However, the presented results show that the shock-bubble interaction in a gas-gas flow is not sensitive to the employed numerical methods and the spatial resolution of the computational mesh, contrary to the shock-bubble interaction in a liquid-gas flow. This puts the informative value of validating numerical schemes using gas-gas shock-bubble interactions into question and strongly suggests that the shock-bubble interaction in liquid-gas flows are generally better suited to scrutinise and compare numerical methods for compressible interfacial flows.


\begin{thebibliography}{85}
\expandafter\ifx\csname natexlab\endcsname\relax\def\natexlab#1{#1}\fi
\expandafter\ifx\csname url\endcsname\relax
  \def\url#1{\texttt{#1}}\fi
\expandafter\ifx\csname urlprefix\endcsname\relax\def\urlprefix{URL }\fi
\providecommand{\eprint}[2][]{\url{#2}}
\providecommand{\bibinfo}[2]{#2}
\ifx\xfnm\relax \def\xfnm[#1]{\unskip,\space#1}\fi
\bibitem[{Abgrall and Karni(2001)}]{Abgrall2001}
\bibinfo{author}{Abgrall, R.}, \bibinfo{author}{Karni, S.},
  \bibinfo{year}{2001}.
\newblock \bibinfo{title}{Computations of compressible multifluids}.
\newblock \bibinfo{journal}{Journal of Computational Physics}
  \bibinfo{volume}{169}, \bibinfo{pages}{594--623}.
\bibitem[{Abgrall and Saurel(2003)}]{Abgrall2003}
\bibinfo{author}{Abgrall, R.}, \bibinfo{author}{Saurel, R.},
  \bibinfo{year}{2003}.
\newblock \bibinfo{title}{Discrete equations for physical and numerical
  compressible multiphase mixtures}.
\newblock \bibinfo{journal}{Journal of Computational Physics}
  \bibinfo{volume}{186}, \bibinfo{pages}{361--396}.
\bibitem[{Allaire et~al.(2002)Allaire, Clerc and Kokh}]{Allaire2002}
\bibinfo{author}{Allaire, G.}, \bibinfo{author}{Clerc, S.},
  \bibinfo{author}{Kokh, S.}, \bibinfo{year}{2002}.
\newblock \bibinfo{title}{A {{Five}}-{{Equation Model}} for the {{Simulation}}
  of {{Interfaces}} between {{Compressible Fluids}}}.
\newblock \bibinfo{journal}{Journal of Computational Physics}
  \bibinfo{volume}{181}, \bibinfo{pages}{577--616}.
\bibitem[{Anderson(2003)}]{Anderson2003}
\bibinfo{author}{Anderson, J.D.}, \bibinfo{year}{2003}.
\newblock \bibinfo{title}{Modern {{Compressible Flow}}: {{With}} a {{Historical
  Perspective}}}.
\newblock \bibinfo{publisher}{{McGraw-Hill New York}}.
\bibitem[{Ando et~al.(2012)Ando, Liu and Ohl}]{Ando2012}
\bibinfo{author}{Ando, K.}, \bibinfo{author}{Liu, A.Q.}, \bibinfo{author}{Ohl,
  C.D.}, \bibinfo{year}{2012}.
\newblock \bibinfo{title}{Homogeneous {{Nucleation}} in {{Water}} in
  {{Microfluidic Channels}}}.
\newblock \bibinfo{journal}{Physical Review Letters} \bibinfo{volume}{109}.
\bibitem[{Baer and Nunziato(1986)}]{Baer1986}
\bibinfo{author}{Baer, M.}, \bibinfo{author}{Nunziato, J.},
  \bibinfo{year}{1986}.
\newblock \bibinfo{title}{A two-phase mixture theory for the
  deflagration-to-detonation transition (ddt) in reactive granular materials}.
\newblock \bibinfo{journal}{International Journal of Multiphase Flow}
  \bibinfo{volume}{12}, \bibinfo{pages}{861--889}.
\bibitem[{Bagabir and Drikakis(2001)}]{Bagabir2001}
\bibinfo{author}{Bagabir, A.}, \bibinfo{author}{Drikakis, D.},
  \bibinfo{year}{2001}.
\newblock \bibinfo{title}{Mach number effects on shock-bubble interaction}.
\newblock \bibinfo{journal}{Shock Waves} \bibinfo{volume}{11},
  \bibinfo{pages}{209--218}.
\bibitem[{Balay et~al.(2017)Balay, Abhyankar, Adams, Brown, Brune, Buschelman,
  Dalcin, Eijkhout, Kaushik, Knepley, May, McInnes, Gropp, Rupp, Sanan, Smith,
  Zampini, Zhang and Zhang}]{petsc-user-ref}
\bibinfo{author}{Balay, S.}, \bibinfo{author}{Abhyankar, S.},
  \bibinfo{author}{Adams, M.F.}, \bibinfo{author}{Brown, J.},
  \bibinfo{author}{Brune, P.}, \bibinfo{author}{Buschelman, K.},
  \bibinfo{author}{Dalcin, L.}, \bibinfo{author}{Eijkhout, V.},
  \bibinfo{author}{Kaushik, D.}, \bibinfo{author}{Knepley, M.G.},
  \bibinfo{author}{May, D.A.}, \bibinfo{author}{McInnes, L.C.},
  \bibinfo{author}{Gropp, W.D.}, \bibinfo{author}{Rupp, K.},
  \bibinfo{author}{Sanan, P.}, \bibinfo{author}{Smith, B.F.},
  \bibinfo{author}{Zampini, S.}, \bibinfo{author}{Zhang, H.},
  \bibinfo{author}{Zhang, H.}, \bibinfo{year}{2017}.
\newblock \bibinfo{title}{{{PETSc Users Manual}}}.
\newblock \bibinfo{type}{Technical Report} \bibinfo{number}{ANL-95/11 -
  Revision 3.8}. {Argonne National Laboratory}.
\bibitem[{Bartholomew et~al.(2018)Bartholomew, Denner, {Abdol-Azis}, Marquis
  and {van Wachem}}]{Bartholomew2018}
\bibinfo{author}{Bartholomew, P.}, \bibinfo{author}{Denner, F.},
  \bibinfo{author}{{Abdol-Azis}, M.}, \bibinfo{author}{Marquis, A.},
  \bibinfo{author}{{van Wachem}, B.}, \bibinfo{year}{2018}.
\newblock \bibinfo{title}{Unified formulation of the momentum-weighted
  interpolation for collocated variable arrangements}.
\newblock \bibinfo{journal}{Journal of Computational Physics}
  \bibinfo{volume}{375}, \bibinfo{pages}{177--208}.
\bibitem[{Bo and Grove(2014)}]{Bo2014}
\bibinfo{author}{Bo, W.}, \bibinfo{author}{Grove, J.W.}, \bibinfo{year}{2014}.
\newblock \bibinfo{title}{A volume of fluid method based ghost fluid method for
  compressible multi-fluid flows}.
\newblock \bibinfo{journal}{Computers \& Fluids} \bibinfo{volume}{90},
  \bibinfo{pages}{113--122}.
\bibitem[{Brouillette(2002)}]{Brouillette2002}
\bibinfo{author}{Brouillette, M.}, \bibinfo{year}{2002}.
\newblock \bibinfo{title}{The {{Richtmyer}}-{{Meshkov}} instability}.
\newblock \bibinfo{journal}{Annual Review of Fluid Mechanics}
  \bibinfo{volume}{34}, \bibinfo{pages}{445--468}.
\bibitem[{Chang and Liou(2007)}]{Chang2007}
\bibinfo{author}{Chang, C.H.}, \bibinfo{author}{Liou, M.S.},
  \bibinfo{year}{2007}.
\newblock \bibinfo{title}{A robust and accurate approach to computing
  compressible multiphase flow: {{Stratified}} flow model and {{AUSM}}+-up
  scheme}.
\newblock \bibinfo{journal}{Journal of Computational Physics}
  \bibinfo{volume}{225}, \bibinfo{pages}{840--873}.
\bibitem[{Chorin(1967)}]{Chorin1967}
\bibinfo{author}{Chorin, A.J.}, \bibinfo{year}{1967}.
\newblock \bibinfo{title}{A numerical method for solving incompressible viscous
  flow problems}.
\newblock \bibinfo{journal}{Journal of Computational Physics}
  \bibinfo{volume}{2}, \bibinfo{pages}{12--26}.
\bibitem[{Chorin and Marsden(1993)}]{Chorin1993}
\bibinfo{author}{Chorin, A.J.}, \bibinfo{author}{Marsden, J.E.},
  \bibinfo{year}{1993}.
\newblock \bibinfo{title}{A {{Mathematical Introduction}} to {{Fluid
  Mechanics}}}.
\newblock \bibinfo{publisher}{{Springer Verlag}}.
\bibitem[{Coralic and Colonius(2014)}]{Coralic2014}
\bibinfo{author}{Coralic, V.}, \bibinfo{author}{Colonius, T.},
  \bibinfo{year}{2014}.
\newblock \bibinfo{title}{Finite-volume {{WENO}} scheme for viscous
  compressible multicomponent flows}.
\newblock \bibinfo{journal}{Journal of Computational Physics}
  \bibinfo{volume}{274}, \bibinfo{pages}{95--121}.
\bibitem[{Cordier et~al.(2012)Cordier, Degond and Kumbaro}]{Cordier2012}
\bibinfo{author}{Cordier, F.}, \bibinfo{author}{Degond, P.},
  \bibinfo{author}{Kumbaro, A.}, \bibinfo{year}{2012}.
\newblock \bibinfo{title}{An {{Asymptotic}}-{{Preserving}} all-speed scheme for
  the {{Euler}} and {{Navier}}\textendash{{Stokes}} equations}.
\newblock \bibinfo{journal}{Journal of Computational Physics}
  \bibinfo{volume}{231}, \bibinfo{pages}{5685--5704}.
\bibitem[{Delale(2013)}]{Delale2013}
\bibinfo{editor}{Delale, C.F.} (Ed.), \bibinfo{year}{2013}.
\newblock \bibinfo{title}{Bubble {{Dynamics}} and {{Shock Waves}}}.
\newblock \bibinfo{publisher}{{Springer Berlin Heidelberg}},
  \bibinfo{address}{{Berlin, Heidelberg}}.
\bibitem[{Demird{\v z}i{\'c} et~al.(1993)Demird{\v z}i{\'c}, Lilek and
  Peri{\'c}}]{Demirdzic1993}
\bibinfo{author}{Demird{\v z}i{\'c}, I.}, \bibinfo{author}{Lilek, v.},
  \bibinfo{author}{Peri{\'c}, M.}, \bibinfo{year}{1993}.
\newblock \bibinfo{title}{A collocated finite volume method for predicting
  flows at all speeds}.
\newblock \bibinfo{journal}{International Journal for Numerical Methods in
  Fluids} \bibinfo{volume}{16}, \bibinfo{pages}{1029--1050}.
\bibitem[{Denner(2018)}]{Denner2018c}
\bibinfo{author}{Denner, F.}, \bibinfo{year}{2018}.
\newblock \bibinfo{title}{Fully-coupled pressure-based algorithm for
  compressible flows: Linearisation and iterative solution strategies}.
\newblock \bibinfo{journal}{Computers \& Fluids} \bibinfo{volume}{175},
  \bibinfo{pages}{53--65}.
\bibitem[{Denner and {van Wachem}(2014)}]{Denner2014e}
\bibinfo{author}{Denner, F.}, \bibinfo{author}{{van Wachem}, B.},
  \bibinfo{year}{2014}.
\newblock \bibinfo{title}{Compressive {{VOF}} method with skewness correction
  to capture sharp interfaces on arbitrary meshes}.
\newblock \bibinfo{journal}{Journal of Computational Physics}
  \bibinfo{volume}{279}, \bibinfo{pages}{127--144}.
\bibitem[{Denner and {van Wachem}(2015)}]{Denner2015a}
\bibinfo{author}{Denner, F.}, \bibinfo{author}{{van Wachem}, B.},
  \bibinfo{year}{2015}.
\newblock \bibinfo{title}{{{TVD}} differencing on three-dimensional
  unstructured meshes with monotonicity-preserving correction of mesh
  skewness}.
\newblock \bibinfo{journal}{Journal of Computational Physics}
  \bibinfo{volume}{298}, \bibinfo{pages}{466--479}.
\bibitem[{Denner et~al.(2018)Denner, Xiao and {van Wachem}}]{Denner2018b}
\bibinfo{author}{Denner, F.}, \bibinfo{author}{Xiao, C.N.},
  \bibinfo{author}{{van Wachem}, B.}, \bibinfo{year}{2018}.
\newblock \bibinfo{title}{Pressure-based algorithm for compressible interfacial
  flows with acoustically-conservative interface discretisation}.
\newblock \bibinfo{journal}{Journal of Computational Physics}
  \bibinfo{volume}{367}, \bibinfo{pages}{192--234}.
\bibitem[{Fedkiw et~al.(1999a)Fedkiw, Aslam, Merriman and Osher}]{Fedkiw1999}
\bibinfo{author}{Fedkiw, R.}, \bibinfo{author}{Aslam, T.},
  \bibinfo{author}{Merriman, B.}, \bibinfo{author}{Osher, S.},
  \bibinfo{year}{1999}a.
\newblock \bibinfo{title}{A {{Non}}-oscillatory {{Eulerian Approach}} to
  {{Interfaces}} in {{Multimaterial Flows}} (the {{Ghost Fluid Method}})}.
\newblock \bibinfo{journal}{Journal of Computational Physics}
  \bibinfo{volume}{152}, \bibinfo{pages}{457--492}.
\bibitem[{Fedkiw et~al.(1999b)Fedkiw, Aslam and Xu}]{Fedkiw1999a}
\bibinfo{author}{Fedkiw, R.P.}, \bibinfo{author}{Aslam, T.},
  \bibinfo{author}{Xu, S.}, \bibinfo{year}{1999}b.
\newblock \bibinfo{title}{The {{Ghost Fluid Method}} for {{Deflagration}} and
  {{Detonation Discontinuities}}}.
\newblock \bibinfo{journal}{Journal of Computational Physics}
  \bibinfo{volume}{154}, \bibinfo{pages}{393--427}.
\bibitem[{Fuster(2019)}]{Fuster2019}
\bibinfo{author}{Fuster, D.}, \bibinfo{year}{2019}.
\newblock \bibinfo{title}{A {{Review}} of {{Models}} for {{Bubble Clusters}} in
  {{Cavitating Flows}}}.
\newblock \bibinfo{journal}{Flow, Turbulence and Combustion}
  \bibinfo{volume}{102}, \bibinfo{pages}{497--536}.
\bibitem[{Fuster and Popinet(2018)}]{Fuster2018}
\bibinfo{author}{Fuster, D.}, \bibinfo{author}{Popinet, S.},
  \bibinfo{year}{2018}.
\newblock \bibinfo{title}{An all-{{Mach}} method for the simulation of bubble
  dynamics problems in the presence of surface tension}.
\newblock \bibinfo{journal}{Journal of Computational Physics}
  \bibinfo{volume}{374}, \bibinfo{pages}{752--768}.
\bibitem[{Goncalves et~al.(2019)Goncalves, Hoarau and Zeidan}]{Goncalves2019}
\bibinfo{author}{Goncalves, E.}, \bibinfo{author}{Hoarau, Y.},
  \bibinfo{author}{Zeidan, D.}, \bibinfo{year}{2019}.
\newblock \bibinfo{title}{Simulation of shock-induced bubble collapse using a
  four-equation model}.
\newblock \bibinfo{journal}{Shock Waves} \bibinfo{volume}{29},
  \bibinfo{pages}{221--234}.
\bibitem[{Haas and Sturtevant(1987)}]{Haas1987}
\bibinfo{author}{Haas, J.F.}, \bibinfo{author}{Sturtevant, B.},
  \bibinfo{year}{1987}.
\newblock \bibinfo{title}{Interaction of weak shock waves with cylindrical and
  spherical gas inhomogeneities}.
\newblock \bibinfo{journal}{Journal of Fluid Mechanics} \bibinfo{volume}{181},
  \bibinfo{pages}{41}.
\bibitem[{Haimovich and Frankel(2017)}]{Haimovich2017}
\bibinfo{author}{Haimovich, O.}, \bibinfo{author}{Frankel, S.H.},
  \bibinfo{year}{2017}.
\newblock \bibinfo{title}{Numerical simulations of compressible multicomponent
  and multiphase flow using a high-order targeted {{ENO}} ({{TENO}})
  finite-volume method}.
\newblock \bibinfo{journal}{Computers \& Fluids} \bibinfo{volume}{146},
  \bibinfo{pages}{105--116}.
\bibitem[{Harlow and Amsden(1971a)}]{Harlow1971}
\bibinfo{author}{Harlow, F.}, \bibinfo{author}{Amsden, A.},
  \bibinfo{year}{1971}a.
\newblock \bibinfo{title}{Fluid Dynamics}.
\newblock \bibinfo{type}{Monograph} \bibinfo{number}{LA-4700}. {Los Alamos
  National Laboratory}.
\bibitem[{Harlow and Amsden(1971b)}]{Harlow1971a}
\bibinfo{author}{Harlow, F.H.}, \bibinfo{author}{Amsden, A.A.},
  \bibinfo{year}{1971}b.
\newblock \bibinfo{title}{A numerical fluid dynamics calculation method for all
  flow speeds}.
\newblock \bibinfo{journal}{Journal of Computational Physics}
  \bibinfo{volume}{8}, \bibinfo{pages}{197--213}.
\bibitem[{Hauke and Hughes(1998)}]{Hauke1998}
\bibinfo{author}{Hauke, G.}, \bibinfo{author}{Hughes, T.J.},
  \bibinfo{year}{1998}.
\newblock \bibinfo{title}{A comparative study of different sets of variables
  for solving compressible and incompressible flows}.
\newblock \bibinfo{journal}{Computer Methods in Applied Mechanics and
  Engineering} \bibinfo{volume}{153}, \bibinfo{pages}{1--44}.
\bibitem[{Hejazialhosseini et~al.(2013)Hejazialhosseini, Rossinelli and
  Koumoutsakos}]{Hejazialhosseini2013}
\bibinfo{author}{Hejazialhosseini, B.}, \bibinfo{author}{Rossinelli, D.},
  \bibinfo{author}{Koumoutsakos, P.}, \bibinfo{year}{2013}.
\newblock \bibinfo{title}{Vortex dynamics in {{3D}} shock-bubble interaction}.
\newblock \bibinfo{journal}{Physics of Fluids} \bibinfo{volume}{25},
  \bibinfo{pages}{110816}.
\bibitem[{Hirt and Nichols(1981)}]{Hirt1981}
\bibinfo{author}{Hirt, C.}, \bibinfo{author}{Nichols, B.},
  \bibinfo{year}{1981}.
\newblock \bibinfo{title}{Volume of fluid ({{VOF}}) method for the dynamics of
  free boundaries}.
\newblock \bibinfo{journal}{Journal of Computational Physics}
  \bibinfo{volume}{39}, \bibinfo{pages}{201--225}.
\bibitem[{Hou and Floch(1994)}]{Hou1994}
\bibinfo{author}{Hou, T.Y.}, \bibinfo{author}{Floch, P.G.L.},
  \bibinfo{year}{1994}.
\newblock \bibinfo{title}{Why {{Nonconservative Schemes Converge}} to {{Wrong
  Solutions}}: {{Error Analysis}}}.
\newblock \bibinfo{journal}{Mathematics of Computation} \bibinfo{volume}{62},
  \bibinfo{pages}{497--530}.
\bibitem[{Hu and Khoo(2004)}]{Hu2004}
\bibinfo{author}{Hu, X.}, \bibinfo{author}{Khoo, B.}, \bibinfo{year}{2004}.
\newblock \bibinfo{title}{An interface interaction method for compressible
  multifluids}.
\newblock \bibinfo{journal}{Journal of Computational Physics}
  \bibinfo{volume}{198}, \bibinfo{pages}{35--64}.
\bibitem[{Johnsen(2007)}]{Johnsen2007}
\bibinfo{author}{Johnsen, E.}, \bibinfo{year}{2007}.
\newblock \bibinfo{title}{Numerical Simulations of Non-Spherical Bubble
  Collapse: With Applications to Shockwave Lithotripsy}.
\newblock \bibinfo{type}{{{PhD}} thesis}. California Institute of Technology.
  \bibinfo{address}{{Pasadena, California, USA}}.
\bibitem[{Johnsen and Colonius(2006)}]{Johnsen2006}
\bibinfo{author}{Johnsen, E.}, \bibinfo{author}{Colonius, T.},
  \bibinfo{year}{2006}.
\newblock \bibinfo{title}{Implementation of {{WENO}} schemes in compressible
  multicomponent flow problems}.
\newblock \bibinfo{journal}{Journal of Computational Physics}
  \bibinfo{volume}{219}, \bibinfo{pages}{715--732}.
\bibitem[{Johnsen and Colonius(2009)}]{Johnsen2009}
\bibinfo{author}{Johnsen, E.}, \bibinfo{author}{Colonius, T.},
  \bibinfo{year}{2009}.
\newblock \bibinfo{title}{Numerical simulations of non-spherical bubble
  collapse.}
\newblock \bibinfo{journal}{Journal of fluid mechanics} \bibinfo{volume}{629},
  \bibinfo{pages}{231--262}.
\bibitem[{Karimian and Schneider(1994)}]{Karimian1994}
\bibinfo{author}{Karimian, S.M.H.}, \bibinfo{author}{Schneider, G.E.},
  \bibinfo{year}{1994}.
\newblock \bibinfo{title}{Pressure-based computational method for compressible
  and incompressible flows}.
\newblock \bibinfo{journal}{Journal of Thermophysics and Heat Transfer}
  \bibinfo{volume}{8}, \bibinfo{pages}{267--274}.
\bibitem[{Kokh and Lagouti{\`e}re(2010)}]{Kokh2010}
\bibinfo{author}{Kokh, S.}, \bibinfo{author}{Lagouti{\`e}re, F.},
  \bibinfo{year}{2010}.
\newblock \bibinfo{title}{An anti-diffusive numerical scheme for the simulation
  of interfaces between compressible fluids by means of a five-equation model}.
\newblock \bibinfo{journal}{Journal of Computational Physics}
  \bibinfo{volume}{229}, \bibinfo{pages}{2773--2809}.
\bibitem[{Kunz et~al.(1999)Kunz, Cope and Venkateswaran}]{Kunz1999}
\bibinfo{author}{Kunz, R.}, \bibinfo{author}{Cope, W.},
  \bibinfo{author}{Venkateswaran, S.}, \bibinfo{year}{1999}.
\newblock \bibinfo{title}{Development of an implicit method for multi-fluid
  flow simulations}.
\newblock \bibinfo{journal}{Journal of Computational Physics}
  \bibinfo{volume}{152}, \bibinfo{pages}{78--101}.
\bibitem[{Layes et~al.(2003)Layes, Jourdan and Houas}]{Layes2003}
\bibinfo{author}{Layes, G.}, \bibinfo{author}{Jourdan, G.},
  \bibinfo{author}{Houas, L.}, \bibinfo{year}{2003}.
\newblock \bibinfo{title}{Distortion of a {{Spherical Gaseous Interface
  Accelerated}} by a {{Plane Shock Wave}}}.
\newblock \bibinfo{journal}{Physical Review Letters} \bibinfo{volume}{91}.
\bibitem[{Layes et~al.(2005)Layes, Jourdan and Houas}]{Layes2005}
\bibinfo{author}{Layes, G.}, \bibinfo{author}{Jourdan, G.},
  \bibinfo{author}{Houas, L.}, \bibinfo{year}{2005}.
\newblock \bibinfo{title}{Experimental investigation of the shock wave
  interaction with a spherical gas inhomogeneity}.
\newblock \bibinfo{journal}{Physics of Fluids} \bibinfo{volume}{17},
  \bibinfo{pages}{028103}.
\bibitem[{Liu and Hu(2017)}]{Liu2017}
\bibinfo{author}{Liu, C.}, \bibinfo{author}{Hu, C.}, \bibinfo{year}{2017}.
\newblock \bibinfo{title}{Adaptive {{THINC}}-{{GFM}} for compressible
  multi-medium flows}.
\newblock \bibinfo{journal}{Journal of Computational Physics}
  \bibinfo{volume}{342}, \bibinfo{pages}{43--65}.
\bibitem[{Liu et~al.(2003)Liu, Khoo and Yeo}]{Liu2003}
\bibinfo{author}{Liu, T.}, \bibinfo{author}{Khoo, B.}, \bibinfo{author}{Yeo,
  K.}, \bibinfo{year}{2003}.
\newblock \bibinfo{title}{Ghost fluid method for strong shock impacting on
  material interface}.
\newblock \bibinfo{journal}{Journal of Computational Physics}
  \bibinfo{volume}{190}, \bibinfo{pages}{651--681}.
\bibitem[{Michael and Nikiforakis(2019)}]{Michael2019}
\bibinfo{author}{Michael, L.}, \bibinfo{author}{Nikiforakis, N.},
  \bibinfo{year}{2019}.
\newblock \bibinfo{title}{The evolution of the temperature field during cavity
  collapse in liquid nitromethane. {{Part I}}: Inert case}.
\newblock \bibinfo{journal}{Shock Waves} \bibinfo{volume}{29},
  \bibinfo{pages}{153--172}.
\bibitem[{Moguen et~al.(2015)Moguen, Bruel and Dick}]{Moguen2015a}
\bibinfo{author}{Moguen, Y.}, \bibinfo{author}{Bruel, P.},
  \bibinfo{author}{Dick, E.}, \bibinfo{year}{2015}.
\newblock \bibinfo{title}{Solving low {{Mach}} number {{Riemann}} problems by a
  momentum interpolation method}.
\newblock \bibinfo{journal}{Journal of Computational Physics}
  \bibinfo{volume}{298}, \bibinfo{pages}{741--746}.
\bibitem[{Moguen et~al.(2019)Moguen, Bruel and Dick}]{Moguen2019}
\bibinfo{author}{Moguen, Y.}, \bibinfo{author}{Bruel, P.},
  \bibinfo{author}{Dick, E.}, \bibinfo{year}{2019}.
\newblock \bibinfo{title}{A combined momentum-interpolation and advection
  upstream splitting pressure-correction algorithm for simulation of convective
  and acoustic transport at all levels of {{Mach}} number}.
\newblock \bibinfo{journal}{Journal of Computational Physics}
  \bibinfo{volume}{384}, \bibinfo{pages}{16--41}.
\bibitem[{Moguen et~al.(2012)Moguen, Kousksou, Bruel, Vierendeels and
  Dick}]{Moguen2012}
\bibinfo{author}{Moguen, Y.}, \bibinfo{author}{Kousksou, T.},
  \bibinfo{author}{Bruel, P.}, \bibinfo{author}{Vierendeels, J.},
  \bibinfo{author}{Dick, E.}, \bibinfo{year}{2012}.
\newblock \bibinfo{title}{Pressure-velocity coupling allowing acoustic
  calculation in low {{Mach}} number flow}.
\newblock \bibinfo{journal}{Journal of Computational Physics}
  \bibinfo{volume}{231}, \bibinfo{pages}{5522--5541}.
\bibitem[{Moukalled et~al.(2016)Moukalled, Mangani and Darwish}]{Moukalled2016}
\bibinfo{author}{Moukalled, F.}, \bibinfo{author}{Mangani, L.},
  \bibinfo{author}{Darwish, M.}, \bibinfo{year}{2016}.
\newblock \bibinfo{title}{The Finite Volume Method in Computational Fluid
  Dynamics: {{An}} Advanced Introduction with {{OpenFOAM}} and {{Matlab}}}.
\newblock \bibinfo{publisher}{{Springer}}.
\bibitem[{Murrone and Guillard(2005)}]{Murrone2005}
\bibinfo{author}{Murrone, A.}, \bibinfo{author}{Guillard, H.},
  \bibinfo{year}{2005}.
\newblock \bibinfo{title}{A five equation reduced model for compressible two
  phase flow problems}.
\newblock \bibinfo{journal}{Journal of Computational Physics}
  \bibinfo{volume}{202}, \bibinfo{pages}{664--698}.
\bibitem[{Niederhaus et~al.(2008a)Niederhaus, Greenough, Oakley and
  Bonazza}]{Niederhaus2008a}
\bibinfo{author}{Niederhaus, J.H.J.}, \bibinfo{author}{Greenough, J.A.},
  \bibinfo{author}{Oakley, J.G.}, \bibinfo{author}{Bonazza, R.},
  \bibinfo{year}{2008}a.
\newblock \bibinfo{title}{Vorticity evolution in two- and three-dimensional
  simulations for shock\textendash{}bubble interactions}.
\newblock \bibinfo{journal}{Physica Scripta} \bibinfo{volume}{T132},
  \bibinfo{pages}{014019}.
\bibitem[{Niederhaus et~al.(2008b)Niederhaus, Greenough, Oakley, Ranjan,
  Anderson and Bonazza}]{Niederhaus2008}
\bibinfo{author}{Niederhaus, J.H.J.}, \bibinfo{author}{Greenough, J.A.},
  \bibinfo{author}{Oakley, J.G.}, \bibinfo{author}{Ranjan, D.},
  \bibinfo{author}{Anderson, M.H.}, \bibinfo{author}{Bonazza, R.},
  \bibinfo{year}{2008}b.
\newblock \bibinfo{title}{A computational parameter study for the
  three-dimensional shock\textendash{}bubble interaction}.
\newblock \bibinfo{journal}{Journal of Fluid Mechanics} \bibinfo{volume}{594}.
\bibitem[{Nourgaliev et~al.(2006)Nourgaliev, Dinh and
  Theofanous}]{Nourgaliev2006}
\bibinfo{author}{Nourgaliev, R.}, \bibinfo{author}{Dinh, T.},
  \bibinfo{author}{Theofanous, T.}, \bibinfo{year}{2006}.
\newblock \bibinfo{title}{Adaptive characteristics-based matching for
  compressible multifluid dynamics}.
\newblock \bibinfo{journal}{Journal of Computational Physics}
  \bibinfo{volume}{213}, \bibinfo{pages}{500--529}.
\bibitem[{Ohl and Ohl(2016)}]{Ohl2016}
\bibinfo{author}{Ohl, S.W.}, \bibinfo{author}{Ohl, C.D.}, \bibinfo{year}{2016}.
\newblock \bibinfo{title}{Acoustic {{Cavitation}} in a {{Microchannel}}}, in:
  \bibinfo{booktitle}{Handbook of {{Ultrasonics}} and {{Sonochemistry}}}.
  \bibinfo{publisher}{{Springer Singapore}}, \bibinfo{address}{{Singapore}},
  pp. \bibinfo{pages}{99--135}.
\bibitem[{Pan et~al.(2018)Pan, Adami, Hu and Adams}]{Pan2018}
\bibinfo{author}{Pan, S.}, \bibinfo{author}{Adami, S.}, \bibinfo{author}{Hu,
  X.}, \bibinfo{author}{Adams, N.A.}, \bibinfo{year}{2018}.
\newblock \bibinfo{title}{Phenomenology of bubble-collapse-driven penetration
  of biomaterial-surrogate liquid-liquid interfaces}.
\newblock \bibinfo{journal}{Physical Review Fluids} \bibinfo{volume}{3}.
\bibitem[{Park and Munz(2005)}]{Park2005}
\bibinfo{author}{Park, J.H.}, \bibinfo{author}{Munz, C.D.},
  \bibinfo{year}{2005}.
\newblock \bibinfo{title}{Multiple pressure variables methods for fluid flow at
  all {{Mach}} numbers}.
\newblock \bibinfo{journal}{International Journal for Numerical Methods in
  Fluids} \bibinfo{volume}{49}, \bibinfo{pages}{905--931}.
\bibitem[{Quirk and Karni(1996)}]{Quirk1996}
\bibinfo{author}{Quirk, J.J.}, \bibinfo{author}{Karni, S.},
  \bibinfo{year}{1996}.
\newblock \bibinfo{title}{On the dynamics of a shock\textendash{}bubble
  interaction}.
\newblock \bibinfo{journal}{Journal of Fluid Mechanics} \bibinfo{volume}{318},
  \bibinfo{pages}{129}.
\bibitem[{Ranjan et~al.(2007)Ranjan, Niederhaus, Motl, Anderson, Oakley and
  Bonazza}]{Ranjan2007}
\bibinfo{author}{Ranjan, D.}, \bibinfo{author}{Niederhaus, J.},
  \bibinfo{author}{Motl, B.}, \bibinfo{author}{Anderson, M.},
  \bibinfo{author}{Oakley, J.}, \bibinfo{author}{Bonazza, R.},
  \bibinfo{year}{2007}.
\newblock \bibinfo{title}{Experimental {{Investigation}} of {{Primary}} and
  {{Secondary Features}} in {{High}}-{{Mach}}-{{Number Shock}}-{{Bubble
  Interaction}}}.
\newblock \bibinfo{journal}{Physical Review Letters} \bibinfo{volume}{98}.
\bibitem[{Ranjan et~al.(2011)Ranjan, Oakley and Bonazza}]{Ranjan2011}
\bibinfo{author}{Ranjan, D.}, \bibinfo{author}{Oakley, J.},
  \bibinfo{author}{Bonazza, R.}, \bibinfo{year}{2011}.
\newblock \bibinfo{title}{Shock-{{Bubble Interactions}}}.
\newblock \bibinfo{journal}{Annual Review of Fluid Mechanics}
  \bibinfo{volume}{43}, \bibinfo{pages}{117--140}.
\bibitem[{Roe(1986)}]{Roe1986}
\bibinfo{author}{Roe, P.}, \bibinfo{year}{1986}.
\newblock \bibinfo{title}{Characteristic-based schemes for the euler
  equations}.
\newblock \bibinfo{journal}{Annual Review of Fluid Mechanics}
  \bibinfo{volume}{18}, \bibinfo{pages}{337--365}.
\bibitem[{Saurel and Abgrall(1999)}]{Saurel1999a}
\bibinfo{author}{Saurel, R.}, \bibinfo{author}{Abgrall, R.},
  \bibinfo{year}{1999}.
\newblock \bibinfo{title}{A {{Simple Method}} for {{Compressible Multifluid
  Flows}}}.
\newblock \bibinfo{journal}{SIAM Journal on Scientific Computing}
  \bibinfo{volume}{21}, \bibinfo{pages}{1115--1145}.
\bibitem[{Saurel et~al.(2007)Saurel, Le~M{\'e}tayer, Massoni and
  Gavrilyuk}]{Saurel2007}
\bibinfo{author}{Saurel, R.}, \bibinfo{author}{Le~M{\'e}tayer, O.},
  \bibinfo{author}{Massoni, J.}, \bibinfo{author}{Gavrilyuk, S.},
  \bibinfo{year}{2007}.
\newblock \bibinfo{title}{Shock jump relations for multiphase mixtures with
  stiff mechanical relaxation}.
\newblock \bibinfo{journal}{Shock Waves} \bibinfo{volume}{16},
  \bibinfo{pages}{209--232}.
\bibitem[{Saurel and Pantano(2018)}]{Saurel2018}
\bibinfo{author}{Saurel, R.}, \bibinfo{author}{Pantano, C.},
  \bibinfo{year}{2018}.
\newblock \bibinfo{title}{Diffuse-{{Interface Capturing Methods}} for
  {{Compressible Two}}-{{Phase Flows}}}.
\newblock \bibinfo{journal}{Annual Review of Fluid Mechanics}
  \bibinfo{volume}{50}, \bibinfo{pages}{105--130}.
\bibitem[{Shukla(2014)}]{Shukla2014}
\bibinfo{author}{Shukla, R.K.}, \bibinfo{year}{2014}.
\newblock \bibinfo{title}{Nonlinear preconditioning for efficient and accurate
  interface capturing in simulation of multicomponent compressible flows}.
\newblock \bibinfo{journal}{Journal of Computational Physics}
  \bibinfo{volume}{276}, \bibinfo{pages}{508--540}.
\bibitem[{Shukla et~al.(2010)Shukla, Pantano and Freund}]{Shukla2010}
\bibinfo{author}{Shukla, R.K.}, \bibinfo{author}{Pantano, C.},
  \bibinfo{author}{Freund, J.B.}, \bibinfo{year}{2010}.
\newblock \bibinfo{title}{An interface capturing method for the simulation of
  multi-phase compressible flows}.
\newblock \bibinfo{journal}{Journal of Computational Physics}
  \bibinfo{volume}{229}, \bibinfo{pages}{7411--7439}.
\bibitem[{Shyue(2006)}]{Shyue2006}
\bibinfo{author}{Shyue, K.M.}, \bibinfo{year}{2006}.
\newblock \bibinfo{title}{A volume-fraction based algorithm for hybrid
  barotropic and non-barotropic two-fluid flow problems}.
\newblock \bibinfo{journal}{Shock Waves} \bibinfo{volume}{15},
  \bibinfo{pages}{407--423}.
\bibitem[{Terashima and Tryggvason(2009)}]{Terashima2009}
\bibinfo{author}{Terashima, H.}, \bibinfo{author}{Tryggvason, G.},
  \bibinfo{year}{2009}.
\newblock \bibinfo{title}{A front-tracking/ghost-fluid method for fluid
  interfaces in compressible flows}.
\newblock \bibinfo{journal}{Journal of Computational Physics}
  \bibinfo{volume}{228}, \bibinfo{pages}{4012--4037}.
\bibitem[{Tian et~al.(2011)Tian, Toro and Castro}]{Tian2011}
\bibinfo{author}{Tian, B.}, \bibinfo{author}{Toro, E.},
  \bibinfo{author}{Castro, C.}, \bibinfo{year}{2011}.
\newblock \bibinfo{title}{A path-conservative method for a five-equation model
  of two-phase flow with an {{HLLC}}-type {{Riemann}} solver}.
\newblock \bibinfo{journal}{Computers \& Fluids} \bibinfo{volume}{46},
  \bibinfo{pages}{122--132}.
\bibitem[{Tokareva and Toro(2010)}]{Tokareva2010}
\bibinfo{author}{Tokareva, S.}, \bibinfo{author}{Toro, E.},
  \bibinfo{year}{2010}.
\newblock \bibinfo{title}{{{HLLC}}-type {{Riemann}} solver for the
  {{Baer}}\textendash{{Nunziato}} equations of compressible two-phase flow}.
\newblock \bibinfo{journal}{Journal of Computational Physics}
  \bibinfo{volume}{229}, \bibinfo{pages}{3573--3604}.
\bibitem[{Toro et~al.(1994)Toro, Spruce and Speares}]{Toro1994}
\bibinfo{author}{Toro, E.F.}, \bibinfo{author}{Spruce, M.},
  \bibinfo{author}{Speares, W.}, \bibinfo{year}{1994}.
\newblock \bibinfo{title}{Restoration of the contact surface in the
  {{HLL}}-{{Riemann}} solver}.
\newblock \bibinfo{journal}{Shock Waves} \bibinfo{volume}{4},
  \bibinfo{pages}{25--34}.
\bibitem[{Turkel(2006)}]{Turkel2006}
\bibinfo{author}{Turkel, E.}, \bibinfo{year}{2006}.
\newblock \bibinfo{title}{Numerical {{Methods}} and {{Nature}}}.
\newblock \bibinfo{journal}{Journal of Scientific Computing}
  \bibinfo{volume}{28}, \bibinfo{pages}{549--570}.
\bibitem[{Turkel et~al.(1993)Turkel, Fiterman and {van Leer}}]{Turkel1993}
\bibinfo{author}{Turkel, E.}, \bibinfo{author}{Fiterman, A.},
  \bibinfo{author}{{van Leer}, B.}, \bibinfo{year}{1993}.
\newblock \bibinfo{title}{Preconditioning and the {{Limit}} to the
  {{Incompressible Flow Equations}}}.
\newblock \bibinfo{type}{Technical Report}. {NASA CR-191500}.
\bibitem[{Ubbink and Issa(1999)}]{Ubbink1999}
\bibinfo{author}{Ubbink, O.}, \bibinfo{author}{Issa, R.}, \bibinfo{year}{1999}.
\newblock \bibinfo{title}{A {{Method}} for {{Capturing Sharp Fluid Interfaces}}
  on {{Arbitrary Meshes}}}.
\newblock \bibinfo{journal}{Journal of Computational Physics}
  \bibinfo{volume}{153}, \bibinfo{pages}{26--50}.
\bibitem[{{van der Heul} et~al.(2003){van der Heul}, Vuik and
  Wesseling}]{vanderHeul2003}
\bibinfo{author}{{van der Heul}, D.}, \bibinfo{author}{Vuik, C.},
  \bibinfo{author}{Wesseling, P.}, \bibinfo{year}{2003}.
\newblock \bibinfo{title}{A conservative pressure-correction method for flow at
  all speeds}.
\newblock \bibinfo{journal}{Computers \& Fluids} \bibinfo{volume}{32},
  \bibinfo{pages}{1113--1132}.
\bibitem[{Van~Doormaal et~al.(1987)Van~Doormaal, Raithby and
  McDonald}]{VanDoormaal1987}
\bibinfo{author}{Van~Doormaal, J.}, \bibinfo{author}{Raithby, G.},
  \bibinfo{author}{McDonald, B.}, \bibinfo{year}{1987}.
\newblock \bibinfo{title}{The {{Segregated Approach}} to {{Predicting Viscous
  Compressible Fluid Flows}}}.
\newblock \bibinfo{journal}{ASME Journal of Turbomachinery}
  \bibinfo{volume}{109}, \bibinfo{pages}{268--277}.
\bibitem[{Wang et~al.(2006)Wang, Liu and Khoo}]{Wang2006a}
\bibinfo{author}{Wang, C.W.}, \bibinfo{author}{Liu, T.G.},
  \bibinfo{author}{Khoo, B.C.}, \bibinfo{year}{2006}.
\newblock \bibinfo{title}{A {{Real Ghost Fluid Method}} for the {{Simulation}}
  of {{Multimedium Compressible Flow}}}.
\newblock \bibinfo{journal}{SIAM Journal on Scientific Computing}
  \bibinfo{volume}{28}, \bibinfo{pages}{278--302}.
\bibitem[{Wesseling(2001)}]{Wesseling2001}
\bibinfo{author}{Wesseling, P.}, \bibinfo{year}{2001}.
\newblock \bibinfo{title}{Principles of {{Computational Fluid Dynamics}}}.
\newblock \bibinfo{publisher}{{Springer}}.
\bibitem[{Wong and Lele(2017)}]{Wong2017}
\bibinfo{author}{Wong, M.L.}, \bibinfo{author}{Lele, S.K.},
  \bibinfo{year}{2017}.
\newblock \bibinfo{title}{High-order localized dissipation weighted compact
  nonlinear scheme for shock- and interface-capturing in compressible flows}.
\newblock \bibinfo{journal}{Journal of Computational Physics}
  \bibinfo{volume}{339}, \bibinfo{pages}{179--209}.
\bibitem[{Xiang and Wang(2017)}]{Xiang2017}
\bibinfo{author}{Xiang, G.}, \bibinfo{author}{Wang, B.}, \bibinfo{year}{2017}.
\newblock \bibinfo{title}{Numerical study of a planar shock interacting with a
  cylindrical water column embedded with an air cavity}.
\newblock \bibinfo{journal}{Journal of Fluid Mechanics} \bibinfo{volume}{825},
  \bibinfo{pages}{825--852}.
\bibitem[{Xiao et~al.(2017)Xiao, Denner and {van Wachem}}]{Xiao2017}
\bibinfo{author}{Xiao, C.N.}, \bibinfo{author}{Denner, F.},
  \bibinfo{author}{{van Wachem}, B.}, \bibinfo{year}{2017}.
\newblock \bibinfo{title}{Fully-coupled pressure-based finite-volume framework
  for the simulation of fluid flows at all speeds in complex geometries}.
\newblock \bibinfo{journal}{Journal of Computational Physics}
  \bibinfo{volume}{346}, \bibinfo{pages}{91--130}.
\bibitem[{Xiao(2004)}]{Xiao2004}
\bibinfo{author}{Xiao, F.}, \bibinfo{year}{2004}.
\newblock \bibinfo{title}{Unified formulation for compressible and
  incompressible flows by using multi-integrated moments {{I}}: One-dimensional
  inviscid compressible flow}.
\newblock \bibinfo{journal}{Journal of Computational Physics}
  \bibinfo{volume}{195}, \bibinfo{pages}{629--654}.
\bibitem[{Yoo and Sung(2018)}]{Yoo2018}
\bibinfo{author}{Yoo, Y.L.}, \bibinfo{author}{Sung, H.G.},
  \bibinfo{year}{2018}.
\newblock \bibinfo{title}{Numerical investigation of an interaction between
  shock waves and bubble in a compressible multiphase flow using a diffuse
  interface method}.
\newblock \bibinfo{journal}{International Journal of Heat and Mass Transfer}
  \bibinfo{volume}{127}, \bibinfo{pages}{210--221}.
\bibitem[{Zhai et~al.(2011)Zhai, Si, Luo and Yang}]{Zhai2011}
\bibinfo{author}{Zhai, Z.}, \bibinfo{author}{Si, T.}, \bibinfo{author}{Luo,
  X.}, \bibinfo{author}{Yang, J.}, \bibinfo{year}{2011}.
\newblock \bibinfo{title}{On the evolution of spherical gas interfaces
  accelerated by a planar shock wave}.
\newblock \bibinfo{journal}{Physics of Fluids} \bibinfo{volume}{23},
  \bibinfo{pages}{084104}.

\end{thebibliography}

\end{document}